\documentclass[11pt,letterpaper]{article}
\usepackage[T1]{fontenc}
\usepackage[utf8]{inputenc}
\usepackage{lmodern}
\usepackage[margin=1in]{geometry}
\usepackage{amsmath,amssymb,amsthm,mathtools}
\usepackage{aliascnt}
\usepackage{booktabs,array,microtype}
\usepackage[backend=biber,style=alphabetic,maxbibnames=99]{biblatex}
\usepackage[colorlinks=true,allcolors=blue]{hyperref}
\usepackage[nameinlink,noabbrev]{cleveref}

\newtheorem{theorem}{Theorem}[section]
\newaliascnt{lemma}{theorem}
\newtheorem{lemma}[lemma]{Lemma}
\aliascntresetthe{lemma}
\newaliascnt{proposition}{theorem}
\newtheorem{proposition}[proposition]{Proposition}
\aliascntresetthe{proposition}
\newaliascnt{corollary}{theorem}
\newtheorem{corollary}[corollary]{Corollary}
\aliascntresetthe{corollary}
\theoremstyle{definition}

\theoremstyle{remark}

\crefname{theorem}{Theorem}{Theorems}
\crefname{lemma}{Lemma}{Lemmas}
\crefname{proposition}{Proposition}{Propositions}
\crefname{corollary}{Corollary}{Corollaries}
\crefname{section}{Section}{Sections}
\crefname{appendix}{Appendix}{Appendices}
\crefname{table}{Table}{Tables}
\crefname{equation}{equation}{equations}
\newcommand{\R}{\mathbb R}
\newcommand{\E}{\mathbb E}
\newcommand{\Prob}{\mathbb P}
\newcommand{\ind}{\mathbf 1}
\newcommand{\Unif}{\operatorname{Unif}}
\newcommand{\Reg}{\operatorname{Reg}}
\newcommand{\Rat}{\operatorname{Rat}}
\newcommand{\softO}{\widetilde O}

\newcommand{\abs}[1]{\left|#1\right|}

\title{Efficient Equilibrium Computation in Colonel Blotto Games}
\author{%
    Martin Bichler\\
    {\small Technical University of Munich, Germany}\\
    {\small\href{mailto:m.bichler@tum.de}{\texttt{m.bichler@tum.de}}}
    \and
    Abheek Ghosh\\
    {\small Technical University of Munich, Germany}\\
    {\small\href{mailto:abheek.ghosh@tum.de}{\texttt{abheek.ghosh@tum.de}}}
}
\date{}

\begin{document}
\maketitle

\begin{abstract}
Colonel Blotto is a fundamental resource-allocation game in which two players divide their budgets across battlefields, and the higher allocation on each battlefield wins its value.
The problem of computing an equilibrium efficiently has remained unresolved for its classical continuous version with arbitrary budgets and with values that may vary across battlefields but are the same for both players.
We give a fully polynomial-time approximation scheme that computes an $\varepsilon$-Nash equilibrium in time polynomial in the input length and $1/\varepsilon$, and nearly linear in the input length for fixed $\varepsilon$.
For sufficiently balanced budgets and battlefield values, we also compute exact equilibria in linear time.
\end{abstract}

\section{Introduction}
\label{sec:intro}

Colonel Blotto is a fundamental model of competition with limited resources, dating back to Borel's work in 1921~\cite{borel1921theorie}.
In the classical game, two players allocate their budgets of a divisible resource across several battlefields.
Each player's total allocation must not exceed its budget.
The players may have different budgets, and values may vary across battlefields, but each battlefield has the same value for both players.
On each battlefield, the higher allocation wins its value, and a tie splits that value equally.
Each player seeks to maximize the total value it wins.

Despite more than a century of study, computing approximate equilibria efficiently has remained an open problem for arbitrary budgets and battlefield values.
We resolve this question with a fully polynomial-time approximation scheme (FPTAS) that computes an $\varepsilon$-Nash equilibrium in time polynomial in the binary input length and $1/\varepsilon$.

One approach to constructing Blotto equilibria starts from the General Lotto relaxation, in which the budget constraint holds only in expectation.
Nash equilibria of General Lotto have been characterized~\cite{kovenock2021generalizations}.
To turn such an equilibrium into Blotto strategies, each player's bid distributions on individual battlefields must be coupled into a joint distribution that respects its budget on every draw.
\cite{perchet2024algorithmic} characterizes exactly when such couplings exist, shows that they yield a Blotto equilibrium, and gives an algorithm for computing and sampling an approximate equilibrium in this regime.
Outside this regime, no pair of feasible strategies has these marginals, and an algorithm must find strategies with different marginals.
Our approximation scheme removes this restriction.

Another natural approach is to discretize allocations and use an algorithm for computing equilibria in the discrete Blotto game~\cite{ahmadinejad2019duels,behnezhad2017faster,beaglehole2023sampling}.
However, a strategy on a grid assigns positive probability to individual bids, and an arbitrarily small increase above such a bid turns ties into wins.
We give a two-battlefield game in which, for arbitrarily fine uniform grids, the discretized game has an exact equilibrium from which one player gains $1/12$ of the total value by deviating to an allocation outside the grid (\cref{prop:general:grid}).
Thus, an exact equilibrium of a discretized game need not be an approximate equilibrium of the continuous game, however fine the grid.

\subsection{Our results}
\label{sec:intro:results}

For a game with $n$ battlefields, normalize the budgets to $1$ and $b \in (0, 1]$, and the positive common battlefield values $v_1, \ldots, v_n$ to sum to one.
The budgets and values are rational, as is the approximation parameter $0 < \varepsilon < 1$.
An $\varepsilon$-Nash equilibrium allows neither player to increase its expected payoff by more than $\varepsilon$ through any feasible deviation in the continuous game.
Our algorithms construct strategy descriptions from which each player can independently sample an allocation satisfying its budget.

We give bounds in both the finite-bit and real-RAM models.
The finite-bit model counts bit operations on binary input and uses independent random bits.
Our real-RAM model allows exact arithmetic and comparisons on real numbers and, for sampling, independent uniform draws from $[0,1]$, each at unit cost.

\paragraph{An FPTAS for arbitrary budgets and battlefield values.}
We give a deterministic algorithm for computing an $\varepsilon$-Nash equilibrium (\cref{thm:general}).
Let $L$ be the total binary encoding length of the game input and $\varepsilon$, and let $n_* = \min\{n, 1/\varepsilon\}$.
The algorithm constructs the strategy descriptions in
\[
    \softO\!\left(L + \frac{n_*^3}{\varepsilon^6}\right)
    \quad\text{bit operations, or}\quad
    \softO\!\left(n + \frac{n_*^2}{\varepsilon^5}\right)
    \quad\text{real-RAM operations}.
\]
Here $\softO$ suppresses polylogarithmic factors.
The construction takes time polynomial in the binary input length and $1/\varepsilon$, and nearly linear in the input length for fixed $\varepsilon$.
Sampling an allocation takes $O(n + \log(1/\varepsilon))$ real-RAM operations.
In the finite-bit model, sampling and writing a rational allocation take expected time polynomial in the input length and $1/\varepsilon$; \cref{thm:general} gives the detailed bound.

The approximation scheme searches for feasible strategies without fixing their marginal distributions in advance.
For each player, we construct a finite set of candidate allocations that can support an approximate equilibrium.
To check deviations against strategies supported on this set, we construct a second finite set of feasible allocations for the opponent.
For every mixture of the player's candidate allocations, this second set contains a best response in the continuous game.
We compute the strategies without enumerating either set, combining multiplicative weights with dynamic programming to select feasible allocations.

A preliminary reduction to $O(1/\varepsilon)$ battlefields gives the dependence on $n_*$.
Replacing $\varepsilon$ by a nearby power of two and rounding the values and budget ratio to numbers with short binary encodings make the dependence on $L$ additive.
Transferring these strategies back to the original game requires care because small bid changes can cause large payoff changes at ties.
We show how to preserve feasibility while controlling the loss in equilibrium accuracy.

We used ChatGPT to help optimize the FPTAS and improve its running-time bounds, including obtaining nearly linear dependence on the input length for fixed $\varepsilon$.

\paragraph{Exact equilibria from General Lotto marginals.}
Our second result gives a faster algorithm that computes an exact Nash equilibrium when
\[
    \max_{i \in [n]} v_i \le b / 2.
\]
This is precisely the condition under which the General Lotto equilibrium marginals can be coupled into feasible Blotto strategies~\cite[Corollary~1]{perchet2024algorithmic}.
For equal battlefield values, it becomes $b \ge 2 / n$.
In this regime, computing an exact Nash equilibrium and sampling each allocation both take $O(n)$ time in the real-RAM model (\cref{thm:prg:exact}).
The sampling bound matches the $\Omega(n)$ time needed to write all $n$ allocation coordinates.

The algorithm of \cite[Theorem~2]{perchet2024algorithmic} takes $O(n^2 + \varepsilon^{-4}\log(1/\varepsilon))$ arithmetic operations to sample from an $\varepsilon$-Nash equilibrium.
Its construction reduces the coupling problem to at most four random variables and then approximates their joint distribution by discretization and tensor scaling.
We give explicit exact couplings for these remaining variables and carry out the reduction in linear time per draw.
This removes both the quadratic dependence on $n$ and the approximation cost in the real-RAM model.

A sampler restricted to rational allocations cannot always sample from an exact equilibrium, even for rational inputs.
For example, with equal budgets and three equally valued battlefields, every equilibrium has continuous uniform marginal distributions~\cite[Theorem~2]{roberson2006blotto}.
Our exact sampler uses the real-RAM model's ability to draw independent uniform samples from $[0,1]$.

Our construction also has a finite-bit implementation that computes an $\varepsilon$-Nash equilibrium and samples rational allocations satisfying the budgets on every draw (\cref{cor:prg:finite}).
With $k = \lceil\log_2(1/\varepsilon)\rceil$, constructing the samplers takes $\softO(n L)$ bit operations, and each draw takes $\softO((n + k)(L + k))$ bit operations.

When the players assign different values to the same battlefield, our coupling construction applies under the sufficient condition given for this setting in \cite[Theorem~1]{perchet2024algorithmic}.
Using their parameter search, we can test this condition and, when it holds, compute an exact equilibrium in $O(n \log n)$ operations, with $O(n)$ operations per allocation sampled (\cref{thm:prg:specific}).
For the parameter search, we augment the real-RAM model with a unit-cost operation that returns the exact real roots of a polynomial of degree at most three.
With each player's values normalized to sum to one, a finite-bit implementation computes an $\varepsilon$-Nash equilibrium and samples allocations from it in time polynomial in the input length and $\log(1/\varepsilon)$.
The FPTAS discussed earlier (\cref{thm:general}) requires each battlefield to have the same value for both players.

\subsection{Further related work}
\label{sec:intro:related}

Classical work constructs and characterizes equilibria in several continuous Blotto models.
Gross and Wagner~\cite{gross1950continuous} construct equilibria for two battlefields with arbitrary budgets and for any number of equally valued battlefields with equal budgets.
Macdonell and Mastronardi~\cite{macdonell2015waging} characterize the full set of two-battlefield equilibria, and Roberson~\cite{roberson2006blotto,roberson2010errata} studies games with equal battlefield values and asymmetric budgets.
Thomas~\cite{thomas2018heterogeneous} studies games with heterogeneous common values and equal budgets.

The coupling approach of \cite{perchet2024algorithmic} combines General Lotto equilibria~\cite{kovenock2021generalizations} with conditions under which given marginal distributions can be coupled to have a constant sum~\cite{wang2016mixability}.
Efficient sampling from continuous Blotto equilibria is also known in certain symmetric multiplayer settings~\cite{boix2021multiplayer}.
Both \cite{perchet2024algorithmic} and \cite{boix2021multiplayer} state that equilibrium existence follows from general theorems for discontinuous games.
The hypotheses of the theorems they cite can fail in our model, so we prove directly the minimax equality used by our approximation scheme (\cref{lem:general:minimax}).

Other approximation results for the continuous game give different guarantees.
The strategies of \cite{vu2020approximate} have an equilibrium error that decreases with the number of battlefields under certain conditions on the battlefield values.
For every constant $c$, \cite[Theorem~3.6]{behnezhad2019beyond} gives a polynomial-time approximation scheme for maximizing a player's guaranteed expected payoff among strategies supported on at most $c$ allocations.
Neither result gives, for every instance and every $\varepsilon > 0$, an $\varepsilon$-Nash equilibrium in time polynomial in the input length and $1/\varepsilon$.

For discrete Blotto, \cite{ahmadinejad2019duels,behnezhad2017faster} give exact equilibrium algorithms polynomial in the number of battlefields and the integer budgets.
The sampling and learning methods of \cite{beaglehole2023sampling} give $\varepsilon$-Nash equilibria with polynomial dependence on $1/\varepsilon$ and only polylogarithmic dependence on the budgets.
These bounds concern integer allocations and do not themselves control deviations in the continuous game.
In a different direction, \cite{bichler2026complexity} proves hardness of approximate equilibrium computation in discrete multiplayer games with player-specific values.
Our FPTAS, in contrast, concerns two players with common values.

\section{Model and preliminaries}
\label{sec:prelim}

There are two players, $A$ and $B$, and $n \ge 1$ battlefields indexed by $[n] = \{1, \ldots, n\}$.
Without loss of generality, players $A$ and $B$ have budgets $1$ and $b \in (0, 1]$, respectively.
Battlefield $i$ has a common value $v_i > 0$ for both players, with $\sum_{i=1}^n v_i = 1$.
For a budget $c \ge 0$, let
\[
    \mathcal X(c) = \left\{x \in \R_{\ge 0}^n : \sum_{i=1}^n x_i \le c\right\}.
\]
The pure strategy spaces of $A$ and $B$ are $\mathcal X(1)$ and $\mathcal X(b)$, respectively.
The player who allocates more resources to a battlefield receives its value.
In a tie, including when both allocations are zero, each player receives half its value.
Let $h(s, t)$ denote the share received by a player who bids $s$ against a bid $t$:
\[
    h(s, t) = \begin{cases}
        1, & s > t, \\
        1 / 2, & s = t, \\
        0, & s < t.
    \end{cases}
\]
The players' payoffs at $x \in \mathcal X(1)$ and $y \in \mathcal X(b)$ are
\begin{equation}
    u_A(x, y) = \sum_{i=1}^n v_i h(x_i, y_i),
    \qquad
    u_B(x, y) = \sum_{i=1}^n v_i h(y_i, x_i).
    \label{eq:prelim:payoffs}
\end{equation}
In particular, $u_A(x, y) + u_B(x, y) = 1$ for every pair of feasible allocations.

\paragraph{Nash equilibrium.}
A mixed strategy is a probability distribution on the player's pure strategy space.
We denote the strategies of $A$ and $B$ by $\mu_A$ and $\mu_B$.
For independent allocations $X \sim \mu_A$ and $Y \sim \mu_B$, the expected payoffs are
\[
    u_j(\mu_A, \mu_B) = \E[u_j(X, Y)],
    \qquad j \in \{A, B\}.
\]
We use $u_A(x, \mu_B)$ and $u_B(\mu_A, y)$ when one player chooses a pure strategy.
Each player's regret is the supremum increase in expected payoff from a unilateral deviation:
\begin{align}
    \Reg_A(\mu_A, \mu_B) &= \sup_{x \in \mathcal X(1)} u_A(x, \mu_B) - u_A(\mu_A, \mu_B), \notag\\
    \Reg_B(\mu_A, \mu_B) &= \sup_{y \in \mathcal X(b)} u_B(\mu_A, y) - u_B(\mu_A, \mu_B).
    \label{eq:prelim:regret}
\end{align}
For $\varepsilon \ge 0$, an $\varepsilon$-Nash equilibrium is a pair $(\mu_A, \mu_B)$ with both regrets at most $\varepsilon$; a Nash equilibrium is the case $\varepsilon = 0$.
The approximation is additive in the normalized payoffs.
Pure deviations suffice because the payoff of a mixed deviation is an average of the payoffs from pure deviations.
We use suprema since the payoff is discontinuous at ties and a best response need not be attained.

To bound these regrets, it suffices to consider deviations that spend the full budget.
Any deviation in $\mathcal X(c)$ can be made to spend $c$ by adding the unused budget to one coordinate, which weakly increases the payoff against every allocation of the opponent.

For a mixed strategy $\mu$, let $U(\mu; c)$ denote the supremum expected payoff of an opponent with budget $c \ge 0$:
\[
    U(\mu; c) = \sup_{z \in \mathcal X(c)}
        \E_{X \sim \mu}\!\left[\sum_{i=1}^n v_i h(z_i, X_i)\right].
\]
Since the game is constant-sum, adding the regret definitions gives
\begin{equation}
    \Reg_A(\mu_A, \mu_B) + \Reg_B(\mu_A, \mu_B) = U(\mu_B; 1) + U(\mu_A; b) - 1.
    \label{eq:prelim:security-sum}
\end{equation}
Since each regret is nonnegative, $U(\mu_B; 1) + U(\mu_A; b) \le 1 + \varepsilon$ implies that $(\mu_A, \mu_B)$ is an $\varepsilon$-Nash equilibrium.

\subsection{Marginal distributions and approximation}
\label{sec:prelim:marginals}

Expected payoffs depend only on the marginal distributions of allocations to individual battlefields, although feasibility still requires a joint distribution satisfying the budget constraint.
If the opponent's allocation to a battlefield is a random variable $Y$ with cumulative distribution function (CDF) $F$, then bidding $s$ on that battlefield gives expected share
\begin{equation}
    \E[h(s, Y)] = \Prob(Y < s) + \frac12 \Prob(Y = s) = \frac{F(s-) + F(s)}{2},
    \qquad F(s-) = \Prob(Y < s).
    \label{eq:prelim:half-tie}
\end{equation}
For a mixed strategy $\mu$, write $F_{\mu, i}(t) = \Prob_{X \sim \mu}(X_i \le t)$ for its marginal CDF on battlefield $i$.
We measure the difference between two mixed strategies $\mu$ and $\nu$ by
\[
    d(\mu, \nu) = \max_{i \in [n]} \sup_{t \in \R}
        |F_{\mu, i}(t) - F_{\nu, i}(t)|.
\]
The following lemma bounds changes in payoffs and regrets in terms of $d$. The proof is given in \cref{app:lem:prelim:cdf:proof}.

\begin{lemma}
\label{lem:prelim:cdf}
For any fixed pure or mixed strategy of one player, replacing the opponent's strategy $\mu$ by $\nu$ changes the expected payoff by at most $d(\mu, \nu)$.
If two strategy profiles $(\mu_A, \mu_B)$ and $(\mu'_A, \mu'_B)$ satisfy $d(\mu_A, \mu'_A) \le \delta_A$ and $d(\mu_B, \mu'_B) \le \delta_B$, then
\[
    \begin{aligned}
        \Reg_A(\mu'_A, \mu'_B) &\le \Reg_A(\mu_A, \mu_B) + \delta_A + 2 \delta_B, \\
        \Reg_B(\mu'_A, \mu'_B) &\le \Reg_B(\mu_A, \mu_B) + \delta_B + 2 \delta_A.
    \end{aligned}
\]
\end{lemma}

\subsection{Input, output, and running time}
\label{sec:prelim:computation}

The computational input consists of positive rational values $v_1, \ldots, v_n$ summing to one and a rational budget ratio $b \in (0, 1]$.
For approximate equilibria, the input also includes a rational approximation parameter $0 < \varepsilon < 1$.
We use a real-RAM model for the main analysis.
\Cref{app:computation:bits} describes the finite-bit model used for bit-complexity analysis.

\paragraph{Real-RAM model.}
Each addition, subtraction, multiplication, division, and comparison of exact real numbers has unit cost.
The model also provides independent exact samples from $\Unif[0, 1]$ at unit cost.
All numbers are stored exactly, including the input numbers and the coordinates of sampled allocations.
Arithmetic running times count these operations without charging for binary encoding lengths.
This follows the approach in \cite{perchet2024algorithmic}, whose analysis counts operations on real numbers and whose sampler uses continuous uniform draws.

\paragraph{Finite-bit model.}
The input is given in binary, and computation uses bit operations.
Algorithms obtain their randomness from independent fair coin flips, each counted as one bit operation.
Encoding each outcome as $0$ or $1$ gives one random bit per flip.
An algorithm returns a finite description of each player's strategy together with a sampler that produces rational allocations.
Any error introduced by finite precision or by using finitely many random bits is included in the stated equilibrium guarantee.

\paragraph{Output and running time.}
In both models, every sampled allocation must satisfy the player's budget constraint.
The equilibrium guarantee applies against every feasible continuous deviation.
We state separately the time to compute a strategy description and the time to sample an allocation from it.
For a Las Vegas algorithm, every returned strategy pair satisfies the stated guarantee; the stated time to compute the pair is an expectation over the algorithm's random choices for each fixed input.
For sampling, we specify whether the running-time and random-bit bounds hold in expectation or for every draw.
We sometimes use $\softO$ to suppress polylogarithmic factors in running-time bounds.

\section{Technical overview}
\label{sec:overview}

The Colonel Blotto game has two important properties.
The hard budget constraint couples a player's bids across battlefields: every sampled allocation must fit within the player's budget.
Expected payoffs, however, depend only on the marginal distributions of the bids on individual battlefields.
Both of our algorithms exploit this property of the expected payoffs while constructing joint distributions that satisfy the budget on every draw.
When the General Lotto marginals admit a feasible coupling, we construct one efficiently from these marginals.
For arbitrary budgets and common values, such a coupling need not exist.
Our general algorithm instead computes suitable marginals and feasible strategies together.

\subsection{From Lotto marginals to an exact sampler}
\label{sec:overview:coupling}

In the General Lotto relaxation, the budget constraints need to hold only in expectation.
Its equilibrium marginals have a particularly simple form~\cite{kovenock2021generalizations}:
\[
    X_i \sim \Unif[0, 2 v_i],
    \qquad
    Y_i \sim (1 - b)\delta_0 + b\Unif[0, 2 v_i],
\]
where $\delta_0$ is the point mass at zero.
We first discuss why feasible couplings of these marginals would form a Blotto equilibrium.
Since expected payoffs depend only on marginal distributions, the deviation bounds below hold regardless of how each player's bids are coupled.
Against $A$'s uniform bid on battlefield $i$, a bid $y_i$ by $B$ wins with probability at most $y_i / (2 v_i)$ and therefore earns at most $y_i / 2$.
Every feasible deviation by $B$ thus earns at most $b / 2$.
A similar calculation shows that every deviation by $A$ earns at most $1 - b / 2$.
The prescribed marginals attain these upper bounds against each other, so any feasible couplings of the two families form an exact equilibrium.

The remaining task is to couple each player's marginals so that every sampled allocation satisfies the budget.
Under these marginals, each player's expected total spending equals its budget, but sampling its bids independently across battlefields makes their total exceed the budget with positive probability.
In fact, a feasible coupling must spend the budget exactly, since a random total that never exceeds the budget and has the budget as its mean must equal it with probability one.
\cite{perchet2024algorithmic} shows that the required couplings exist precisely when $\max_i v_i \le b / 2$.
Their reduction shows that, to couple a player's $n$ bids so that they sum to its budget, it suffices to couple at most four random variables with that same sum.
They then compute an approximate coupling of these remaining variables.
We adapt this reduction to run in linear time per draw and replace this approximate coupling by explicit exact constructions.

\paragraph{Reducing to at most four variables.}
Consider one player with budget $c$ and marginals $(1 - p_i)\delta_0 + p_i\Unif[0, \ell_i]$, where $\ell_i \le c$ and $\sum_i p_i \ell_i = 2 c$.
The factor of two comes from the mean $\ell_i / 2$ of a uniform bid.
For the Lotto marginals, $\ell_i = 2 v_i$, so the condition $\ell_i \le c$ for player $B$ is the coupling condition $\max_i v_i \le b / 2$.
We first simplify these distributions by randomly rounding the mixture probabilities $p_i$ until all but at most one are zero or one.
At this stage, we are choosing conditional marginal distributions; the bids themselves will be sampled later.
Each step takes two probabilities $p_i$ and $p_j$ strictly between zero and one.
We can increase $p_i$ and decrease $p_j$, or do the reverse, keeping $p_i \ell_i + p_j \ell_j$ fixed.
In either direction, we stop as soon as one probability reaches zero or one, and randomly choose between the two outcomes so that each probability keeps its expectation.
Every step preserves $\sum_i p_i \ell_i = 2c$, so the conditional expected total spending remains $c$.
After at most $n - 1$ steps, each bid is either identically zero or uniform on its interval, except for at most one bid whose distribution still has both components.
Because the rounding preserves each probability in expectation, averaging these conditional distributions over the random choices restores the original marginals.

Next, we combine the uniformly distributed bids into groups, each represented by one variable.
If the bids in a group are uniform on intervals whose lengths sum to $L$, the variable representing the group is uniform on $[0,L]$.
We recover the original bids by splitting this variable's value in proportion to their interval lengths.
Each bid then has its required uniform distribution, and the bids sum to the group variable.

We form the groups greedily, starting a new group whenever adding the next interval would make its total length exceed $c$.
Because the interval lengths being grouped sum to at most $2c$, this produces at most three groups.
The group variables, together with any remaining variable whose distribution has both a uniform component and an atom at zero, give at most four variables to couple.
Grouping the bids and recovering them from the group variables take $O(n)$ operations.

\paragraph{Coupling the last few variables.}
Suppose three variables remain: two are uniformly distributed, and the third may also have an atom at zero.
We construct a mixture of uniform distributions on a constant number of line segments in the plane of allocations with sum $c$.
Every point on such a segment is feasible.
If we choose a point uniformly on a segment, each coordinate that varies along the segment is uniform between its two endpoint values.
A coordinate that stays zero contributes an atom at zero instead.
We choose the segments and their selection probabilities so that the resulting distribution of each coordinate is its prescribed marginal (\cref{lem:prg:three}).
Sampling then takes constant time: choose a segment with its assigned probability, then choose a point uniformly on it.

For four variables, a natural approach is to seek a similar mixture of uniform distributions on line segments in the hyperplane of allocations with sum $c$.
We do not have such a direct construction for the full parameter range, but we believe it is likely possible.
Instead, we build on the three-variable construction.
One approach is to split the four variables into two pairs and couple each pair with an auxiliary uniform variable so that each triple has a constant sum.
If we can construct both couplings with auxiliary variables uniform on the same interval, say $[0,d]$, we can make them complementary by taking them to be $Z$ and $d-Z$.
Their total is then fixed, so removing them from the two triples leaves the four desired variables with a constant sum.

The atom at zero can make this approach fail: the variable can be small on average while still taking large values.
Keeping the triple's total fixed requires the other two variables to compensate while retaining their uniform distributions.
For some marginals, this is impossible no matter which interval $[0,d]$ we choose for the auxiliary uniform variable.

We handle this remaining case by first choosing between the uniform component and the atom at zero.
The distributions of the other three variables can then depend on this choice, provided that averaging over the two cases restores their original uniform marginals.
When the uniform component is chosen, this variable that can have an atom at zero contributes more on average than under its original distribution.
We compensate by making the other three variables uniform on shorter intervals starting at zero, so that the expected total remains $c$.
All four conditional distributions are now uniform, so grouping reduces this case to the three-variable construction.

When the atom at zero is chosen, the other three variables must supply the entire budget.
To restore their original uniform marginals after averaging over both cases, their conditional densities must be higher near the upper ends of the intervals, where the shortened uniform distributions in the previous case contribute no probability.
The resulting densities have two constant levels.
Reflecting each variable about the midpoint of its interval moves the higher density next to zero.
This reflected distribution is a mixture of a uniform distribution on the full interval and another on a shorter interval starting at zero.
We apply the earlier rounding procedure to these mixture probabilities, preserving the sum of the expectations until at most one variable still has two uniform components.
The other two variables are uniform, and we give an explicit three-variable coupling for this form.
Reflecting the sampled values back then gives a coupling of the three conditional marginals with sum $c$ (\cref{sec:prg:nested,sec:prg:four}).

This conditioning construction can also fail: the three distributions obtained when the atom at zero is chosen may have no coupling with sum $c$.
However, it works whenever the auxiliary-variable construction fails, so the two constructions together cover all cases.

The complete construction computes the strategy descriptions and samples each allocation in $O(n)$ time in the real-RAM model.
For the finite-bit implementation, we replace the continuous uniform draws inside the sampler by draws from a fine rational grid.
The coupling formulas make the sampled bids sum to the budget for every choice of these random values, so this replacement preserves exact spending.
We bound the changes in marginal CDFs and apply \cref{lem:prelim:cdf} to obtain an approximate equilibrium.
Finally, the coupling construction depends only on the budget, interval endpoints, and mixture probabilities.
It therefore also applies to player-specific values whenever their Lotto marginals satisfy the sufficient condition in \cref{sec:prg:specific}.

\subsection{Computing strategies without prescribed marginals}
\label{sec:overview:general}

We now seek an approximate equilibrium for arbitrary budgets and common battlefield values.
Our algorithm constructs each strategy as a distribution on feasible allocations, without specifying its marginals in advance.
The constant-sum structure lets us compute the two players' strategies separately: for each player, we seek a strategy that makes the opponent's supremum payoff as small as possible.
The two best achievable bounds sum to one, so approximating both optimization problems gives an approximate equilibrium by \eqref{eq:prelim:security-sum}.
Payoff discontinuities at ties prevent a direct application of the usual minimax theorem.
We extend the argument to our setting using suitable perturbations of $A$'s allocations that respect its budget and restore the required continuity (\cref{lem:general:minimax}).

\paragraph{Reducing to a finite game without losing deviations.}
We describe the computation for one fixed player, whom we call \emph{the player}, and call the other \emph{the opponent}.
A natural first step is to restrict bids to a fine uniform grid.
As the example in \cref{prop:general:grid} shows, however, an equilibrium on such a grid can remain far from equilibrium in the continuous game.
An off-grid deviation can turn ties into wins on several battlefields by adding arbitrarily small amounts, whereas a deviation on the grid must pay a full grid step for each such win.
We therefore need a finite representation that preserves which combinations of ties and wins a continuous deviation can afford.

Suppose the player's strategy uses finitely many bids on each battlefield, and call these bids and zero the \emph{thresholds} on that battlefield.
Each bid of the opponent is equivalent, for payoffs, to bidding exactly at a threshold or just above one.
If the chosen thresholds are $t_1, \ldots, t_n$ and the opponent's budget is $c'$, these choices are affordable precisely when
\[
    \sum_i t_i \le c',
    \qquad\text{with strict inequality if any threshold must be exceeded.}
\]
Any positive slack suffices for all the required increments, but when the thresholds already sum to the budget, no increment is possible.
One feasible allocation for each affordable combination therefore represents every continuous deviation of the opponent, as far as its payoff against this strategy is concerned.

This gives a finite set of responses for the opponent once we have fixed the bids allowed in the player's strategy.
We next show that a finite set of bids, specified from the input alone, can support a nearly optimal strategy.
To see why, start from an unknown nearly optimal strategy and round each bid down to a quantile of its marginal distribution, using probability levels spaced $\Theta(\varepsilon)$ apart.
Rounding every coordinate down preserves the budget on each allocation, and only $O(1 / \varepsilon)$ bids remain on each battlefield.
The useful notion of accuracy here is probability mass, not distance between bid amounts: each marginal CDF changes by at most $O(\varepsilon)$.
By \cref{lem:prelim:cdf}, the opponent's payoff from every deviation consequently increases by at most $O(\varepsilon)$.

The quantile bids still depend on the unknown strategy and need not have short rational encodings.
To replace them by rationals, we preserve their order on each battlefield, including ties, and the comparison of every sum of one bid per battlefield with each budget.
These comparisons keep the player's allocations feasible and, by the affordability test, preserve exactly which combinations of ties and wins the opponent can afford.
Thus the bid amounts can change without changing the opponent's supremum payoff.
The comparisons form a system of linear equalities and inequalities with small integer coefficients.
A vertex argument gives a rational solution whose encoding length is polynomial in $n$, $1 / \varepsilon$, and the encoding length of the budget ratio.
We take a finite set of \emph{candidate bids} that includes all rationals in $[0, 1]$ up to the resulting denominator bound.
This set can be specified from the input without finding the unknown strategy or its quantiles (\cref{lem:general:candidates}).

The finite game now has two different sets of allocations.
The player's \emph{candidate allocations} are all feasible allocations made from candidate bids; some distribution on them is nearly optimal.
The opponent's \emph{response allocations} contain one allocation for every affordable combination of tying or just exceeding candidate bids.
They include a best response in the continuous game to every distribution on candidate allocations.
Solving this finite game approximately therefore produces the desired strategy for the player.
Repeating the computation with the roles reversed gives the other player's strategy.

\paragraph{Learning while enforcing the budget.}
Both sets of allocations are too large to enumerate.
Running multiplicative weights directly on them would require maintaining a weight for every complete allocation.
Payoffs are easier to evaluate: they depend only on bid distributions on individual battlefields.
This suggests running multiplicative weights on those distributions instead, but the resulting marginals need not admit a feasible coupling.

We use such marginals to guide the computation while building the output from feasible allocations.
For each side of the finite game, the algorithm maintains \emph{auxiliary marginals} with no budget constraint linking the battlefields.
In each iteration, it also selects a candidate allocation for the player and a response allocation for the opponent.
It returns the uniform distribution on the player's selected candidate allocations, so every draw satisfies the budget.
The challenge is to connect the auxiliary marginals, where payoffs are evaluated, to these selected allocations.

We make this connection through penalties based on CDF differences.
A basic test on one battlefield compares one side's auxiliary CDF at a threshold with the corresponding indicator for that side's selected bid: one if the bid is at most the threshold, and zero otherwise.
The test can take either sign, detecting an excess or a deficit of probability at or below that threshold.
A penalty averages such tests on each battlefield and combines the results with the battlefield values as weights.
Maximizing over the tests gives the value-weighted sum of the largest absolute CDF differences, which bounds the change in payoff when one family of marginals replaces the other.

In each iteration, each side selects a feasible allocation that approximately minimizes its current penalty.
The auxiliary marginals are then updated by multiplicative weights, with each side trading off a higher payoff against a smaller penalty.
The payoff terms in these updates are evaluated between the two families of auxiliary marginals.
Separate multiplicative-weights updates shift the penalty weights toward thresholds and signs that detect larger CDF differences.
Thus the payoff updates act on individual battlefields, while the allocation choices enforce the budget across battlefields.

Because the tests are linear in the CDFs, averaging any fixed choice of tests over iterations compares each side's average auxiliary CDFs with those of the uniform distribution on its selected allocations.
The combined regret bounds account for the payoff changes when we replace the average auxiliary marginals by the marginals of these feasible distributions.

This replacement is needed on both sides.
For the player, it transfers the guarantee to the feasible strategy that we actually return.
For the opponent, it lets us use the nearly optimal strategy from the quantile argument as a comparison in the regret analysis.
That strategy's guarantee holds only against feasible deviations, so it need not hold against the opponent's auxiliary marginals.
With both replacements, the regret bounds show that the returned strategy nearly minimizes the opponent's supremum payoff over response allocations.
Since these allocations represent all continuous deviations, this gives the desired guarantee in the original game (\cref{sec:general:source}).

The number of learning iterations depends only logarithmically on the number of possible bids.
To obtain an efficient algorithm, we must also select allocations and update weights without enumerating all bids or allocations.
Selecting an allocation is a knapsack problem: the penalty is a sum of battlefield terms, and the budget limits the total bid.
We solve this problem approximately by a dynamic program that rounds objective values to keep its table small but tests budgets exactly.

For the weight updates, we approximate the relevant CDFs by step functions with few breakpoints.
Together with the selected bids, these breakpoints partition the possible bids into intervals that receive the same approximate update.
Updating each interval at once avoids visiting every bid.
The analysis controls the errors from both approximate allocation choices and approximate updates, giving an FPTAS.

\subsection{Obtaining the sharper running time}
\label{sec:overview:efficiency}

Two preliminary reductions improve the dependence on the number of battlefields and on the input length.
Applied before the learning computation, they give the nearly linear computation time when $\varepsilon$ is fixed.
Each reduction must later be undone, so that the final strategies satisfy the original budgets and remain nearly optimal against deviations on the original battlefields.

\paragraph{Reducing the number of battlefields.}
Only $O(1 / \varepsilon)$ battlefields have values above a suitable multiple of $\varepsilon$.
We keep these and replace the others, of combined value $V$, by $k = \Theta(1 / \varepsilon)$ representative battlefields of value $V / k$ each, provided this reduces the number of battlefields.
The reduction leaves $O(\min\{n, 1 / \varepsilon\})$ battlefields for the main computation.
Afterwards, we must reconstruct bids on the replaced battlefields, while bids on the battlefields kept individually stay unchanged.
A natural reconstruction gives each replaced battlefield the bid on a uniformly chosen representative battlefield, scaled by the ratio of their values.
This gets the expected spending right but can violate the budget on a particular draw.

We enforce the budget with a step function on the circle $[0, 1)$ whose values on $k$ equal subintervals are the sorted representative bids, each multiplied by $k$.
Its integral is their total bid.
Partition the circle into arcs, giving each replaced battlefield $i$ an arc of length $v_i / V$, and rotate the partition by one uniformly random shift.
Each replaced battlefield receives its arc length times the minimum of the step function on its arc.
This bid is at most the integral of the step function over the arc, so the reconstructed bids sum to at most the total bid on the representative battlefields.

Taking the step function's value at the start of each arc instead would give the natural reconstruction, since that point is uniformly distributed on the circle.
Because the step function is sorted, the points where it is at most a given threshold form a single arc.
The minimum on a battlefield's arc is at most the threshold exactly when the arc overlaps this lower arc.
Compared with requiring the start of the arc to lie in the lower arc, this enlarges the set of favorable shifts by at most the battlefield's own arc length.
Thus each reconstructed bid's CDF differs from that of the natural reconstruction by at most the arc length, however large the gaps between representative bids.
Weighting these errors by battlefield values gives a payoff change of at most $\sum_i v_i^2 / V$ over the replaced battlefields, which is at most their largest value and hence $O(\varepsilon)$.

We must also compare the opponent's deviations in the two games.
On the replaced battlefields, form a distribution of the deviation's bids divided by their battlefield values, choosing each battlefield in proportion to its value.
This distribution determines the deviation's payoff against the natural reconstruction on those battlefields.
We round this distribution down to $k$ equally likely quantiles, assign them in random order to the representative battlefields, and rescale them by the representative values.
Rounding down preserves the budget, and the CDF approximation changes the payoff by only $O(\varepsilon)$.
Together, the two comparisons show that reconstruction increases the opponent's supremum payoff by only $O(\varepsilon)$ (\cref{lem:general:lift}).

\paragraph{Keeping long encodings out of the learning iterations.}
The values and the budget ratio may have long encodings, even after the number of battlefields is reduced.
Rounding the values is straightforward, since their $\ell_1$ error bounds the payoff change for every pair of allocations.
Rounding the budget ratio is more delicate.
Two very close budgets can afford different combinations of strict wins, so simply rescaling a strategy computed for a nearby ratio need not preserve its guarantee.

We therefore compute strategies for budgets $1$ and an auxiliary ratio $\rho$ with a short encoding, and then transfer them to the original budgets.
As in the candidate argument, we first round each computed strategy down to $O(1 / \varepsilon)$ quantile bids per battlefield.
We keep fixed which quantiles occur together in each allocation and change only their bid amounts.
The new amounts must preserve feasibility and the order of bids on each battlefield.
We also impose lower bounds on sums of quantile bids to keep the opponent from affording combinations of wins that would exceed its payoff bound in the auxiliary game, allowing an $O(\varepsilon)$ rounding error.
The rounded strategy satisfies these constraints at the auxiliary budgets.

These requirements are linear inequalities in the quantile bids, with small integer coefficients.
Determinant bounds show that, as the budget ratio varies, their feasibility can change only at rational ratios whose denominators are bounded in terms of the reduced number of battlefields and $1 / \varepsilon$.
Between these ratios, a feasible solution can be continued as the budget changes.
The denominator bound holds uniformly over all strategies, so we can choose $\rho$ with a short encoding before computing either strategy.
We choose $\rho$ so that any necessary adjustment can raise the lower bounds on these sums strictly above the opponent's target budget while keeping the constraints feasible.
Even tying the corresponding quantile bids is then unaffordable, which preserves the payoff guarantee (\cref{sec:general:transfer}).

The denominator bound for candidate bids depends on the denominator of the budget ratio, so using $\rho$ also reduces the number of learning iterations, even in the real-RAM model.
Together with the smaller bit cost per iteration, this leaves only an additive dependence on the original input length.

\section{Fast equilibrium construction from Lotto marginals}
\label{sec:prg}

We first consider games in which no battlefield accounts for too much of the total value relative to the smaller budget:
\begin{equation}
    \max_{i\in[n]} v_i \le b/2.
    \label{eq:prg:promise}
\end{equation}
For equal battlefield values, this condition becomes $b \ge 2/n$.
Under this condition, we construct an exact equilibrium by coupling the marginal distributions of a General Lotto equilibrium.
Given one distribution for each battlefield, a coupling is a joint distribution over allocations with those battlefield marginals.
The General Lotto game imposes the budget constraint only on expected total spending.
To obtain a Blotto strategy from its equilibrium marginals, we must couple them so that every sampled allocation satisfies the budget.

For every positive value vector and budget ratio, \cite[Corollary~1]{kovenock2021generalizations} gives the unique equilibrium marginals of General Lotto, recalled in \eqref{eq:prg:marginals} below.
This uniqueness concerns the distributions on individual battlefields; the joint distributions across battlefields need not be unique.
The result holds whether or not \eqref{eq:prg:promise} is satisfied.

The additional requirement for Blotto is feasibility under the hard budgets.
\cite[Corollary~1]{perchet2024algorithmic} proves that these Lotto marginals admit couplings satisfying the hard budgets if and only if \eqref{eq:prg:promise} holds, and such couplings form a Blotto equilibrium.
The proof uses joint mixability results to establish the existence of constant-sum couplings \cite{wang2016mixability}.
When the condition fails, these marginals cannot be realized by feasible Blotto strategies; equilibria with different marginals may still exist.

Under condition~\eqref{eq:prg:promise}, the algorithm of \cite{perchet2024algorithmic} reduces the coupling problem to at most four variables, discretizes their distributions, and computes an approximate coupling by tensor scaling.
It computes an $\varepsilon$-Nash equilibrium and draws an allocation using a total of $O(n^2+\varepsilon^{-4}\log(1/\varepsilon))$ arithmetic operations \cite[Theorem~2]{perchet2024algorithmic}.
We replace the discretization and tensor-scaling stage with explicit exact couplings of the at most four variables that remain after the reduction.
We also carry out the random choices in the reduction as we draw each allocation, so that computing the strategy descriptions and drawing each allocation take $O(n)$ operations.

\begin{theorem}
\label{thm:prg:exact}
Suppose \eqref{eq:prg:promise} holds.
In the real-RAM model of \cref{sec:prelim:computation}, an exact Nash equilibrium can be computed in $O(n)$ time, with $O(n)$ time per allocation draw.
Every sampled allocation spends the player's budget exactly.
The equilibrium payoffs are $1-b/2$ for player $A$ and $b/2$ for player $B$.
\end{theorem}

\Cref{cor:prg:finite} gives samplers for an $\varepsilon$-Nash equilibrium with rational allocations satisfying the hard budgets on every draw.
Computing the samplers takes $\softO(nL)$ bit operations, and generating each allocation takes $\softO((n+k)(L+k))$ bit operations in the worst case.
Here $L$ is the total binary input length, including the approximation parameter, and $k=\lceil\log_2(1/\varepsilon)\rceil$.

\subsection{Algorithm and proof of Theorem~\ref{thm:prg:exact}}
\label{sec:prg:proof}

We first recall why feasible couplings of the Lotto marginals form an equilibrium.
We then adapt the reduction of \cite{perchet2024algorithmic} to obtain a coupling problem with at most three uniform variables and one variable whose distribution is a mixture of a uniform and an atom at zero.
The main task is to couple these remaining variables so that their sum is always the budget.
For three variables, we construct a finite collection of line segments, each joining two allocations that spend the budget exactly.
We select a segment with a prescribed probability and sample a point uniformly along it.
For four variables, we either combine two such couplings or condition on whether the fourth variable is zero.
\Cref{sec:prg:sampler} puts these constructions together and proves the linear running time.

\subsubsection{Marginal distributions and equilibrium payoffs}
\label{sec:prg:security}

The unique equilibrium marginals of the common-value Lotto game are \cite[Corollary~1]{kovenock2021generalizations}:
\begin{equation}
    X_i\sim\Unif[0,\ell_i],\qquad
    Y_i\sim(1-b)\delta_0+b\Unif[0,\ell_i],\qquad
    \ell_i=2v_i,
    \label{eq:prg:marginals}
\end{equation}
where $\delta_0$ denotes the point mass at zero.
The expected total spending is $1$ for player $A$ and $b$ for player $B$.
Condition~\eqref{eq:prg:promise} says that every endpoint $\ell_i$ is at most $b$.
It is necessary for a feasible coupling of $B$'s marginals: if $\ell_i>b$, that marginal assigns positive probability to an infeasible bid.
The construction below gives an explicit sampler whenever this known coupling condition holds.

We recall why any pair of feasible strategies with these marginals forms a Nash equilibrium.
The resulting deviation bounds are also used in the finite-bit analysis in \cref{app:prg:bits}.
Consider any deviation $y\in\mathcal X(b)$ against $A$'s marginals.
Since they are atomless, its payoff is
\[
    \sum_i v_i\min\{y_i/\ell_i,1\}
    \le \frac12\sum_i y_i\le b/2.
\]
Against $B$'s marginals, a bid $x_i\ge0$ on battlefield $i$ earns at most
\[
    v_i(1-b)+\frac b2 x_i.
\]
For $0<x_i\le\ell_i$, this is the exact expected payoff.
At $x_i=0$, player $A$ ties with probability $1-b$ and otherwise loses, giving expected payoff $v_i(1-b)/2$.
For $x_i>\ell_i$, player $A$ wins with probability one and receives $v_i$, which is at most the displayed bound.
Summing over any $x\in\mathcal X(1)$ gives a deviation payoff at most $1-b/2$.
These two bounds sum to one, so \eqref{eq:prelim:security-sum} proves that any pair of feasible couplings of \eqref{eq:prg:marginals} forms an exact Nash equilibrium.
The same upper bounds apply to the players' expected payoffs, since these are averages over feasible pure allocations.
Since these payoffs sum to one, they must equal $1-b/2$ for $A$ and $b/2$ for $B$.

\subsubsection{Reducing the coupling to at most four variables}
\label{sec:prg:reduction}

We give the construction for a general budget $B>0$ and marginal distributions
\begin{equation}
    Z_i\sim(1-p_i)\delta_0+p_i\Unif[0,\ell_i],\qquad
    0\le p_i\le1,\qquad
    0<\ell_i\le B,\qquad \sum_i p_i\ell_i=2B.
    \label{eq:prg:coupling-condition}
\end{equation}
Both players in \eqref{eq:prg:marginals} satisfy these conditions.
Here $p_i$ is the probability of using the uniform component.
The reduction of \cite{perchet2024algorithmic} first expresses these marginals as a mixture in which all but at most one coordinate are either identically zero or uniform.
It then combines the uniform coordinates into at most three groups.
Their algorithm computes the mixture components and their weights in advance.
We implement the reduction in $O(n)$ operations per allocation draw by using pairwise randomized rounding to select the conditional marginals directly and a single scan to form the groups.
After coupling the resulting at most four variables, we split each group back into its original coordinates.

\paragraph{Rounding the mixture probabilities.}
Set $r_i=p_i\ell_i$, so $0\le r_i\le\ell_i$ and $\sum_i r_i=2B$.
We randomly change the $r_i$ while preserving their sum and each coordinate's expectation, until all but at most one belong to $\{0,\ell_i\}$.
While $0<r_i<\ell_i$ and $0<r_j<\ell_j$ for two indices $i,j$, put
\[
    t=r_i+r_j,\qquad a=\max\{0,t-\ell_j\},\qquad c=\min\{\ell_i,t\}.
\]
Set $r_i'=c$ with probability $(r_i-a)/(c-a)$ and $r_i'=a$ otherwise, and put $r_j'=t-r_i'$.
The strict inequalities for $r_i$ and $r_j$ imply $a<r_i<c$, so the denominator is positive.
Conditional on the current vector $r$, we have
\[
    \begin{aligned}
        \E[r_i'\mid r]&=\frac{r_i-a}{c-a}\,c+\frac{c-r_i}{c-a}\,a=r_i,\\
        \E[r_j'\mid r]&=t-\E[r_i'\mid r]=r_j.
    \end{aligned}
\]
We replace $(r_i,r_j)$ by $(r_i',r_j')$, keeping the notation $r$ for the updated vector.
This update preserves the sum and each conditional expectation, and sends at least one of the two values to an endpoint of its interval.
Maintaining a list of indices with $0<r_i<\ell_i$ gives at most $n-1$ updates and $O(n)$ operations.

Given the final vector $r$, we will couple the conditional marginals $(1-r_i/\ell_i)\delta_0+(r_i/\ell_i)\Unif[0,\ell_i]$ with sum $B$.
We construct these couplings below, first grouping the uniform coordinates and then coupling the resulting at most four variables.
Since $\E[r_i]=p_i\ell_i$, averaging over the rounding choices recovers each original marginal in \eqref{eq:prg:coupling-condition}.
The sum will therefore be $B$ for every draw, while the unconditional marginals remain as required.

\paragraph{Grouping uniform coordinates.}
We set coordinates with $r_i=0$ to zero and set aside the coordinate with $0<r_i<\ell_i$, if there is one.
The remaining coordinates, with $r_i=\ell_i$, are uniform on $[0,\ell_i]$.
We group these coordinates in their original order, keeping the sum of interval lengths in each group at most $B$.
Whenever adding the next coordinate would exceed this bound, we start a new group with that coordinate.
This produces at most three groups.
The first coordinate in each new group did not fit in the preceding group, so these two groups have combined length greater than $B$.
If there were four groups, the first two and the next two would together have total length greater than $2B$.
This is impossible because all the lengths being grouped sum to at most $2B$.

We replace each group $G$ by one variable $Y_G\sim\Unif[0,\sum_{j\in G}\ell_j]$.
After sampling $Y_G$, we recover the original coordinates by setting
\[
    Z_i=\frac{\ell_i}{\sum_{j\in G}\ell_j}Y_G\qquad(i\in G).
\]
Each $Z_i$ is uniform on $[0,\ell_i]$, and the coordinates in $G$ sum to $Y_G$.
Grouping and splitting take $O(n)$ operations.

If the coordinate set aside has mixture probability $p\in(0,1)$ and endpoint $\ell$, the grouped lengths sum to $2B-p\ell>B$.
There are therefore two or three uniform variables, together with the remaining mixture $(1-p)\delta_0+p\Unif[0,\ell]$.
If no coordinate was set aside, the grouped lengths sum to $2B$.
If there are two groups, both must have length $B$, and we sample them as $(BU,B(1-U))$, where $U\sim\Unif[0,1]$.
If there are three groups, we use the next construction with $p=1$.
In \cref{sec:prg:three}, we construct an $O(1)$-time sampler for three variables.
We then develop an auxiliary coupling in \cref{sec:prg:nested} and combine these constructions in \cref{sec:prg:four} to obtain an $O(1)$-time sampler for four variables.

\subsubsection{A coupling for three variables}
\label{sec:prg:three}

We begin with the three-variable case of the reduction, where each interval length is at most the budget $B$.
The first two variables are uniform; the third has an atom at zero and a uniform component.
We construct a mixture of uniform distributions on line segments in the plane of allocations with sum $B$.
A segment with endpoints $P,Q$ and weight $\alpha$ is sampled by choosing it with probability $\alpha$ and then returning $(1-U)P+UQ$ for an independent $U\sim\Unif[0,1]$.
Segments with zero weight are discarded.
We choose the segment weights to obtain the marginal distributions in the following lemma.
Setting the atom's weight to zero also gives a coupling of three uniform variables, used later in \cref{sec:prg:four}.

\begin{lemma}
\label{lem:prg:three}
Let $B>0$, $0<a,d,c\le B$, and $0<p\le1$ satisfy $a+d+pc=2B$.
There is an explicit mixture of uniform distributions on at most seven line segments with marginals $\Unif[0,a]$, $\Unif[0,d]$, and $(1-p)\delta_0+p\Unif[0,c]$, supported on allocations with sum $B$.
It can be sampled in $O(1)$ time.
\end{lemma}

\begin{proof}
Swap the first two coordinates if necessary so that $a\le d$.
Put
\[
    u=B-a,\quad v=B-d,\quad w=B-c,\quad
    h=B-u-v-w.
\]
Then $u\ge v\ge0$, $w,h\ge0$, $a=v+w+h$, $d=u+w+h$, $c=u+v+h$, and $p=(u+v)/c$.
The six vertices below have coordinate sum $B$:
\[
\begin{array}{lll}
 P_1=(0,w,c),& P_2=(w,0,c),&P_3=(a,0,u),\\
 P_4=(a,u,0),& P_5=(v,d,0),&P_6=(0,d,v).
\end{array}
\]
Write $E=w+h$, $F=u+h$, and $G=v+h$.
We first handle $E=0$, which implies $w=h=0$, $p=1$, and $a+d=c=B$.
For $U\sim\Unif[0,1]$, return $(aU,dU,c(1-U))$.
This samples the segment from $(0,0,c)$ to $(a,d,0)$, with the required uniform marginals and sum $B$.

Now assume $E>0$.
The lemma statement gives $a,c>0$, and $u+v=pc>0$ together with $u\ge v\ge0$ implies $u>0$, hence $F=u+h>0$.
Thus all denominators in the segment weights in the tables below are positive.
Since $u\ge v$, we distinguish three cases according to the position of $w$ relative to $u$ and $v$.

\paragraph{First case: $w\le v\le u$.}
Set $R=h^2+h(u+w)+w(u+v)$ and $D=acEF$.
We use the following seven segments and weights:
\[
\begin{array}{c|c@{\qquad}c|c}
\text{segment}&\text{weight}&\text{segment}&\text{weight}\\\hline
P_1P_3&wGR/D&P_1P_4&(u-w)GR/D\\
P_2P_5&(v-w)h/(cE)&P_2P_6&wh/(cE)\\
P_3P_6&w(u-v)R/D&P_4P_5&h/c\\
P_4P_6&vw(u+v+2h)/(acF)&&
\end{array}
\]
The weights are nonnegative under the case inequalities.
We will show below that they sum to one and give the required marginal distributions.
Using these weights as probabilities, the sampler chooses a segment, samples uniformly on it, and returns the three coordinates of the sampled point.
Every such point has coordinate sum $B$.
In each case, write $q_{ij}$ for the weight assigned to $P_iP_j$, with weight zero for any segment absent from the relevant table.

When some parameters are zero or equal, some segment weights may vanish.
In interpreting the formulas below, we disregard components of weight zero and omit checks of the continuous density on empty intervals.
We first verify the first marginal.
Reading the endpoints in the first coordinate of all seven segments gives the weighted sum
\[
\begin{aligned}
    &q_{13}\Unif[0,a]+q_{14}\Unif[0,a]+q_{25}\Unif[w,v]\\
    &\quad+q_{26}\Unif[0,w]+q_{36}\Unif[0,a]+q_{45}\Unif[v,a]
    +q_{46}\Unif[0,a].
\end{aligned}
\]
Each term contributes its weight divided by its interval length to the density on that interval.
In particular, the terms on $[0,w]$, $[w,v]$, and $[v,a]$ have equal densities:
\[
    \frac{q_{26}}w=\frac{q_{25}}{v-w}
    =\frac{q_{45}}{a-v}=\frac{h}{cE},
\]
where $a-v=E$.
These intervals cover $[0,a]$ with disjoint interiors, so the three terms combine to $(ah/(cE))\Unif[0,a]$.
Thus the full weighted sum is
\[
    \left(q_{13}+q_{14}+q_{36}+q_{46}+\frac{ah}{cE}\right)\Unif[0,a]
    =\Unif[0,a],
\]
where the last equality follows by substituting the table weights.
This proves the required first marginal.
Each retained uniform distribution has total mass one, so the calculation also shows that the segment weights sum to one.

For the second coordinate, the densities on $(0,w)$, $(w,u)$, and $(u,d)$ are
\[
    \frac{q_{25}+q_{26}+q_{36}}d+\frac{q_{13}}w
    =\frac{q_{25}+q_{26}+q_{36}}d+\frac{q_{14}}{u-w}
    =\frac{q_{25}+q_{26}+q_{36}}d+\frac{q_{45}+q_{46}}E=\frac1d.
\]
For the third coordinate, the densities on $(0,v)$, $(v,u)$, and $(u,c)$ are
\[
    \frac{q_{14}+q_{25}}c+\frac{q_{46}}v
    =\frac{q_{14}+q_{25}}c+\frac{q_{26}}F+\frac{q_{36}}{u-v}
    =\frac{q_{14}+q_{25}}c+\frac{q_{26}}F+\frac{q_{13}}G=\frac pc.
\]
The identities for the third coordinate use $c-u=G$, $c-v=F$, and $R+ah=E(u+v+2h)$.

In the tables for each of the three cases, only $P_4P_5$ can have both a positive weight and a constant coordinate.
Its third coordinate is zero, and its weight $q_{45}=h/c=1-p$ gives the required atom in the third marginal.
There are no other atoms, so the density identities above complete the first case, and the remaining two cases require only the continuous marginal density checks.

The total mass of any marginal equals the sum of the segment weights.
Since the first marginal has no atoms, verifying that its density is $1/a$ on $(0,a)$ also proves that the weights sum to one.

\paragraph{Second case: $v\le w\le u$.}
Set $S=h^3+(u+v+w)h^2+(uv+uw+vw)h+uvw+v^2w$.
We use the following six segments and weights:
\[
\begin{array}{c|c@{\qquad}c|c}
\text{segment}&\text{weight}&\text{segment}&\text{weight}\\\hline
P_1P_3&wG/(aF)&P_1P_4&(u-w)S/(acEF)\\
P_1P_5&vh/(cE)&P_3P_6&w(u-v)/(aF)\\
P_4P_5&h/c&P_4P_6&vw/(aF)
\end{array}
\]
Again the weights are nonnegative, and we use the same sampling rule.
For the first coordinate, the densities on $(0,v)$ and $(v,a)$ are
\[
    \frac{q_{13}+q_{14}+q_{36}+q_{46}}a+\frac{q_{15}}v=\frac{q_{13}+q_{14}+q_{36}+q_{46}}a+\frac{q_{45}}E=\frac1a.
\]
For the second coordinate, the densities on $(0,w)$, $(w,u)$, and $(u,d)$ are
\[
    \frac{q_{36}}d+\frac{q_{13}}w
    =\frac{q_{36}}d+\frac{q_{14}}{u-w}+\frac{q_{15}}F
    =\frac{q_{36}}d+\frac{q_{15}}F+\frac{q_{45}+q_{46}}E=\frac1d.
\]
For the third coordinate, the densities on $(0,v)$, $(v,u)$, and $(u,c)$ are
\[
    \frac{q_{14}+q_{15}}c+\frac{q_{46}}v
    =\frac{q_{14}+q_{15}}c+\frac{q_{36}}{u-v}
    =\frac{q_{14}+q_{15}}c+\frac{q_{13}}G=\frac pc.
\]

\paragraph{Third case: $v\le u\le w$.}
We use the following five segments and weights, all nonnegative in this case:
\[
\begin{array}{c|c@{\qquad}c|c}
\text{segment}&\text{weight}&\text{segment}&\text{weight}\\\hline
P_1P_4&v(w-u)/(cE)&P_1P_5&vF/(cE)\\
P_2P_4&uG/(cE)&P_2P_5&u(w-v)/(cE)\\
P_4P_5&h/c&&
\end{array}
\]
We use the same sampling rule and verify each marginal as before.
For the first coordinate, the densities on $(0,v)$, $(v,w)$, and $(w,a)$ are
\[
    \frac{q_{14}}a+\frac{q_{15}}v
    =\frac{q_{14}}a+\frac{q_{25}}{w-v}+\frac{q_{45}}E
    =\frac{q_{14}}a+\frac{q_{24}}G+\frac{q_{45}}E=\frac1a.
\]
For the second coordinate, the densities on $(0,u)$, $(u,w)$, and $(w,d)$ are
\[
    \frac{q_{25}}d+\frac{q_{24}}u
    =\frac{q_{25}}d+\frac{q_{14}}{w-u}+\frac{q_{45}}E
    =\frac{q_{25}}d+\frac{q_{15}}F+\frac{q_{45}}E=\frac1d.
\]
For the third coordinate, the density on $(0,c)$ is
\[
    \frac{q_{14}+q_{15}+q_{24}+q_{25}}c=\frac pc.
\]

Each case uses at most seven segments, and computing their weights and sampling a point take $O(1)$ time.
\end{proof}

\subsubsection{A mixture of uniforms on nested intervals}
\label{sec:prg:nested}

The construction in \cref{sec:prg:four} will also require a three-variable coupling with two uniform marginals and a third marginal that is a mixture of uniforms on $[0,t]$ and $[0,c]$, where $t<c$.
Under the condition below, the joint mixability theorem of \cite[Theorem 3.2]{wang2016mixability} for nonincreasing densities already gives the existence of a coupling with constant sum.
We give an explicit sampler that takes $O(1)$ time.

\begin{lemma}
\label{lem:prg:nested}
Let $a,d,c>0$, $0<t<c$, and $0<q<1$.
If $M=(a+d+qt+(1-q)c)/2\ge\max\{a,d,c\}$, then $\Unif[0,a]$, $\Unif[0,d]$, and $q\Unif[0,t]+(1-q)\Unif[0,c]$ have an explicit coupling with sum $M$ that can be sampled in $O(1)$ time.
\end{lemma}

\begin{proof}
Swap the first two coordinates if necessary so that $a\le d$.
Set
\[
    u=M-a,\quad v=M-d,\quad h=q(c-t)=c-u-v.
\]
The assumption $M\ge d\ge a$ gives $u\ge v\ge0$.
Also, $u+v=qt+(1-q)c$, so $t<u+v<c$ and $h>0$.
We will express the target third marginal as a mixture of distributions that are easier to couple with the first two uniforms.
To keep the sum equal to $M$ in each component, its third marginal must have mean $M-(a+d)/2=(u+v)/2$.
We therefore consider the family
\[
    F_x=\frac{h}{c-x}\Unif[0,x]
          +\left(1-\frac{h}{c-x}\right)\Unif[0,c],
    \qquad 0<x\le u+v.
\]
The range $x\le u+v=c-h$ ensures that $0<h/(c-x)\le1$, so each $F_x$ is a probability distribution.
Its mean is
\[
    \frac12\left(\frac{hx}{c-x}
        +c\left(1-\frac{h}{c-x}\right)\right)
    =\frac{c-h}{2}=\frac{u+v}{2}.
\]
Thus varying $x$ changes the shorter interval while preserving the required mean, and $F_t$ is the desired third marginal because $h/(c-t)=q$.
We consider three cases according to the position of $t$ relative to $v$ and $u$.

\paragraph{First case: $t\le v$.}
We sample a location $S\in[0,t]$, then use \cref{lem:prg:three} to couple the first two uniforms with a third distribution that has an atom at $S$ and a uniform component on $[S,c]$.
The distribution of $S$ is chosen so that averaging these conditional distributions gives $F_t$.
Specifically, draw
\[
    S=\frac{htR}{h+t-tR},\qquad R\sim\Unif[0,1].
\]
For a sampled value $S=s$, we use the conditional third marginal $(1-p_s)\delta_s+p_s\Unif[s,c]$, where $\delta_s$ denotes the point mass at $s$.
Its mean is $s+p_s(c-s)/2$, so the required mean $(u+v)/2$ determines the weight
\[
    p_s=\frac{u+v-2s}{c-s}=\frac{c-h-2s}{c-s}.
\]
When $p_s>0$, apply \cref{lem:prg:three} with lengths $a,d,c-s$, mixture probability $p_s$, and sum $M-s$, then add $s$ to the third coordinate.
This gives the specified conditional third marginal and restores the sum to $M$.

To check that this application is valid, first note that $s\le t\le v\le u$ makes $u+v-2s$ nonnegative.
Moreover, $1-p_s=(h+s)/(c-s)>0$, so $0\le p_s<1$.
The same bounds $s\le v\le u$ give $a,d\le M-s$, while $c\le M$ gives $c-s\le M-s$.
Finally, the lengths and mixture weight satisfy
\[
    a+d+p_s(c-s)=a+d+u+v-2s=2(M-s).
\]
If $p_s=0$, then $u=v=s$ and $a=d=M-s$; return $(aU,a(1-U),s)$ for $U\sim\Unif[0,1]$.

We now verify the third marginal; the first two conditional marginals are $\Unif[0,a]$ and $\Unif[0,d]$ for every $s$.
For $0\le s\le t$, solving the inequality $S\le s$ for $R$ gives
\[
    \Pr(S\le s)
    =\Pr\!\left(R\le\frac{s(h+t)}{t(h+s)}\right)
    =\frac{s(h+t)}{t(h+s)},
\]
since $R$ is uniform on $[0,1]$.
Differentiating gives the density of $S$ on $(0,t)$:
\[
    f_S(s)=\frac{d}{ds}\left(\frac{s(h+t)}{t(h+s)}\right)
    =\frac{h(h+t)}{t(h+s)^2}.
\]

Fix $0<z<t$.
For $S=s$, the atom at $s$ has mass $1-p_s$.
After averaging over $S$, the mass contributed by these atoms to $[0,z]$ is $\int_0^z(1-p_s)f_S(s)\,ds$.
Differentiating with respect to $z$ gives their contribution to the density:
\[
    (1-p_z)f_S(z)
    =\frac{h+z}{c-z}\cdot\frac{h(h+t)}{t(h+z)^2}
    =\frac{h(h+t)}{t(h+z)(c-z)}.
\]
The conditional uniform component has weight $p_s$ and density $1/(c-s)$ on $(s,c)$.
Thus its contribution to the density at $z$ is $p_s/(c-s)$ when $s\le z$, and zero otherwise.
Averaging over the sampled location $S$ gives
\[
\begin{aligned}
    \int_0^z\frac{p_s}{c-s}f_S(s)\,ds
    &=\frac{h(h+t)}{t}\int_0^z\frac{c-h-2s}{(h+s)^2(c-s)^2}\,ds\\
    &=\frac{h(h+t)}{t}\left[-\frac{1}{(h+s)(c-s)}\right]_{s=0}^{s=z}\\
    &=\frac{h+t}{ct}\left(1-\frac{hc}{(h+z)(c-z)}\right).
\end{aligned}
\]
Adding the two contributions cancels the terms involving $1/((h+z)(c-z))$ and leaves $(h+t)/(ct)$.
This is the density of the target mixture $F_t$ on $(0,t)$, since $h=q(c-t)$ gives
\[
    \frac qt+\frac{1-q}c
    =\frac{t+q(c-t)}{ct}
    =\frac{h+t}{ct}.
\]
For $t<z<c$, only the uniform components contribute, and every sampled location $s\in[0,t]$ satisfies $s<z$.
Setting the upper integration limit to $t$ therefore gives density
\[
    \frac{c-h-t}{c(c-t)}=\frac{1-q}{c},
\]
which is the density of $F_t$ on $(t,c)$.

The densities on $(0,t)$ and $(t,c)$ account for total probability
\[
    t\cdot\frac{h+t}{ct}
    +(c-t)\cdot\frac{c-h-t}{c(c-t)}
    =\frac{h+t}{c}+\frac{c-h-t}{c}=1.
\]
Thus there are no additional point masses, and the third marginal is $F_t$.

\paragraph{Second case: $t\ge u$.}
We express $F_t$ as a mixture of a distribution covered by the first case and a single uniform distribution.
Set $T=u+v-t$.
The inequality $t<u+v$ gives $T>0$, while the present assumption $t\ge u$ gives
\[
    T=v-(t-u)\le v\le t.
\]
Thus the first case applies to $F_T$.
If $T=t$, this already gives the required coupling, so assume $T<t$.
The choice of $T$ also makes $\Unif[T,t]$ have mean $(T+t)/2=(u+v)/2$, the same as $F_t$ and $F_T$.

To determine the mixture weight, compare the densities of $F_t$ and $F_T$ on $(0,T)$, which are $(h+t)/(ct)$ and $(h+T)/(cT)$, respectively.
Set
\[
    \alpha=\frac{(h+t)/(ct)}{(h+T)/(cT)}
    =\frac{T(h+t)}{t(c-t)},
\]
where $h+T=c-t$.
Then $\alpha F_T$ and $F_t$ have the same density on $(0,T)$.
Their densities also agree on $(t,c)$: both equal $T/(c(c-t))$, using $T+t=c-h$.
On the remaining interval $(T,t)$, subtracting the density of $\alpha F_T$ from that of $F_t$ gives
\[
    \frac{h+t}{ct}-\frac{T}{c(c-t)}
    =\frac{h}{t(c-t)}.
\]
This constant density has total mass
\[
    1-\alpha=\frac{h(t-T)}{t(c-t)}>0.
\]
Since $\alpha>0$, these are valid mixture weights, and we obtain
\[
    F_t=\alpha F_T+(1-\alpha)\Unif[T,t].
\]

With probability $\alpha$, use the coupling from the first case for $F_T$.
With probability $1-\alpha$, we need to couple the first two uniforms with $\Unif[T,t]$.
For this branch, apply \cref{lem:prg:three} with lengths $a,d,t-T$, mixture probability $1$, and sum $M-T$, then add $T$ to the third coordinate.
The length bounds hold because $T\le v\le u$ gives $a,d\le M-T$, and $t<c\le M$ gives $t-T\le M-T$.
Also, $T+t=u+v$ gives
\[
    a+d+(t-T)=2M-(u+v)+t-T=2(M-T).
\]
After adding $T$, this branch has third marginal $\Unif[T,t]$ and sum $M$.
Both branches preserve the first two marginals, and the mixture identity gives the required third marginal $F_t$.

\paragraph{Third case: $v<t<u$.}
We again separate a component covered by the first case, then construct a coupling for the remaining distribution.
When $v>0$, we use $F_v$, whose shorter interval ends at the largest value allowed in the first case.
As in the second case, choose its weight to match the density of $F_t$ below this endpoint:
\[
    \alpha=\frac{(h+t)/(ct)}{(h+v)/(cv)}
    =\frac{v(h+t)}{t(h+v)},
    \qquad 1-\alpha=\frac{h(t-v)}{t(h+v)}.
\]
Both weights are positive because $0<v<t$.
Subtracting $\alpha F_v$ leaves no mass on $(0,v)$.
On each of the intervals $(v,t)$ and $(t,c)$, we subtract the density of $\alpha F_v$ on $(v,c)$ and divide by $1-\alpha$.
The resulting densities are, respectively,
\[
\begin{aligned}
    \frac{1}{1-\alpha}\left(\frac{h+t}{ct}-\alpha\frac{c-h-v}{c(c-v)}\right)
        &=\frac{h+t}{(c-v)(t-v)},\\
    \frac{1}{1-\alpha}\left(\frac{c-h-t}{c(c-t)}-\alpha\frac{c-h-v}{c(c-v)}\right)
        &=\frac{u-t}{(c-v)(c-t)}.
\end{aligned}
\]
These are the densities of the mixture
\[
    \frac{v+h}{c-t}\Unif[v,t]
      +\frac{u-t}{c-t}\Unif[v,c].
\]
Its weights are positive and sum to one because $(v+h)+(u-t)=c-t$.
This is therefore the normalized residual distribution $(F_t-\alpha F_v)/(1-\alpha)$.
If $v=0$, the displayed mixture is already $F_t$, since $h/(c-t)=q$; we couple it directly without introducing $F_v$.

We now couple the displayed mixture with the first two uniforms using three segments, all with coordinate sum $M$.
Set
\[
    W=M-c,\quad s=u-t,\quad H=v+h=a-W,
    \qquad K=Hd+Ws.
\]
The values $W$ and $s$ make the points $(W,0,c)$ and $(a,s,t)$ have sum $M$; the point $(0,d,v)$ has the same sum because $v=M-d$.
Here $W\ge0$, $s,H>0$, and $d-s=a+t-v>0$.
Thus the points have nonnegative coordinates, and $K>0$.
We use the three segments joining these points, with weights normalized by $K$:
\[
\begin{array}{c|c}
\text{endpoints}&\text{weight}\\\hline
(0,d,v)\ --\ (W,0,c)&Ws/K\\
(0,d,v)\ --\ (a,s,t)&H(a+t-v)/K\\
(W,0,c)\ --\ (a,s,t)&Hs/K
\end{array}
\]
Every point on these segments has nonnegative coordinates and sum $M$.
The weights are nonnegative and sum to one because $Ws+H(a+t-v)+Hs=Ws+Hd=K$.
Choose a segment with these probabilities and sample uniformly on it.
When $W=0$, the first segment has weight zero and is omitted.

For the first coordinate, the first and third segments project onto the adjacent intervals $[0,W]$ and $[W,a]$.
Since $a-W=H$, their combined contribution has density $s/K$ on $[0,a]$, while the second segment contributes on all of $[0,a]$.
The resulting density is
\[
    \frac{s}{K}+\frac{H(a+t-v)}{aK}
    =\frac{as+H(d-s)}{aK}=\frac1a,
\]
where $as+H(d-s)=Hd+(a-H)s=Hd+Ws=K$.
For the second coordinate, the first segment contributes on $[0,d]$, while the other two contribute on $[s,d]$ and $[0,s]$.
Since $d-s=a+t-v$, the density on both intervals is
\[
    \frac{Ws}{dK}+\frac H K=\frac{Ws+Hd}{dK}=\frac1d.
\]
For the third coordinate, the first segment contributes on $[v,c]$, the second on $[v,t]$, and the third on $[t,c]$.
The densities on $(v,t)$ and $(t,c)$ are therefore, respectively,
\[
    \frac{Ws}{(c-v)K}+\frac{H(a+t-v)}{(t-v)K}
    =\frac{h+t}{(c-v)(t-v)},
    \qquad
    \frac{Ws}{(c-v)K}+\frac{Hs}{(c-t)K}
    =\frac{u-t}{(c-v)(c-t)}.
\]
These match the two densities of the residual distribution computed above.
Every retained segment has nonconstant projections onto all three coordinates, so there are no atoms.
For $v>0$, choose the coupling from the first case for $F_v$ with probability $\alpha$ and this residual coupling with probability $1-\alpha$; for $v=0$, use the residual coupling directly.
This gives the required marginals and sum $M$ in the third case.

Each case uses a bounded number of arithmetic operations, comparisons, and uniform draws, proving the claimed $O(1)$ sampling cost.
\end{proof}

\subsubsection{A coupling for four variables}
\label{sec:prg:four}

We now combine the preceding two constructions to couple three uniform variables and a fourth whose distribution is a mixture of a uniform and an atom at zero.

\begin{lemma}
\label{lem:prg:four}
Let $0<u\le v\le w$, $\ell>0$, and $0<p<1$.
If $B=(u+v+w+p\ell)/2\ge\max\{w,\ell\}$, then $\Unif[0,u]$, $\Unif[0,v]$, $\Unif[0,w]$, and $(1-p)\delta_0+p\Unif[0,\ell]$ have an explicit coupling with sum $B$ that can be sampled in $O(1)$ time.
\end{lemma}

\begin{proof}
Write $X_1,X_2,X_3,X_4$ for the four output coordinates.
There are two cases.
When $w\ge(1-p)\ell$, we combine two three-variable couplings using auxiliary uniform variables whose sum is fixed.
When this inequality fails, we choose between the uniform component and the atom of $X_4$, and use \cref{lem:prg:nested} in the latter branch.

\paragraph{Combining two couplings: $w\ge(1-p)\ell$.}

We introduce two auxiliary variables $Z_1,Z_2$, both uniform on $[0,d]$, and construct separate couplings for $(X_1,X_2,Z_1)$ and $(Z_2,X_3,X_4)$.
We will combine them so that $Z_1+Z_2=d$, then discard the auxiliary coordinates.
The length $d$ must allow both applications of \cref{lem:prg:three}.
For the three uniforms with lengths $(u,v,d)$, the lemma's length bounds are equivalent to
\[
    v-u\le d\le u+v.
\]
For lengths $(d,w,\ell)$ with mixture probability $p$ on the last coordinate, they are equivalent to
\[
    \max\{w-p\ell,(2-p)\ell-w\}\le d\le w+p\ell.
\]
We therefore choose
\[
    d=\max\{v-u,w-p\ell,(2-p)\ell-w,0\}.
\]
To check the upper bounds on $d$, the budget conditions $B\ge w$ and $B\ge\ell$ give
\[
    w-p\ell\le u+v,
    \qquad (2-p)\ell-w\le u+v.
\]
The case assumption $w\ge(1-p)\ell$ gives $(2-p)\ell-w\le w+p\ell$.
The other bounds on the entries of the maximum follow from $0<u\le v\le w$ and $p\ell>0$.
Thus $d\le\min\{u+v,w+p\ell\}$, as required.
Also, $(w-p\ell)+((2-p)\ell-w)=2(1-p)\ell>0$, so at least one entry of the maximum is positive and $d>0$.

The two couplings supplied by \cref{lem:prg:three} satisfy
\[
    X_1+X_2+Z_1=\frac{u+v+d}{2},
    \qquad Z_2+X_3+X_4=\frac{d+w+p\ell}{2}.
\]
Sample the second coupling, then sample the first conditional on $Z_1=d-Z_2$.
Since $d-Z_2$ is uniform on $[0,d]$, averaging these conditional samples recovers the first coupling; the second coupling is unchanged.
Discarding $Z_1,Z_2$ therefore leaves the four required marginals, with sum
\[
    X_1+X_2+X_3+X_4
    =\frac{u+v+d}{2}+\frac{d+w+p\ell}{2}-d=B.
\]

We first explain how to sample $(X_1,X_2)$ given $Z_1=z$, using the segments and weights supplied by \cref{lem:prg:three} for the first coupling.
The $Z_1$ coordinate is nonconstant on every segment of positive weight, since otherwise its uniform marginal would have an atom.
Fixing $Z_1=z$ therefore leaves at most one point on each segment, and the conditional sample is a choice among these points.
Suppose segment $j$ of the first coupling has weight $q_j$ and its $Z_1$ coordinate has endpoint values $r_j,s_j$.
This segment contributes density $q_j/|r_j-s_j|$ between those values, while the total density of $Z_1$ is $1/d$.
For $z$ away from all endpoint values $r_j,s_j$, the conditional probability of choosing segment $j$ is therefore
\[
    \frac{q_j/|r_j-s_j|}{1/d}
    =\frac{d q_j}{|r_j-s_j|}
\]
when $z$ lies between $r_j,s_j$, and zero otherwise.
Once the segment is chosen, the value $Z_1=z$ fixes its interpolation parameter to $(z-r_j)/(s_j-r_j)$, so $X_1,X_2$ are affine functions of $z$.

At an endpoint value $z$, use the conditional segment probabilities from an adjacent interval between consecutive endpoint values in $[0,d]$, then evaluate the selected segment at $Z_1=z$.
The resulting point remains on that segment, so feasibility is preserved; this convention does not affect the marginals because $Z_1$ is uniform.
Sampling the second coupling and making this conditional choice among at most seven first segments takes $O(1)$ time.
For the finite-bit implementation, \cref{app:prg:bits} also expresses this coupling as a finite mixture of uniform distributions on line segments.

\paragraph{Conditioning on the fourth coordinate: $w<(1-p)\ell$.}
Let $S=u+v+w$.
With probability $p$, we will sample $X_4$ uniformly on $[0,\ell]$ and make the first three coordinates uniform on intervals shortened by a common factor $c$.
To make their four means sum to $B$, we need $cS+\ell=2B=S+p\ell$.
Set
\[
    d=\frac{(1-p)\ell}{S},\qquad c=1-d.
\]
Since $B\ge\ell$, we have $S\ge(2-p)\ell>\ell$ and hence
\[
    c-p=(1-p)\left(1-\frac{\ell}{S}\right)>0.
\]
In particular, $0<d<1$ and $p<c<1$.
The inequality $c>p$ will ensure that the other branch has nonnegative marginal densities.

In the branch where $X_4$ is uniform, couple four uniforms with lengths $(cu,cv,cw,\ell)$.
These lengths sum to $2B$, and each is at most $B$ because $c<1$ and $u\le v\le w\le B$, $\ell\le B$.
Merge the two smallest lengths, whose sum is at most half the total and hence at most $B$.
Apply \cref{lem:prg:three} with mixture probability $1$ to the resulting three lengths, then split the merged coordinate proportionally, as in \cref{sec:prg:reduction}.
This gives the required four conditional uniform marginals with sum $B$.

With probability $1-p$, set $X_4=0$.
For a coordinate whose target marginal is $\Unif[0,L]$, where $L\in\{u,v,w\}$, write $f_L$ for its density in this branch.
Averaging over the two branches must give density $1/L$, so
\[
    \frac{p}{cL}\ind_{[0,cL]}(x)+(1-p)f_L(x)=\frac1L,
    \qquad 0<x<L.
\]
Solving for $f_L$ gives
\[
    f_L(x)=
    \begin{cases}
        \dfrac{c-p}{c(1-p)L},&0<x<cL,\\[1mm]
        \dfrac{1}{(1-p)L},&cL<x<L.
    \end{cases}
\]
Both values are positive because $c>p$, and their integrals over the two intervals sum to one.

To construct a coupling with these conditional marginals, we first construct one for the reflected coordinates $Y=L-X$.
Reflection moves the larger density to $(0,dL)$ and the smaller density to $(dL,L)$.
This gives a mixture of uniforms on $[0,dL]$ and $[0,L]$.
Indeed, only the latter uniform contributes density on $(dL,L)$, so its weight must be the density there multiplied by $L$.
The two weights are therefore
\[
    1-q=\frac{c-p}{c(1-p)},\qquad
    q=\frac{pd}{c(1-p)}=\frac{p\ell}{S-(1-p)\ell}.
\]
They lie strictly between zero and one, and on $(0,dL)$ the mixture has density
\[
    \frac{q}{dL}+\frac{1-q}{L}=\frac{1}{(1-p)L},
\]
as required.
Thus the reflected marginal is
\[
    Y\sim q\Unif[0,dL]+(1-q)\Unif[0,L].
\]
The original three coordinates must sum to $B$, so the reflected coordinates must sum to
\[
    M=S-B=\frac{S-p\ell}{2}.
\]
The case assumption $w<(1-p)\ell$ and the bound $S\ge(2-p)\ell$ give
\[
    2(M-w)=S-p\ell-2w
    \ge 2(1-p)\ell-2w>0.
\]
Thus every reflected interval endpoint is at most $w<M$.

All three reflected marginals are mixtures of uniforms, whereas \cref{lem:prg:nested} allows only one such mixture.
We remove the extra mixtures by rounding their probabilities as in \cref{sec:prg:reduction}, while preserving the sum of their means.
Write $(L_1,L_2,L_3)=(u,v,w)$, so the reflected coordinates are $Y_j=L_j-X_j$, and initialize $r_j=qcL_j$ for each $j\in\{1,2,3\}$.
For any $0\le r_j\le cL_j$, associate the conditional marginal
\[
    \frac{r_j}{cL_j}\Unif[0,dL_j]
    +\left(1-\frac{r_j}{cL_j}\right)\Unif[0,L_j].
\]
Its mean is $(L_j-r_j)/2$.
Initially $\sum_j r_j=qcS=p\ell$, so preserving this sum keeps the total conditional mean equal to $(S-p\ell)/2=M$.
Apply the pairwise updates with bounds $0\le r_j\le cL_j$, preserving $\sum_j r_j$ at every update and each $r_j$ in expectation.
At most two updates leave at most one $r_j$ strictly between zero and $cL_j$.
At $r_j=0$ the marginal is $\Unif[0,L_j]$, and at $r_j=cL_j$ it is $\Unif[0,dL_j]$.
Thus the resulting marginals are either three uniforms or two uniforms and one mixture on nested intervals.
Their means sum to $M$ and their endpoints are at most $w<M$, so \cref{lem:prg:three} with mixture probability $1$ handles the first possibility and \cref{lem:prg:nested} handles the second.

Sample this conditional coupling with sum $M$, then return $X_j=L_j-Y_j$ for $j\in\{1,2,3\}$ and $X_4=0$.
These coordinates are nonnegative and sum to $S-M=B$.
Averaging over the rounding choices restores the reflected marginals because $\E[r_j/(cL_j)]=q$.
Reflection therefore gives the densities $f_{L_j}$, and averaging the two branches restores each original uniform marginal by the defining equation for $f_L$.
The fourth marginal is $(1-p)\delta_0+p\Unif[0,\ell]$ by the initial branch choice.
This case uses at most two rounding choices followed by a coupling sampled in $O(1)$ time.
This proves the claimed sampling cost and completes both cases.
\end{proof}

\subsubsection{The complete sampler and its running time}
\label{sec:prg:sampler}

We now assemble the reduction and the three- and four-variable couplings into a sampler for any marginals satisfying \eqref{eq:prg:coupling-condition}.
The strategy description stores the budget $B$, lengths $\ell_i$, and mixture probabilities $p_i$, together with the following procedure for drawing an allocation.
\begin{enumerate}
    \item Initialize $r_i=p_i\ell_i$ and apply the pairwise rounding rule from \cref{sec:prg:reduction} until at most one $r_i$ lies strictly between $0$ and $\ell_i$.
    \item Set aside that coordinate, if present, and group the coordinates with $r_i=\ell_i$ as in \cref{sec:prg:reduction}.
    Each group $G$ is represented by a uniform variable $Y_G$ on $[0,\sum_{j\in G}\ell_j]$, with interval length at most $B$.
    \item Couple the two or three group variables and the coordinate set aside, if present.
    With no coordinate set aside, two groups both have length $B$ and can be sampled as $(BU,B(1-U))$, where $U\sim\Unif[0,1]$; for three groups, use \cref{lem:prg:three} with $p=1$.
    With a coordinate set aside, use \cref{lem:prg:three} for two groups or \cref{lem:prg:four} for three groups, taking $p=r_i/\ell_i$ and interval length $\ell_i$ for that coordinate.
    \item For each group $G$, return $Z_i=\ell_iY_G/\sum_{j\in G}\ell_j$ for every $i\in G$.
    Return the sampled value for the coordinate set aside, if present, and set coordinates with $r_i=0$ to zero.
\end{enumerate}
The reduction ensures that step 3 always falls into one of the listed cases and that the hypotheses of the indicated lemma hold.
The sampled group variables and any coordinate set aside sum to $B$, and splitting a group preserves its total allocation.
Thus every output is nonnegative and spends budget $B$ exactly.
Conditional on the final vector $r$, coordinate $i$ has distribution $(1-r_i/\ell_i)\delta_0+(r_i/\ell_i)\Unif[0,\ell_i]$.
Since $\E[r_i/\ell_i]=p_i$, averaging over the rounding choices recovers each original marginal.

Storing the strategy description takes $O(n)$ operations.
For each allocation, the rounding, grouping, and splitting steps take $O(n)$ operations, and the coupling step takes $O(1)$ operations.
Each random choice among a constant number of alternatives takes $O(1)$ time using an independent uniform draw and comparisons with the cumulative probabilities.
Hence the complete sampler takes $O(n)$ time per allocation in the real-RAM model.

\begin{proof}[Completion of the proof of \cref{thm:prg:exact}]
For player $A$, apply the sampler with $p_i=1$, $\ell_i=2v_i$, and total allocation $1$.
For player $B$, use $p_i=b$, the same lengths, and total allocation $b$.
These parameters can be computed in $O(n)$ operations.
The normalization $\sum_i v_i=1$ and condition~\eqref{eq:prg:promise} verify \eqref{eq:prg:coupling-condition} for both players.
Apply the construction separately to the two players.
The preceding argument proves exact feasibility, the desired marginals, and the computation and sampling bounds.
The payoff bounds in \cref{sec:prg:security} prove equilibrium and the stated payoffs against all deviations that respect the player's budget.
\end{proof}

\subsection{Finite-bit implementation}
\label{sec:prg:finite}

The exact sampler in \cref{thm:prg:exact} uses continuous uniform draws.
We now implement it with finitely many random bits and rational output, obtaining an $\varepsilon$-Nash equilibrium.
We bound the error in each marginal CDF and apply \cref{lem:prelim:cdf} to bound the resulting regret.

\begin{corollary}
\label{cor:prg:finite}
Under \eqref{eq:prg:promise}, for every rational $0<\varepsilon<1$, two finite-bit samplers can be computed whose output distributions form an $\varepsilon$-Nash equilibrium.
Their allocations are rational and spend the budgets exactly on every draw.
With $k=\lceil\log_2(1/\varepsilon)\rceil$, computing the samplers takes $\softO(nL)$ bit operations, and the worst-case cost per allocation is $\softO((n+k)(L+k))$.
Each allocation uses $O(n+k)$ random bits.
Here $L$ is the total binary encoding length of the values, budget ratio, and approximation parameter.
\end{corollary}

We handle the continuous uniform draws separately from the discrete random choices in the reduction and couplings.
Each allocation uses at most two continuous uniform draws: one to sample a point on a segment and, when needed, one to generate the random atom location $S$ in \cref{lem:prg:nested}.
The implementation replaces these draws by uniformly chosen midpoints of $2^K$ equal subintervals of $[0,1]$, using $K=O(k)$ random bits per draw.
The coupling formulas enforce the required sum for every choice of these values, so this replacement preserves exact spending.

For the discrete choices, we impose a bound on the total number of random bits used in one allocation draw.
If another bit is needed after this bound has been reached, the sampler returns a fixed allocation that spends the player's budget exactly.
The probability of using this fallback is included in the marginal CDF error.
The bounds also hold when $\max_i v_i=b/2$ and require no lower bound on $b/2-\max_i v_i$.
The error estimates and bit-complexity analysis are given in \cref{app:prg:bits}.

\subsection{Extension to player-specific values}
\label{sec:prg:specific}

The same coupling construction applies when the players assign different values to a battlefield.
In this subsection player $A$ has values $v_i>0$, player $B$ has values $w_i>0$, and both value vectors sum to one.
The budgets remain $1$ and $b$.
Each player's payoff uses their own values, and a tie awards each player half their value for that battlefield.

We use the marginal formulas of \cite{kovenock2021generalizations}, also used by \cite{perchet2024algorithmic}.
For parameters $\gamma,\lambda>0$, define
\begin{equation}
\begin{aligned}
    \ell_i&=\frac{\min\{v_i,\gamma w_i\}}{\lambda},\\
    p_{A,i}&=\min\{1,v_i/(\gamma w_i)\},\qquad
    p_{B,i}=\min\{1,\gamma w_i/v_i\},\\
    X_i&\sim(1-p_{A,i})\delta_0+p_{A,i}\Unif[0,\ell_i],\\
    Y_i&\sim(1-p_{B,i})\delta_0+p_{B,i}\Unif[0,\ell_i].
\end{aligned}
\label{eq:prg:specific-marginals}
\end{equation}
To match the budgets and apply the coupling construction, we seek parameters satisfying
\begin{equation}
    \frac12\sum_i p_{A,i}\ell_i=1,\qquad
    \frac12\sum_i p_{B,i}\ell_i=b,\qquad
    \max_i\ell_i\le b.
    \label{eq:prg:specific-promise}
\end{equation}
The two equalities require each player's expected total spending to equal their budget, and the inequality bounds every interval length by the smaller budget.
Under these conditions, \cite[Theorem~1]{perchet2024algorithmic} proves that the marginals can be coupled into a Blotto equilibrium.
They also imply \eqref{eq:prg:coupling-condition} for each player, so we can construct these couplings using \cref{sec:prg:sampler}.

The parameter enumeration of \cite[Proposition~1]{perchet2024algorithmic} finds all pairs satisfying the two mean equations in $O(n\log n)$ operations by sorting the ratios $v_i/w_i$ and solving a polynomial of degree at most three on each interval they delimit in $(0,\infty)$.
For this search, we augment the real-RAM model with a unit-cost operation that returns the exact real roots of a polynomial of degree at most three.
Combining this enumeration with the coupling criterion $\max_i\ell_i\le b$ of \cite[Theorem~1]{perchet2024algorithmic} and our construction gives the following result.

\begin{theorem}
\label{thm:prg:specific}
Given positive rational input, we can decide in $O(n\log n)$ operations in this augmented real-RAM model whether there exists a positive pair $(\gamma,\lambda)$ satisfying \eqref{eq:prg:specific-promise}.
If such a pair exists, an exact Nash equilibrium can be computed within the same bound, with $O(n)$ operations per allocation draw.

When such a pair exists, for every rational $0<\varepsilon<1$, finite-bit samplers of rational allocations can instead be computed in time polynomial in $n,L,\log(1/\varepsilon)$.
Their worst-case sampling cost and random-bit count are polynomial in the same parameters.
Their output distributions form an $\varepsilon$-Nash equilibrium against all continuous deviations, and every allocation satisfies the player's budget constraint.
Here $L$ includes both value vectors, the budget ratio, and the approximation parameter.
\end{theorem}

\Cref{app:prg:specific} explains how to test all candidates within the stated bound and apply the coupling construction.
If no parameter pair passes the test, this rules out only parameters satisfying \eqref{eq:prg:specific-promise}; the Blotto game may still have equilibria outside this family.

Even for rational input, the parameters $\gamma$ and $\lambda$ can be irrational.
The finite-bit implementation represents these parameters exactly as algebraic numbers, approximates the random draws, and rounds each output coordinate down to a rational number.
Rounding down preserves the budget constraint, although the output may no longer spend the budget exactly.
The marginal CDF bounds control the resulting regret for each player using their own normalized values.
\Cref{app:prg:specific-bits} gives the construction and its bit-complexity analysis.

\section{A general approximation scheme}
\label{sec:general}

We give a fully polynomial-time approximation scheme (FPTAS) for computing an $\varepsilon$-Nash equilibrium for arbitrary common battlefield values and budget ratio.
Unlike the construction in \cref{sec:prg}, this algorithm also applies when the General Lotto marginals cannot be coupled to satisfy the hard budgets.

Recall that $n$ is the number of battlefields and the players' budgets are $1$ and $b \in (0, 1]$.
Let $L_0$ be the binary length of the game input and $E$ that of $\varepsilon$, and let $n_* = \min\{n, 1/\varepsilon\}$.
In the finite-bit model, we charge $\softO(W)$ bit operations for each exact rational arithmetic operation on operands encoded with at most $W$ bits (see \cref{app:computation:bits} for details).

\begin{theorem}
\label{thm:general}
For every Colonel Blotto game with common rational battlefield values and a rational budget ratio, and every rational $0 < \varepsilon < 1$, a deterministic algorithm computes finite descriptions of strategies forming an $\varepsilon$-Nash equilibrium.
In the real-RAM model, the computation uses
\[
    \softO\!\left(n + \frac{n_*^2}{\varepsilon^5}\right)
\]
operations; in the finite-bit model, it uses
\[
    \softO\!\left(L_0 + E + \frac{n_*^3}{\varepsilon^6}\right)
    \quad\text{bit operations and}\quad
    \softO\!\left(L_0 + E + \frac{n_*^3}{\varepsilon^4}\right)
    \quad\text{working bits}.
\]
The two strategy descriptions together use $\softO(L_0 + E + n + n_*^2/\varepsilon^3)$ bits.

Sampling an allocation takes $O(n + \log(1/\varepsilon))$ real-RAM operations for every draw.

If $n > 1$ and $b < 1$, sampling uses $O(\log(n/\varepsilon))$ random bits and $\softO(n(L_0 + n_*/\varepsilon))$ bit operations for every draw, including writing all $n$ rational coordinates.

If $n > 1$ and $b = 1$, the expected number of random bits is $O(\log(n/\varepsilon))$, and the expected cost of sampling and writing all $n$ rational coordinates is $\softO(n(L_0 + n_*/\varepsilon))$ bit operations.

For $n = 1$, both players spend their full budgets, giving an exact equilibrium.
\end{theorem}

For $n > 1$ and $b < 1$, each strategy in \cref{thm:general} is the uniform distribution on an implicitly represented list of $\softO(n_*/\varepsilon^4)$ allocations whose length is a power of two.
Sampling chooses a uniformly random list entry and computes its coordinates from the description without enumerating the list.
The number of random bits is exactly the base-two logarithm of the list length.

\paragraph{Why a fine uniform grid does not suffice.}
A natural approach is to restrict every bid to a multiple of a small $\delta > 0$ and compute an equilibrium of the resulting discrete Blotto game.
After scaling by $1/\delta$, the integer budgets are at most $1/\delta$, so an exact equilibrium can be computed in time polynomial in $n$, $1/\delta$, and the input bit length \cite[Theorem~1]{behnezhad2017faster}.
This would give an FPTAS if choosing $1/\delta$ polynomial in $n$ and $1/\varepsilon$ ensured that the returned strategies formed an $\varepsilon$-Nash equilibrium of the original game.
The following example shows that an exact equilibrium of the restricted game can have regret bounded away from zero in the original game, even as the discretization is refined.
Both budgets are exact multiples of $\delta$, so the failure does not come from rounding the budgets.

\begin{proposition}
\label{prop:general:grid}
Fix two battlefields of value $1/2$ each and budgets $1$ and $b = 2/3$.
For every integer $q \ge 3$, the game restricted to bids in $\delta\mathbb Z_{\ge 0}$, where $\delta = 1/(3q)$, has an exact Nash equilibrium for which player $A$ has zero regret and player $B$ has regret at least $1/12$ in the original game.
\end{proposition}
\begin{proof}
Measure bids in units of $\delta$ and let both players spend their full budgets.
Their actual allocations are $(\delta T, \delta(3q - T))$ and $(\delta Y, \delta(2q - Y))$, where $T$ and $Y$ are independent and specify the players' bids on the first battlefield in these units.
Player $B$ chooses $Y$ uniformly from $\{0, q, 2q\}$, and player $A$ uses
\[
\begin{array}{c|cccc}
t&q - 1&q + 1&2q&2q + 1\\\hline
\Prob(T = t)&1/3&1/6&1/3&1/6.
\end{array}
\]
All these allocations lie on the grid and satisfy the original budgets.
It suffices to check deviations that spend the full budget: unused budget can be added to one coordinate without reducing payoff, and this operation preserves the grid when the initial allocation is on it.

Suppose $A$ bids $t \in [0, 3q]$ on the first battlefield and $3q - t$ on the second.
It wins the first battlefield when $Y < t$ and the second when $3q - t > 2q - Y$, or equivalently $Y > t - q$.
Thus $A$ wins both battlefields when $t - q < Y < t$, giving payoff one, and exactly one battlefield when $Y$ lies outside $[t - q, t]$, giving payoff $1/2$.
At either endpoint it wins one battlefield and ties the other, giving payoff $3/4$.
Starting from the baseline payoff $1/2$, its expected payoff is therefore
\[
    \frac12 + \frac12\left(\Prob(t - q < Y < t)
        + \frac12\Prob(Y \in \{t - q, t\})\right).
\]
The open interval $(t - q, t)$ has length $q$, so it contains at most one of $B$'s possible bids $0, q, 2q$.
If it contains one, neither endpoint is a possible bid of $B$, so the quantity in parentheses is $1/3$.
Otherwise, each endpoint contributes at most $\frac12 \cdot \frac13 = \frac16$, again giving a total of at most $1/3$.
Thus $A$'s expected payoff from any bid $t$ is at most $\frac12 + \frac12 \cdot \frac13 = \frac23$.

All four bids in $A$'s distribution attain this payoff.
For $t = q - 1, q + 1, 2q + 1$, the open interval contains exactly one of $B$'s possible bids: respectively $0, q, 2q$.
For $t = 2q$, its endpoints $q$ and $2q$ each contribute $1/6$ to the quantity in parentheses.
Hence $A$'s distribution gives payoff $2/3$, so $A$ has no profitable deviation in the original game; $B$'s payoff is $1 - 2/3 = 1/3$.

Similarly, if $B$ bids $y \in [0, 2q]$ on the first battlefield and $2q - y$ on the second, it wins the first battlefield when $T < y$ and the second when $T > y + q$.
Its payoff is zero for $y < T < y + q$, $1/4$ at either endpoint, and $1/2$ outside the closed interval.
Subtracting these losses from the baseline $1/2$ gives expected payoff
\[
    \frac12 - \frac12 W(y),
    \qquad
    W(y) = \Prob(y < T < y + q) + \frac12\Prob(T \in \{y, y + q\}).
\]
We show that $W(k) \ge 1/3$ for every integer $0 \le k \le 2q$.
If $k \le q - 2$, the interval contains the atom $q - 1$ strictly; if $q + 1 \le k \le 2q - 1$, it contains the atom $2q$ strictly.
Each atom has mass $1/3$.
For the remaining integers, the only atoms in $[k, k + q]$ are
\[
\begin{array}{c|cc}
k&\text{interior atom}&\text{endpoint atom}\\\hline
q - 1&q + 1&q - 1\\
q&q + 1&2q\\
2q&2q + 1&2q.
\end{array}
\]
In each row, the interior atom has probability $1/6$ and the endpoint atom has probability $1/3$.
The definition of $W$ counts the latter with weight $1/2$, so in all three cases
\[
    W(k) = \frac16 + \frac12 \cdot \frac13 = \frac13.
\]
Thus every deviation on the grid gives $B$ payoff at most $1/3$, proving that the two strategies form an exact equilibrium of the restricted game.

In the original game, $B$ can instead choose $y = q - 1/2$, between adjacent grid points.
Its interval $(y, y + q)$ contains only the atom $q + 1$, of mass $1/6$, and neither endpoint is an atom.
It gives payoff $1/2 - 1/12 = 5/12$, a gain of $1/12$.
\end{proof}

For the game in \cref{prop:general:grid}, taking $q = 2^s$ for $s \ge 2$ gives nested discretizations with $\delta = 1/(3 \cdot 2^s)$ tending to zero, while $B$'s regret in the original game remains at least $1/12$.
Thus refining the discretization does not make the regret of an arbitrary exact equilibrium of the restricted game tend to zero in the original game.
A suitable choice or modification of the equilibrium in the restricted game could still give small regret in the original game, but this is not obvious and would need a specialized construction and proof.

\paragraph{Algorithm overview.}
For the construction below assume $n > 1$ and $b < 1$; \cref{app:general:boundary} treats the remaining cases.
The main step computes the two strategy distributions by multiplicative weights (\cref{sec:general:source}).
The reductions that precede it improve the dependence on the number of battlefields and on the input encoding length; afterwards, we convert the strategies for the reduced game back to strategies for the original game.

We first reduce the game to $O(n_*)$ battlefields: when sufficiently many battlefields have small values, we replace them by $O(1/\varepsilon)$ representative battlefields with equal values.
The reduction also rounds the values.
We also replace $b$ by a nearby rational budget ratio $\rho$.
The encoding lengths of the rounded values and $\rho$ are bounded in terms of the number of battlefields in the reduced game and $1/\varepsilon$, so the computation avoids repeated arithmetic on the potentially long original encodings.
We compute strategies for this reduced game with budgets $1$ and $\rho$.

For each player separately, we seek a feasible distribution that approximately minimizes the opponent's supremum expected payoff.
The infima of these two minimization problems sum to one by a minimax argument (\cref{lem:general:minimax}).
Thus the sum of the players' regrets is at most the sum of the two optimization errors, by \eqref{eq:prelim:security-sum}.
We restrict each player to a finite set of candidate allocations that supports such a distribution.
On each battlefield, the payoff from an opposing bid depends only on the candidate bids it ties or exceeds.
By using bids equal to or just above candidate bids, we obtain a second finite set of feasible allocations for the opponent that contains a best response to every distribution on the candidate allocations, even among all continuous deviations of the opponent.
Since these sets are too large to enumerate, we use their implicit descriptions in a multiplicative-weights procedure.
For each player, it returns a list of complete feasible candidate allocations; the strategy is the uniform distribution on this list.
The number of iterations depends logarithmically on the number of candidate bids, whose bound depends on the denominator of the budget ratio $\rho$.
Using $\rho$ therefore prevents a long encoding of $b$ from increasing the number of iterations.

We next undo the changes to the input, first transferring these strategies to the original budgets $1$ and $b$.
A small budget change can turn ties into wins, so closeness of $\rho$ to $b$ alone does not control the opponent's supremum payoff.
We adjust the bid amounts in the two lists so that every allocation satisfies the original budget and each opponent's supremum payoff increases by at most $O(\varepsilon)$.
Finally, we reconstruct allocations on the original battlefields from the allocations in these lists.
The reconstruction respects the budget in every draw and adds $O(\varepsilon)$ to the bound on the opponent's supremum payoff.

To sample from a player's strategy, choose a complete allocation on the reduced battlefields uniformly from that player's list and, independently, a string of random bits of a fixed length for its reconstruction.
These two are put together to sample uniformly from an implicit list of allocations on the original battlefields.
We give the main construction and correctness argument below.
\Cref{app:general:catalogue,app:general:source,app:general:transfer,app:general:cost} supply the deferred proofs, implementation details, and full cost analysis, including the cost of writing rational output.

\subsection{Reducing the battlefields and values}
\label{sec:general:reduction}

The construction uses the parameter $e = 2^{-a}$, where $a \ge 1$ is an integer selected to ensure $\varepsilon / 2 < e \le \varepsilon$.
Working with $e$ instead of $\varepsilon$ loses at most a factor of two and makes several parameters below, such as $8 / e$ and $128 / e$, powers of two.
Before computing strategies, we simplify the game in three steps, each of which changes payoffs by $O(e)$: we round the values, combine the battlefields with small values, and round the values of the resulting smaller game.

Both roundings use the following rule.
For positive numbers $\alpha_1, \ldots, \alpha_r$ summing to one and a scale $M > 0$, let $a_j = \lceil M \alpha_j \rceil$ and $A = \sum_j a_j$, and replace each $\alpha_j$ by $a_j / A$.
The new numbers are positive, sum to one, and have the common denominator $A$.
Since $M \alpha_j \le a_j < M \alpha_j + 1$, we have $M \le A < M + r$.
To bound the change in the values, add and subtract $M\alpha_j/A$ and apply the triangle inequality:
\[
    \begin{aligned}
        \sum_j \abs{\frac{a_j}{A} - \alpha_j}
        &\le \sum_j \abs{\frac{a_j}{A} - \frac{M\alpha_j}{A}}
            + \sum_j \abs{\frac{M\alpha_j}{A} - \alpha_j} \\
        &= \frac{\sum_j (a_j - M\alpha_j)}{A}
            + \frac{A-M}{A}\sum_j \alpha_j \\
        &= \frac{2(A-M)}{A}
        < \frac{2r}{M}.
    \end{aligned}
\]
The final inequality uses $A-M < r$ and $A \ge M$.
Each player's share of a battlefield lies in $[0, 1]$, so changing the values changes the payoff of every pair of allocations by at most the $\ell_1$ distance between the old and new values.
The same bound then holds for expected payoffs and for the opponent's supremum payoff against any fixed strategy.
Since several value vectors appear below, we write $U_w(\mu; c)$ for the opponent's supremum payoff, defined as in \cref{sec:prelim} with $w$ in place of the original value vector $v$.

First, round the original values with scale $256 n / e$.
This gives integer weights $w'_j = \lceil 256 n v_j / e \rceil \ge 1$ with sum $Q' < 256 n / e + n$, and values $v'_j = w'_j / Q'$ with $\sum_j \abs{v'_j - v_j} \le e / 128$.

Next, combine the battlefields with small values.
Call battlefield $j$ \emph{light} if $v'_j \le e / 10$.
If more than $k = 8 / e$ battlefields are light, replace all of them by $k$ \emph{representative battlefields}, each with value $V / k$, where $V$ is the sum of $v'_j$ over the light battlefields.
Battlefields that are not light remain unchanged.
If at most $k$ battlefields are light, no battlefields are combined.
Let $m$ be the number of battlefields in the resulting reduced game and $v^{\rm red}$ its value vector: each battlefield kept individually retains its value $v'_j$, and each representative battlefield has value $V / k$.
Since the values sum to one, fewer than $10 / e$ battlefields have value greater than $e / 10$, and the reduced game contains at most $k$ further battlefields.
Thus $m < 18 / e$ and $m \le n$, so $m = O(n_*)$.
Moreover, $m \ge 2$, since either $m = n$ or $m \ge k \ge 16$.
\Cref{sec:general:lift} converts allocations on the reduced battlefields back to the original battlefields at a payoff cost of at most $e / 10 + e / 8$ for each player (\cref{lem:general:lift}); the two terms come from the threshold $e / 10$ and the number $k = 8 / e$.

Finally, round the reduced values with scale $512 m / e$.
This gives integer weights $\widehat w_j = \lceil 512 m v^{\rm red}_j / e \rceil$ with sum $Q_v < 512 m / e + m$, and values $\widehat v_j = \widehat w_j / Q_v$ with $\sum_j \abs{\widehat v_j - v^{\rm red}_j} \le e / 256$.
\Cref{sec:general:source,sec:general:transfer} compute strategies for the reduced game with values $\widehat v$, whose common denominator depends on $m$ rather than $n$.
The conversion in \cref{sec:general:lift} uses the individual values $v'_j$ of the light battlefields and their small common denominator, which is why we do the first rounding before combining the light battlefields.

\subsection{Computing the strategy distributions}
\label{sec:general:source}

In this subsection, players $A$ and $B$ have budgets $1$ and $\rho$, where $\rho \in (0, 1)$ is the auxiliary budget ratio chosen in \cref{sec:general:transfer}, and the battlefield values are $\widehat v$.
We use the notation of \cref{sec:prelim} for this game, with $m$ battlefields in place of $n$.

We compute the two players' strategies separately.
For each player, the goal is a feasible strategy that nearly minimizes the opponent's supremum payoff.
Let
\[
    C_1 = \inf_P U_{\widehat v}(P; \rho)
    \qquad\text{and}\qquad
    C_\rho = \inf_P U_{\widehat v}(P; 1),
\]
where the first infimum is over distributions on $\mathcal X(1)$ and the second over distributions on $\mathcal X(\rho)$.
For a fixed strategy $P$ of $A$, the opponent's supremum payoff is $U_{\widehat v}(P; \rho)$, so $A$'s worst-case expected payoff is $1 - U_{\widehat v}(P; \rho)$.
Taking the supremum over $A$'s choice of $P$ gives $1 - C_1$.
For a fixed strategy $P$ of $B$, the quantity $U_{\widehat v}(P; 1)$ is $A$'s supremum payoff; taking the infimum over $B$'s choice of $P$ gives $C_\rho$.
The next lemma shows that $1 - C_1 = C_\rho$.

\begin{lemma}
\label{lem:general:minimax}
For $m \ge 2$ battlefields, positive values summing to one, and budgets $1$ and $\rho \in (0, 1)$, we have $C_1 + C_\rho = 1$.
\end{lemma}

An exact Nash equilibrium would imply this equality.
Earlier papers state that equilibrium existence for Colonel Blotto games follows from general theorems for discontinuous games \cites[Section~2.1]{perchet2024algorithmic}[Section~2]{boix2021multiplayer}.
However, the hypotheses of the theorems they cite can fail under our budget constraints and equal sharing of ties, as a two-battlefield example in \cref{app:general:minimax} shows.
We therefore prove the minimax equality directly in that appendix.

To handle the payoff discontinuities at ties, we first show that, against every fixed strategy of $B$, restricting $A$ to distributions whose marginals have no atoms leaves its supremum payoff unchanged.
We do so by randomly perturbing $A$'s allocations in a way that respects its budget and gives atomless marginals.
Against a fixed allocation of $B$, as the perturbation shrinks, $A$ wins with probability tending to $1/2$ on each battlefield where the original bids are equal and positive, and $A$ wins on each battlefield where both original bids are zero; on the other battlefields, the outcome is eventually unchanged.
Thus $A$'s limiting expected payoff is no smaller than its original payoff; averaging over $B$'s fixed strategy preserves this conclusion.
For a strategy of $A$ with atomless marginals, ties have probability zero against every allocation of $B$, so $A$'s expected payoff is continuous in $B$'s strategy.
The Kneser--Fan minimax theorem \cite[Theorem~4.2$'$]{sion1958general} imposes no topological condition on the maximizing player's strategies, so it applies to these strategies of $A$ and completes the proof.
The same perturbation verifies the condition on $A$'s strategies required by \cite[Remark~2.10(3)]{barelli2014majority}, which also gives the minimax equality; we explain this connection in \cref{app:general:minimax}.

The lemma reduces the computation to two separate problems.
Suppose that distributions $P_A$ on $\mathcal X(1)$ and $P_B$ on $\mathcal X(\rho)$ satisfy $U_{\widehat v}(P_A; \rho) \le C_1 + \delta$ and $U_{\widehat v}(P_B; 1) \le C_\rho + \delta$.
Then \eqref{eq:prelim:security-sum} and \cref{lem:general:minimax} give
\[
    \Reg_A(P_A, P_B) + \Reg_B(P_A, P_B)
    = U_{\widehat v}(P_B; 1) + U_{\widehat v}(P_A; \rho) - 1
    \le 2 \delta,
\]
so $(P_A, P_B)$ is a $2 \delta$-Nash equilibrium of this game.

For the rest of this subsection, fix one of the two players, whose strategy we compute; we refer to it as \emph{the player} and to the other as \emph{the opponent}.
Let $c \in \{1, \rho\}$ be its budget and $c'$ the opponent's budget, so that $\{c, c'\} = \{1, \rho\}$.
We will optimize over two finite sets: a set $X$ of feasible allocations for the player that supports a nearly optimal strategy, and a set $Y$ of feasible allocations for the opponent that does as well as every feasible deviation against strategies supported on $X$.
Both sets rest on the following observation.

\paragraph{Deviations against finitely many bids.}
Suppose that on each battlefield $j$, the player's bids lie in a finite set $\mathcal T_j \subseteq [0, \infty)$ that contains zero; we call its elements thresholds.
Each bid $y_j \ge 0$ of the opponent on battlefield $j$ then makes exactly one of the following choices: the \emph{closed choice} at $t \in \mathcal T_j$ if $y_j = t$, or the \emph{open choice} at $t$ if $y_j \notin \mathcal T_j$ and $t$ is the largest threshold below $y_j$.
Against a bid $x_j \in \mathcal T_j$ of the player, the closed choice at $t$ gives the opponent a tie if $x_j = t$, a win if $x_j < t$, and a loss if $x_j > t$; the open choice at $t$ gives a win if $x_j \le t$ and a loss if $x_j > t$.
Thus the opponent's share of each battlefield depends only on its choices, not on its exact bids.
The next lemma tests whether a tuple of choices can be made within the budget; one feasible deviation for each tuple that passes this test suffices to maximize the opponent's expected payoff.

\begin{lemma}
\label{lem:general:responses}
Let $c' > 0$, and fix one choice on each battlefield $j$, at a threshold $t_j \in \mathcal T_j$; some choices may be open and others closed.
There is a deviation $y \in \mathcal X(c')$ making all the specified choices if and only if
\begin{equation}
    \sum_j t_j \le c',
    \quad\text{with strict inequality if at least one choice is open.}
    \label{eq:general:affordable}
\end{equation}
Consequently, fix battlefield values $w$ and a distribution $P$ on allocations whose bids on each battlefield $j$ lie in $\mathcal T_j$.
For each tuple of choices satisfying \eqref{eq:general:affordable}, choose one feasible deviation that makes those choices.
The largest expected payoff against $P$ among these finitely many deviations is $U_w(P; c')$; in particular, the supremum is attained.
\end{lemma}

\begin{proof}
If $y \in \mathcal X(c')$ makes the choices, then $t_j \le y_j$ for every $j$, with strict inequality at every open choice.
Hence $\sum_j t_j \le \sum_j y_j \le c'$, and the first inequality is strict if some choice is open.
Conversely, if every choice is closed and $\sum_j t_j \le c'$, the deviation $(t_1, \ldots, t_m)$ makes the choices.
If $\sum_j t_j < c'$, choose $\eta > 0$ smaller than every gap between consecutive thresholds on every battlefield and with $m \eta < c' - \sum_j t_j$.
Bidding $t_j + \eta$ at each open choice and $t_j$ at each closed choice then makes the choices and respects the budget.
To show that one feasible deviation per tuple suffices to attain $U_w(P; c')$, note that deviations making the same choices have the same payoff against every allocation whose bids lie in the sets $\mathcal T_j$, and hence the same expected payoff against $P$.
Every feasible deviation makes some tuple of choices, which satisfies \eqref{eq:general:affordable}, and every such tuple is made by some feasible deviation.
The supremum is therefore a maximum over finitely many tuples, each attained by a feasible deviation.
\end{proof}

This lemma explains the failure in \cref{prop:general:grid}.
Measured in units of $\delta$, $B$'s profitable deviation bids $q - 1/2$ and $q + 1/2$, making open choices at the thresholds $q - 1$ and $q$ formed by $A$'s bids on the two battlefields.
The two thresholds sum to $(q - 1) + q = 2 q - 1$, which is less than $B$'s budget $2 q$, so these choices satisfy \eqref{eq:general:affordable}.
On the grid, however, each open choice costs at least one unit more than its threshold, so no grid deviation within the budget makes both choices.

\paragraph{Candidate allocations.}
We restrict the player's bids to a finite set $\mathcal S \subseteq [0, 1]$ of rationals, the \emph{candidate bids}, and let
\[
    X = \{x \in \mathcal S^m : \textstyle\sum_j x_j \le c\}
\]
be the set of \emph{candidate allocations}.
Bids in $X$ must be compared exactly with both budgets, so $\mathcal S$ depends on the denominator of $\rho$.
\Cref{sec:general:transfer} chooses $\rho$ with denominator at most $\overline q$, where $\overline q$ depends only on $m$ and $e$ and satisfies $\log \overline q = O((m / e) \log(m + 2))$; we construct $\mathcal S$ from this bound before computing either strategy.
Let $z = e / 2048$; the computation in this subsection loses $O(z)$ in payoff.

\begin{lemma}
\label{lem:general:candidates}
There is a finite set $\mathcal S \subseteq [0, 1] \cap \mathbb Q$ containing $0$ and $1$ such that, for every rational $\rho \in (0, 1)$ with denominator at most $\overline q$, both choices of $c$, and every distribution $P$ on $\mathcal X(c)$, there is a distribution $P^*$ on $X$ with
\[
    U_{\widehat v}(P^*; c') \le U_{\widehat v}(P; c') + z / 64.
\]
Every element of $\mathcal S$ has binary encoding length $O(\Lambda)$, where $\Lambda$ is an integer specified in \cref{app:general:parameters} with $\Lambda = O(\log \overline q + (m / z) \log(m + 2)) = \softO(m / z)$.
\end{lemma}

We sketch the proof here; \cref{app:general:comparator} gives the complete proof.
The distributions $P$ and $P^*$ appear only in the analysis; the algorithm uses the set $\mathcal S$, but not these distributions.
The proof modifies a random allocation with distribution $P$ in two steps, both of which keep every allocation feasible; $P^*$ is the distribution of the resulting allocation.
The first step rounds bids down so that at most $K' = 64 / z$ distinct bids remain on each battlefield; this increases the opponent's supremum payoff by at most $z / 64$.
The second step replaces these bids by elements of $\mathcal S$ without changing the opponent's supremum payoff.

Moving a bid by a small amount can turn a tie into a win or a loss, so the first step controls the marginal CDFs rather than distances between bids.
\Cref{lem:prelim:cdf} then bounds the change in expected payoffs by the largest change in a marginal CDF.
This step rounds each bid down to a quantile of its own marginal distribution.
It finds the bid's position in this distribution on a scale from $0$ to $1$, rounds this position down to a multiple of $1 / K'$, and takes the quantile at the rounded position.

Let $F_j$ be the CDF of $P$ on battlefield $j$, and let $Q_j(\tau) = \inf\{t \ge 0 : F_j(t) \ge \tau\}$ be its quantile at $\tau \in [0, 1]$.
Note that $Q_j(0) = 0$.
Let $x$ be a random allocation with distribution $P$.
Given $x$, let $\tau_1, \ldots, \tau_m$ be independent, with $\tau_j$ uniform on $[F_j(x_j-), F_j(x_j)]$.
We call $\tau_j$ the \emph{rank} of $x_j$, since it is the position of $x_j$ in the marginal distribution on battlefield $j$, scaled to $[0, 1]$.
If $F_j$ has no atom at $x_j$, the rank is $F_j(x_j)$.
Otherwise, it is a uniformly random position within the atom at $x_j$.
Each rank is uniform on $[0, 1]$.
When $F_j$ has no atoms, this is the familiar fact that $F_j(x_j)$ is uniform.
Give the bid $x_j$ one of the $K'$ \emph{labels} $0, \ldots, K' - 1$, namely $l_j = \min\{\lfloor K' \tau_j \rfloor, K' - 1\}$, so that $l_j / K'$ is the rank rounded down to a multiple of $1 / K'$ unless $\tau_j = 1$.
The \emph{anchor} of label $l$ on battlefield $j$ is the quantile $a_{jl} = Q_j(l / K')$, so $a_{j0} = 0$.
Replace the bid $x_j$ by the anchor $a_{j l_j}$ of its label, the quantile at the rounded-down rank $l_j / K'$.

Since $l_j / K' \le \tau_j \le F_j(x_j)$, the anchor $a_{j l_j}$ is at most $x_j$.
All coordinates are rounded down from the same allocation $x$, so every rounded allocation still spends at most $c$.
Because each rank is uniform, each label has probability $1 / K'$ on every battlefield.
By the definition of $Q_j$ and the right continuity of $F_j$, for every label $l$ and every $t \ge 0$, the anchor $a_{jl}$ is at most $t$ exactly when $l / K' \le F_j(t)$.
Hence the CDF of the rounded bid at $t \ge 0$ is the fraction of labels $l$ with $l / K' \le F_j(t)$, which lies between $F_j(t)$ and $F_j(t) + 1 / K'$; for $t < 0$, both CDFs are zero.
Thus every marginal CDF increases by at most $1 / K'$, and \cref{lem:prelim:cdf} shows that the opponent's expected payoff from every deviation increases by at most $1 / K' = z / 64$; so does its supremum payoff.

The anchors depend on $P$ and may be irrational or have long encodings, whereas the lemma needs one finite set $\mathcal S$ of candidate bids for all the ratios $\rho$ and distributions $P$ that it covers.
The second step therefore keeps the joint distribution of the labels fixed and replaces the anchors by rationals with bounded denominators.
After the first step, an allocation is specified by its tuple of labels $(l_1, \ldots, l_m)$: its bid on battlefield $j$ is $a_{j l_j}$.
We call $\sum_j a_{j l_j}$ the \emph{anchor sum} of the tuple.

Choose new anchors in $[0, 1]$ with $a_{j0} = 0$ that preserve the order of the anchors on each battlefield, including equalities.
The open and closed choices at a new anchor beat and tie the same labels as the corresponding choices at its old value, so their expected payoffs are unchanged.
Also preserve the comparison ($<$, $=$ or $>$) of every anchor sum with each budget $c$ and $c'$.
The comparisons with $c$ keep the player's allocations feasible.
By \cref{lem:general:responses}, applied with the anchors as thresholds, the comparisons with $c'$ preserve exactly which tuples of choices the opponent can afford.
The opponent's supremum payoff is therefore unchanged.

These conditions form a finite system of linear equalities and inequalities with coefficients in $\{0, \pm 1\}$ and right-hand sides $0$, $1$ or $\rho$.
To obtain anchors with small denominators, introduce a common margin for the strict inequalities and maximize it by a linear program.
A positive margin is feasible because the finitely many strict inequalities all have positive slack at the original anchors.
Thus an optimal vertex satisfies all the required comparisons.

At this vertex, the anchors and the margin solve a nonsingular linear system with the stated coefficients and right-hand sides.
Cramer's rule and Hadamard's inequality therefore bound their denominators by a number $D$ depending only on $m$, $z$ and $\overline q$, with $\log D = O(\Lambda)$.
The set $\mathcal S$ contains all rationals in $[0, 1]$ with denominator at most $D$; \cref{app:general:comparator} gives its exact definition and the remaining proof details.
The new anchors therefore give a distribution $P^*$ on $X$.
Only the first step increased the opponent's supremum payoff, by at most $z / 64$, which proves the lemma.

Let $C_c$ be the infimum of $U_{\widehat v}(P; c')$ over distributions $P$ on $\mathcal X(c)$.
Applying \cref{lem:general:candidates} to a distribution $P$ on $\mathcal X(c)$ with $U_{\widehat v}(P; c') \le C_c + z / 64$ gives a distribution $P^*$ on $X$ with
\begin{equation}
    U_{\widehat v}(P^*; c') \le C_c + z / 32.
    \label{eq:general:comparator}
\end{equation}

\paragraph{Response allocations.}
To construct $Y$, apply \cref{lem:general:responses} with the threshold set $\mathcal T_j = \mathcal S$ on each battlefield $j$.
The set $\mathcal S$ contains zero and all bids of candidate allocations.
We use one positive rational increment $\Delta$ for all tuples of choices and both possible opponent budgets, $1$ and $\rho$.
For each tuple satisfying \eqref{eq:general:affordable}, with thresholds $t_1, \ldots, t_m$, include in $Y$ the allocation that bids $t_j$ at a closed choice on battlefield $j$ and $t_j + \Delta$ at an open choice.
We call the allocations in $Y$ \emph{response allocations}.

Choose $\Delta$ smaller than every gap between consecutive candidate bids.
Then $t_j + \Delta$ exceeds $t_j$ without reaching any larger candidate bid, so it makes the open choice at $t_j$.
For each opponent budget $c' \in \{1, \rho\}$, also require $m \Delta < c' - \sum_j t_j$ whenever $t_1, \ldots, t_m \in \mathcal S$ and $\sum_j t_j < c'$.
An affordable tuple containing an open choice has threshold sum strictly below $c'$.
Adding $\Delta$ on at most $m$ battlefields therefore keeps the response allocation within the opponent's budget.
\Cref{app:general:codes} gives an explicit $\Delta$ with $O(m\Lambda)$ bits.

By \cref{lem:general:responses}, every distribution $P$ on $X$ satisfies
\begin{equation}
    U_{\widehat v}(P; c') = \max_{y \in Y} \E_{x \sim P}\Bigl[\sum_j \widehat v_j h(y_j, x_j)\Bigr].
    \label{eq:general:responses}
\end{equation}
We therefore seek a distribution $P$ on $X$ against which every response allocation $y \in Y$ gives the opponent an expected payoff of at most $C_c + O(z)$.
The distribution $P^*$ on $X$ from \eqref{eq:general:comparator} satisfies this bound.
We compare the algorithm's output with $P^*$ in the proof of \cref{lem:general:source}.

\paragraph{Computing a distribution on candidate allocations.}
The following lemma states what the algorithm achieves.
We then describe the algorithm and prove the lemma.

\begin{lemma}
\label{lem:general:source}
For each budget $c \in \{1, \rho\}$, the algorithm below computes a list of $T$ candidate allocations, where $T = \Theta(\Lambda / z^2)$ is a power of two specified in \cref{app:general:parameters}.
Let $P_A$ and $P_B$ be the uniform distributions on the lists for $c = 1$ and $c = \rho$, respectively, counting repeated allocations.
Then $U_{\widehat v}(P_A; \rho) \le C_1 + 3 z / 64$ and $U_{\widehat v}(P_B; 1) \le C_\rho + 3 z / 64$, so by \cref{lem:general:minimax},
\begin{equation}
    U_{\widehat v}(P_A; \rho) + U_{\widehat v}(P_B; 1) \le 1 + 3 z / 32.
    \label{eq:general:source}
\end{equation}
The computation is deterministic and uses $\softO(m \Lambda^2 / z^2 + m \Lambda / z^4)$ real-RAM operations, or $\softO(m \Lambda^3 / z^2 + m \Lambda^2 / z^4)$ bit operations and $\softO(m \Lambda^2 / z^2)$ working bits.
\end{lemma}

Expected payoffs depend only on the marginal distributions on the battlefields.
This suggests optimizing over marginals, as in the General Lotto game, rather than over distributions on the exponentially many candidate allocations.
As recalled in \cref{sec:prg}, however, marginals need not be realizable by allocations that satisfy the budget in every realization.

For each side, the algorithm maintains \emph{auxiliary marginals}, one distribution per battlefield, with no budget constraint linking these distributions.
Each side's penalty assigns weights to the CDF differences between its auxiliary marginals and a feasible allocation for that side, as described below.
In each iteration, budget-constrained optimization selects a candidate allocation for the player and a response allocation for the opponent, each approximately minimizing the corresponding side's current penalty.
Multiplicative weights then updates the auxiliary marginals using both payoffs and penalties.
It also updates the weights in these penalties using the CDF differences.
The algorithm returns the uniform distribution on the player's list of $T$ candidate allocations, one selected in each iteration.

To bound the opponent's payoff against this output, the proof also considers the uniform distribution on the opponent's list of $T$ response allocations generated by the algorithm.
Both distributions are feasible by construction.
Their marginals may differ from the corresponding auxiliary marginals averaged over the $T$ iterations.
In the regret analysis, the CDF penalties compensate for the payoff changes when the averaged auxiliary marginals are replaced by the marginals of these feasible distributions.
This yields the payoff bound for the returned distribution, as the proof of \cref{lem:general:source} shows.

We first put the player's and opponent's bids on a common ordered domain.
Each candidate bid, and each candidate bid plus $\Delta$, receives an integer code, so that the order of the codes is the order of the bids.
For given bids on a battlefield, comparing their codes determines the winner or a tie, and hence the payoff on that battlefield.
Budget feasibility still depends on the numerical bid amounts.
To support efficient encoding and updates, the construction reserves separate positions for candidate bids and candidate bids plus $\Delta$, and pads the code range to obtain the full code set $I = \{0, \ldots, 2^\Lambda - 1\}$ (\cref{app:general:codes}).
Some of these integer positions are unused and represent no bid.
The auxiliary marginals are distributions on $I$ and may assign mass to unused codes, but selected allocations use only codes that represent actual bids.

We now introduce CDF notation for auxiliary marginals, fixed allocations, and distributions on allocations.
Here $j \in [m]$ is the battlefield index, and $a \in I$ is the integer code at which we evaluate a CDF.
Each CDF gives the probability of a code at most $a$ on battlefield $j$.
Write $p = (p_j)_{j \in [m]}$ for a family of auxiliary distributions on $I$, and identify the bids of an allocation $x \in X \cup Y$ with their codes.
The three cases are
\[
\begin{array}{l|l}
    \text{object} & \text{CDF on battlefield } j \\ \hline
    \text{auxiliary marginals } p & F_{p, j}(a) = \sum_{a' \le a} p_j(a') \\
    \text{fixed allocation } x & F_{x, j}(a) = \ind\{x_j \le a\} \\
    \text{distribution } \pi \text{ on } X \text{ or } Y & F_{\pi, j}(a) = \Pr_{x \sim \pi}[x_j \le a]
\end{array}
\]
For a fixed allocation, the bid is deterministic, so $F_{x, j}(a)$ is either $0$ or $1$.
A distribution on allocations averages these values, giving $F_{\pi, j}(a) = \E_{x \sim \pi}[F_{x, j}(a)]$.
We write $F_p = (F_{p, j})_{j \in [m]}$ for the full family of CDFs, and use $F_x$ and $F_\pi$ in the same way.

To relate CDF differences to payoffs, let $p$ and $q$ be the player's and opponent's auxiliary marginals and define the opponent's payoff
\[
    f(p, q) = \sum_j \widehat v_j \sum_{a, a' \in I} p_j(a) \, q_j(a') \left(\ind\{a' > a\} + \tfrac12 \ind\{a' = a\}\right).
\]
Thus, on each battlefield $j$, $f$ applies the win-and-tie rule to two independently drawn codes: the player's code from $p_j$ and the opponent's code from $q_j$.
Hence, if $p$ and $q$ are the marginals of distributions on $X$ and $Y$, then $f(p, q)$ is the opponent's expected payoff under these distributions.
We also allow allocations and distributions on allocations as arguments of $f$, evaluating $f$ using their marginals.
The function $f$ is linear in each marginal family separately.

To measure the effect of changing marginals, define the weighted CDF distance between two marginal families $p$ and $p'$ by
\[
    d_{\widehat v}(p, p') = \sum_j \widehat v_j \max_{a \in I} \abs{F_{p, j}(a) - F_{p', j}(a)}.
\]
This distance bounds the change in the opponent's payoff when one side's marginals change.
By \eqref{eq:prelim:half-tie} and the identity $h(s, t) = 1 - h(t, s)$, which use only the order of bids, replacing $p$ by $p'$ while keeping $q$ fixed changes $f(p, q)$ by at most $d_{\widehat v}(p, p')$.
Likewise, replacing $q$ by $q'$ while keeping $p$ fixed changes $f(p, q)$ by at most $d_{\widehat v}(q, q')$.
Using the CDFs in the table, we also write $d_{\widehat v}(p, x)$ or $d_{\widehat v}(p, \pi)$ to compare auxiliary marginals with a fixed allocation or a distribution on allocations.

We next express this CDF distance as a maximum of linear penalties.
The penalty weights are represented by \emph{signed multipliers}.
A signed multiplier is a family $\lambda = (\lambda_j)_{j \in [m]}$ in which each $\lambda_j$ is a distribution on pairs $(s, a) \in \{-1, 1\} \times I$.
The code $a$ specifies where the CDFs are compared, and the sign $s$ allows a difference in either direction to contribute positively.
A signed multiplier is paired with a family of CDFs by
\[
    \lambda \cdot F_p = \sum_j \widehat v_j \sum_{s, a} \lambda_j(s, a) \, s \, F_{p, j}(a),
\]
and similarly for $F_x$.

For auxiliary marginals $p$ and a fixed allocation $x$, the penalty $\lambda \cdot (F_p - F_x)$ averages signed CDF differences on each battlefield and sums these averages with weights $\widehat v_j$.
With $p$ and $x$ fixed, each average is at most the largest absolute CDF difference on that battlefield.
To attain this bound, put all the mass of $\lambda_j$ on a pair $(s, a)$ whose code $a$ maximizes $\abs{F_{p, j}(a) - F_{x, j}(a)}$ and whose sign $s$ makes the signed difference nonnegative.
Maximizing independently on each battlefield therefore gives
\begin{equation}
    \max_\lambda \lambda \cdot (F_p - F_x) = d_{\widehat v}(p, x).
    \label{eq:general:cdf-duality}
\end{equation}
The identity also holds with a distribution $\pi$ in place of the fixed allocation $x$.
Each fixed multiplier thus gives a penalty that is linear in the CDFs and that the algorithm can use to select an allocation.
Since each $\lambda_j$ is a distribution, the algorithm can update these penalty weights by multiplicative weights.

The algorithm maintains auxiliary marginals $p_t$ for the player and $q_t$ for the opponent, together with signed multipliers $\lambda_t$ for the player's penalty and $\mu_t$ for the opponent's.
Each battlefield distribution starts uniform on its own domain: $I$ for the auxiliary marginals, and $\{-1, 1\} \times I$ for the multipliers.
Each iteration has two steps: select feasible allocations using the current multipliers, then update the four families of distributions.

The first step in iteration $t = 1, \ldots, T$ selects a candidate allocation $x_t \in X$ for the player and a response allocation $y_t \in Y$ for the opponent.
These allocations approximately maximize $\lambda_t \cdot F_x$ over $x \in X$ and $\mu_t \cdot F_y$ over $y \in Y$, respectively.
The allowed additive error is specified in \cref{lem:general:updates}.
Since $p_t$ and $\lambda_t$ are fixed at this step, $\lambda_t \cdot F_{p_t}$ does not depend on the choice of $x$.
Hence $x_t$ also approximately minimizes the player's current penalty $\lambda_t \cdot (F_{p_t} - F_x)$ over $X$, with the same additive approximation error.
The opponent's term $\mu_t \cdot F_{q_t}$ is likewise independent of $y$, so $y_t$ approximately minimizes the opponent's current penalty $\mu_t \cdot (F_{q_t} - F_y)$ over $Y$.

The second step updates the four families by multiplicative weights to obtain the families for iteration $t + 1$.
All four updates use $p_t$, $q_t$, $\lambda_t$ and $\mu_t$ from the start of iteration $t$.
The multiplier updates also use the allocations $x_t$ and $y_t$ selected in the first step.

The following table gives the objective of each update.
In each row, the family in the first column is the variable being updated, and all quantities with subscript $t$ are held fixed.
Each objective is linear in the family being updated.
\[
\begin{array}{c|l|c}
    \text{family} & \text{objective in iteration } t & \text{direction} \\ \hline
    p & f(p, q_t) + \lambda_t \cdot F_p & \text{minimize} \\
    q & f(p_t, q) - \mu_t \cdot F_q & \text{maximize} \\
    \lambda & \lambda \cdot (F_{p_t} - F_{x_t}) & \text{maximize} \\
    \mu & \mu \cdot (F_{q_t} - F_{y_t}) & \text{maximize}
\end{array}
\]
The first two rows omit the terms $-\lambda_t \cdot F_{x_t}$ and $+\mu_t \cdot F_{y_t}$ from the objectives for updating $p$ and $q$, respectively.
These terms are constant during those updates, so omitting them changes neither the updates nor the regrets defined below.

For each row of the table and each battlefield $j$, obtain the loss vector by taking the coefficients of that battlefield's probabilities in the objective and dividing them by $\widehat v_j$.
For the three maximized objectives, negate the resulting values so that a smaller loss always corresponds to a better objective value.
Multiplicative weights reweights codes for $p$ and $q$, and pairs $(s, a)$ for $\lambda$ and $\mu$, giving greater relative weight to elements with smaller loss and then normalizing each battlefield distribution.
The update of $p$ favors codes that lower the opponent's payoff plus the player's penalty.
The update of $q$ favors codes that raise the opponent's payoff minus the opponent's penalty.
The multiplier updates give more weight to pairs $(s, a)$ with larger signed CDF differences.
For the player, these differences compare $F_{p_t, j}$ with $F_{x_t, j}$; for the opponent, they compare $F_{q_t, j}$ with $F_{y_t, j}$.
After $T$ iterations, the algorithm outputs the list $x_1, \ldots, x_T$ of the player's candidate allocations, representing the uniform distribution on this list.

To analyze the output, we will combine regret bounds for the four families.
Here regret measures how much better a fixed family, used in every iteration, would have done in hindsight.
It differs from the players' regrets in \eqref{eq:prelim:regret}.
For the family $p$, let $\phi_t$ denote its objective in the table in iteration $t$.
Its average regret relative to a fixed comparison family $p'$ is
\[
    \frac1T \sum_{t = 1}^T \bigl(\phi_t(p_t) - \phi_t(p')\bigr).
\]
Both terms use the same $q_t$ and $\lambda_t$; the comparison changes only the player's auxiliary marginals.
The regrets of $q$, $\lambda$ and $\mu$ are defined in the same way, with the difference reversed for maximized objectives.

To choose the update rate $\eta$ and the number of iterations $T$, first suppose that the losses and updates are computed exactly.
For each battlefield distribution in any of the four families, the standard multiplicative-weights regret bound depends logarithmically on the number of codes or pairs $(s, a)$ in its domain.
This number is at most $2^{\Lambda + 1}$, and all losses have magnitude at most $2$, giving an average regret bound of
\[
    O\!\left(\frac{\Lambda}{\eta T} + \eta\right).
\]

The regret in a family's objective is the $\widehat v_j$-weighted sum of its battlefield regrets, undoing the division by $\widehat v_j$ in the loss definition.
Since $\sum_j \widehat v_j = 1$, this weighted average satisfies the same bound.
Choosing $\eta = \Theta(z)$ and $T = \Theta(\Lambda / z^2)$ makes both terms $O(z)$, so each family has average regret $O(z)$.
There are four families, so the sum of their average regrets is also $O(z)$.

We must implement the two steps without enumerating $X$, $Y$, or the full code set $I$.
We need both a small sum of the four average regrets and feasible allocations that nearly maximize $\lambda_t \cdot F_x$ and $\mu_t \cdot F_y$ in each iteration.
For the updates, the implementation approximates CDFs using summaries of the maintained distributions and evaluates the weights assigned to codes and pairs $(s, a)$ only to a prescribed numerical precision.
These approximations introduce additional errors, controlled in \cref{app:general:source}.

For either side, the objective used to select an allocation is a sum of battlefield terms, each depending only on the bid on that battlefield.
Maximizing this sum subject to that side's budget is a knapsack problem.
An additive error in the objective is acceptable, so we use a dynamic program that combines bid choices across battlefields (\cref{app:general:profit}).
The program groups bid choices on subsets of battlefields by rounding their total objective contributions to a finite grid, keeping the number of table entries small.
The program nevertheless tests the budget exactly.
For the player's candidate allocation $x_t$, the program uses only candidate bids and checks that the numerical bid amounts sum to at most $c$.
For the opponent's response allocation $y_t$, it records the exact sum of the chosen thresholds and whether any choice is open, as \eqref{eq:general:affordable} requires.
On each battlefield, we partition $I$ into intervals of consecutive codes so that the rounded objective contribution has the same value throughout each interval.
From every interval containing a bid allowed for that side, the program keeps the smallest allowed bid: larger allowed bids in that interval give the same contribution and use more budget.
Different intervals may have different objective contributions, and the dynamic program chooses among all the retained bids.
Quantile summaries of the multiplier distributions, described in \cref{app:general:updates}, keep the number of these intervals small.

For the multiplicative-weights updates, the algorithm uses summaries of all four families to approximate the losses.
Each summary approximates a battlefield distribution by a small weighted list of codes, or of pairs $(s, a)$ for a multiplier.
The summary weights are updated in every iteration, and the lists of codes or pairs are chosen afresh periodically.
For each maintained battlefield distribution, a binary tree groups consecutive elements of its domain: codes for $p$ and $q$, and pairs $(s, a)$ ordered by sign and then by code for $\lambda$ and $\mu$.
The tree stores accumulated updates and applies any change shared by all elements of an interval to that interval at once.
The following lemma bounds both the four average regrets and the additive error in each allocation choice, including the effects of the summaries and the prescribed numerical precision.

\begin{lemma}
\label{lem:general:updates}
With the parameters of \cref{app:general:parameters}, there are numbers $R, \beta_{\rm opt} \ge 0$ with $R + 2 \beta_{\rm opt} < z / 64$ such that the implementation in \cref{app:general:source} has the following properties.
\begin{enumerate}
    \item For every choice of fixed comparison families for $p$, $q$, $\lambda$ and $\mu$, the sum of the four average regrets is at most $R$.
    Here $p_t$, $q_t$, $\lambda_t$ and $\mu_t$ are the families maintained by the implementation, and the regrets are computed with the exact objectives in the table above.
    \item In every iteration, $x_t \in X$, $y_t \in Y$, and
    \[
        \lambda_t \cdot F_{x_t} \ge \max_{x \in X} \lambda_t \cdot F_x - \beta_{\rm opt},
        \qquad
        \mu_t \cdot F_{y_t} \ge \max_{y \in Y} \mu_t \cdot F_y - \beta_{\rm opt}.
    \]
    \item For both budgets together, the implementation satisfies the computation bounds of \cref{lem:general:source}.
\end{enumerate}
\end{lemma}

\Cref{app:general:regret} proves the first two properties and \cref{app:general:ledger} the third.

\begin{proof}[Proof of \cref{lem:general:source}]
Fix a budget $c \in \{1, \rho\}$.
Let $\overline x$ be the algorithm's output distribution, uniform on the player's list of candidate allocations $x_1, \ldots, x_T$.
Let $\overline y$ be the uniform distribution on the opponent's list of response allocations $y_1, \ldots, y_T$.
Fix any distribution $\pi$ on the player's candidate allocations $X$ and any distribution $\sigma$ on the opponent's response allocations $Y$.
We first show that
\[
    f(\overline x, \sigma) < f(\pi, \overline y) + z / 64.
\]
Taking $\pi = P^*$ from \eqref{eq:general:comparator} will then give the desired payoff bound, because $\overline y$ is a feasible opponent strategy.

We use Part~1 of \cref{lem:general:updates} to bound the sum of the four regrets.
Fix signed comparison multipliers $\lambda$ and $\mu$, held constant across all iterations.
Using the marginals of $\pi$ and $\sigma$ and these multipliers as comparison families, the cumulative regrets for $p$, $q$, $\lambda$ and $\mu$, respectively, are
\begin{align*}
    & \textstyle\sum_t \bigl[f(p_t, q_t) + \lambda_t \cdot F_{p_t} - f(\pi, q_t) - \lambda_t \cdot F_\pi\bigr], \\
    & \textstyle\sum_t \bigl[f(p_t, \sigma) - \mu_t \cdot F_\sigma - f(p_t, q_t) + \mu_t \cdot F_{q_t}\bigr], \\
    & \textstyle\sum_t \bigl[\lambda \cdot (F_{p_t} - F_{x_t}) - \lambda_t \cdot (F_{p_t} - F_{x_t})\bigr], \\
    & \textstyle\sum_t \bigl[\mu \cdot (F_{q_t} - F_{y_t}) - \mu_t \cdot (F_{q_t} - F_{y_t})\bigr].
\end{align*}
Since \cref{lem:general:updates} bounds the sum of the average regrets by $R$, these cumulative regrets sum to at most $T R$.
In this sum, the terms $f(p_t, q_t)$ cancel, as do the terms $\lambda_t \cdot F_{p_t}$ and $\mu_t \cdot F_{q_t}$.
The remaining terms involving the current multipliers $\lambda_t$ and $\mu_t$ compare the selected allocations with $\pi$ and $\sigma$.
For the player, $\lambda_t \cdot F_\pi$ is an average of the values $\lambda_t \cdot F_x$ over $x \in X$, each of which is at most $\lambda_t \cdot F_{x_t} + \beta_{\rm opt}$ by Part~2 of \cref{lem:general:updates}.
The same argument applies to $\sigma$ on $Y$ and the opponent's multiplier $\mu_t$.
Each of the two allocation choices therefore contributes at least $-\beta_{\rm opt}$ per iteration, giving
\[
    \textstyle\sum_t \bigl[\lambda_t \cdot (F_{x_t} - F_\pi) + \mu_t \cdot (F_{y_t} - F_\sigma)\bigr] \ge -2 T \beta_{\rm opt}.
\]
Removing these terms from the regret sum costs at most $2 T \beta_{\rm opt}$, so the remaining terms sum to at most $T (R + 2 \beta_{\rm opt})$.

Let $\widetilde p$ and $\widetilde q$ be the auxiliary marginals averaged over the $T$ iterations, so $\widetilde p = T^{-1}\sum_t p_t$ and $\widetilde q = T^{-1}\sum_t q_t$.
The marginal CDFs of $\overline x$ and $\overline y$ are the averages of $F_{x_t}$ and $F_{y_t}$ over these same iterations.
Dividing the bound on the remaining regret terms by $T$ and using linearity in each marginal family gives
\[
    f(\widetilde p, \sigma) - f(\pi, \widetilde q)
    + \lambda \cdot (F_{\widetilde p} - F_{\overline x})
    + \mu \cdot (F_{\widetilde q} - F_{\overline y})
    \le R + 2 \beta_{\rm opt} < z / 64.
\]
The inequality holds for every fixed choice of the comparison multipliers $\lambda$ and $\mu$.
Maximizing over these multipliers and applying \eqref{eq:general:cdf-duality} replaces the two pairings by $d_{\widehat v}(\widetilde p, \overline x)$ and $d_{\widehat v}(\widetilde q, \overline y)$.
The CDF distances bound the changes in payoff when we replace the auxiliary marginals by the marginals of $\overline x$ and $\overline y$.
Thus
\begin{align*}
    f(\overline x, \sigma) - f(\pi, \overline y)
    &\le f(\widetilde p, \sigma) - f(\pi, \widetilde q)
        + d_{\widehat v}(\widetilde p, \overline x)
        + d_{\widehat v}(\widetilde q, \overline y) \\
    &< z / 64
\end{align*}
for all $\pi$ and $\sigma$.
In particular, $(\overline x, \overline y)$ is a $z / 64$-Nash equilibrium of the game restricted to $X$ and $Y$, with payoff $f$ to the opponent.

To obtain the payoff bound in \cref{lem:general:source}, take $\pi = P^*$ from \eqref{eq:general:comparator}, and let $\sigma$ be the point mass at an arbitrary response allocation $y \in Y$.
Since $\overline y$ is supported on allocations within the opponent's budget $c'$, we have $f(P^*, \overline y) \le U_{\widehat v}(P^*; c') \le C_c + z / 32$.
The comparison inequality therefore gives $f(\overline x, y) < C_c + 3 z / 64$ for every $y \in Y$.
By \eqref{eq:general:responses}, the largest payoff over $Y$ equals the opponent's supremum over all feasible deviations, so $U_{\widehat v}(\overline x; c') < C_c + 3 z / 64$.

Both penalties are needed here.
The player's penalty transfers the guarantee from $\widetilde p$ to the feasible distribution $\overline x$.
The opponent's penalty permits the replacement of $\widetilde q$ by $\overline y$.
The feasibility of $\overline y$ then allows the comparison with $P^*$.

The algorithm returns $\overline x$, which is $P_A$ for $c = 1$ and $P_B$ for $c = \rho$.
This proves the two bounds of the lemma.
Part~3 of \cref{lem:general:updates} gives the computation bounds.
\end{proof}

\subsection{Transferring the distributions to the original budgets}
\label{sec:general:transfer}

We now return from the auxiliary budgets $1$ and $\rho$ to the original budgets $1$ and $b$.
\Cref{lem:general:source} gives each player a uniform distribution on a list of $T$ feasible allocations for the auxiliary game.
We change the bids in each player's list while allowing only a small increase in the other player's supremum payoff.
Merely scaling $B$'s bids by $b/\rho$ need not suffice, even when $\rho$ is very close to $b$.
A combination of bids that the opponent could only tie may become cheap enough to beat on every battlefield.
We first replace the bids by a small number of anchors on each battlefield, then adjust those anchors to control what the opponent can afford.

\paragraph{Replacing bids by anchors.}
Store each player's list as a table with one allocation per row, retaining repeated allocations as separate rows.
Let $K = 128/e$; the choice of $T$ in \cref{app:general:parameters} makes $T$ divisible by $K$.
For each player and each battlefield, sort a separate copy of the $T$ bids in increasing order, keeping equal bids in the order of their allocations in the original list.
Split the sorted bids into $K$ consecutive blocks of size $T/K$, numbered $0, \ldots, K - 1$.
Label each bid in the original table by its block number.
Replace bids in block zero by zero, and bids in block $l \ge 1$ by the largest bid in block $l - 1$.
These replacement bids are the \emph{anchors}.
No bid increases, so every row remains feasible.
Fix a battlefield and any bid amount $t \ge 0$.
The marginal CDF at $t$ is the fraction of the $T$ bids that are at most $t$.
Replacing bids by their anchors can move them from above $t$ to at most $t$ only in the first block containing a bid greater than $t$.
Since each block contains $T/K$ bids, each marginal CDF increases by at most $1/K$, as in the first step of the proof of \cref{lem:general:candidates}.
By \cref{lem:prelim:cdf}, the opponent's supremum payoff increases by at most $1/K$.

The resulting distribution is specified by a \emph{label table} $(l_{ij})$ and an \emph{anchor table} $u = (u_{jl})$.
Row $i$ bids $u_{j,l_{ij}}$ on battlefield $j$.
The label table records which labels occur together in each allocation; it will stay fixed while we change the anchor table.
Each label has probability $1/K$ on every battlefield, although several labels may have the same anchor.
We retain these labels separately because the adjustment may give them different anchors.

\paragraph{Which player's anchors must change.}
We first express the anchors as fractions of the budget for which they were computed.
Player $A$'s budget is one, so its anchor table, denoted $u^A$, is unchanged.
For player $B$, divide every anchor by $\rho$ to obtain the table $u^B$.
For either player, the normalized bids selected by each row of its label table sum to at most one.
The same table $u^B$ specifies bids $\rho u^B$ when $B$'s budget is $\rho$, and bids $b u^B$ when its budget is $b$.
Thus keeping $u^B$ fixed means keeping the fractions of $B$'s budget fixed, not keeping its actual bids fixed.

We compare the effect of the budget change on the two strategies separately.
Against $A$'s table $u^A$, player $B$'s budget changes from $\rho$ to $b$.
Against $B$'s strategy, we instead measure all bids in units of $B$'s budget.
Dividing both players' bids by $\rho$ in the auxiliary game leaves $B$'s table as $u^B$ and gives $A$ budget $1/\rho$.
In the original game, dividing both players' bids by $b$ again leaves the same table $u^B$, but gives $A$ budget $1/b$.
Each change of units preserves comparisons of bids, including ties, and hence preserves payoffs.

Suppose $\rho < b$.
Player $B$ can then spend more against the fixed table $u^A$, so we must adjust $A$'s anchors to control $B$'s supremum payoff.
Against the fixed table $u^B$, however, $A$'s budget in these units decreases from $1/\rho$ to $1/b$.
This cannot increase $A$'s supremum payoff, so $B$'s normalized anchors need no adjustment.
If $\rho > b$, the comparison reverses: $B$ can spend less against $u^A$, whereas $A$'s budget against $u^B$ increases from $1/\rho$ to $1/b$.
We therefore adjust only $B$'s normalized anchors in this case.
If $\rho = b$, neither table needs an adjustment.

For an anchor table $u$, write $U_{\widehat v}(u;c)$ for the opponent's supremum payoff with budget $c$ against the uniform distribution specified by $u$ and its label table.
In particular, $U_{\widehat v}(u^A;\rho)$ is $B$'s supremum payoff in the auxiliary game, and $U_{\widehat v}(u^B;1/\rho)$ is $A$'s supremum payoff in that game.
The anchor replacement increased each of these two payoffs by at most $1/K$.
Together with \cref{lem:general:source}, this gives the bound before adjusting either table:
\begin{equation}
    U_{\widehat v}(u^A;\rho) + U_{\widehat v}(u^B;1/\rho)
    \le 1 + 3z/32 + 2/K.
    \label{eq:general:quantized}
\end{equation}

\paragraph{Adjusting the anchors.}
Fix the player whose normalized anchor table $u$ must be adjusted.
Let $\beta_0 < \beta$ be the opponent's old and new budgets, measured in units of that player's budget as above.
We keep the label table fixed and change only the anchor values.
The goal is to limit the increase in the opponent's supremum payoff despite its larger budget.

To compare what the opponent can achieve before and after the adjustment, choose one label $l_j$ on each battlefield $j$.
A bid above the current anchor $u_{j l_j}$ beats every bid carrying a label from $0$ through $l_j$.
Since each label has probability $1/K$, these wins contribute at least $\widehat v_j(l_j + 1)/K$ to the opponent's expected payoff.
We summarize this contribution by the integer score
\[
    s_{j l} = \left\lfloor S\widehat v_j(l + 1)/K \right\rfloor,
    \qquad S = \operatorname{pow2ceil}(128m/e).
\]
Here $S$ is the least power of two at least $128m/e$, and $s_{jl}/S$ underestimates $\widehat v_j(l + 1)/K$ by less than $1/S$.
The total score of the chosen labels is $\sum_j s_{j l_j}$.
If their anchors sum to strictly less than $\beta_0$, the opponent can bid slightly above each of them within its old budget.
Let $h_*$ be the largest total score among choices satisfying this strict budget test; a dynamic program computes it.
Thus $h_*/S$ is a lower bound on the opponent's old supremum payoff.

We now seek new anchors $u'$ with the following properties.
The anchor of label zero stays at zero, and the anchors on each battlefield remain nondecreasing in the label.
The bids selected by each row of the label table must still sum to at most one.
In addition, for every choice of labels with total score greater than $h_*$, we require
\[
    \sum_j u'_{j l_j} > \beta.
\]
Such a choice is then unaffordable even if the opponent only tries to tie the selected anchors.

Consider any deviation with budget $\beta$ against the new anchors.
On each battlefield $j$, let $l_j$ be the largest label whose anchor is at most the deviation's bid on that battlefield.
These selected anchors sum to at most $\beta$, so the constraint above forces their total score to be at most $h_*$.
Counting every tied bid as a win gives an upper bound on the deviation's expected payoff.
On battlefield $j$, this counts exactly the bids with labels $0,\ldots,l_j$.
Allowing an error of less than $1/S$ per battlefield from rounding the scores therefore bounds the payoff by $h_*/S + m/S$.
The opponent's new supremum payoff is consequently at most its old supremum plus $m/S \le e/128$.
\Cref{lem:general:transfer} formalizes this argument.

It remains to choose $\rho$ so that such an adjustment is possible.
To obtain the strict inequalities, we first use weak inequalities: every choice of labels with score greater than $h_*$ must have anchor sum at least a common lower bound.
The old anchors satisfy these inequalities when the lower bound is $\beta_0$, by the definition of $h_*$.
We seek feasible anchors as this lower bound increases past $\beta$, preserving the order of the anchors and the budget of every allocation.

The argument in \cref{app:general:transfer} shows that, as the opposing budget increases, these weak inequalities can lose feasibility only at certain \emph{critical budget ratios}.
A determinant argument using the bounded integer coefficients of the inequalities shows that every critical ratio is rational, with a denominator bound depending only on $m$ and $e$.
If $b$ itself has a denominator within this bound, we take $\rho = b$ and no adjustment is needed.
When an adjustment is needed, we choose a noncritical $\rho$ and a target ratio on the other side of $b$, with no critical ratio strictly between them.
This allows the common lower bound to exceed $\beta$ while the constraints remain feasible, giving the required strict inequalities.
Because the bound is uniform over all label tables and scores, we can choose $\rho$ before computing either strategy.
\Cref{lem:general:endpoint} gives this choice using only $b$, $m$ and $e$.
The chosen $\rho$ has denominator at most $\overline q$, where
\[
    \log\overline q = O((m/e)\log(m + 2)).
\]
The lemma also gives the algorithm for computing the new anchors once the label tables are known.

Let $\widetilde P_A$ and $\widetilde P_B$ be the uniform distributions specified by the label tables and the final anchor tables, still with each player's budget normalized to one.
At most one anchor table is adjusted, and the opponent's normalized budget against an unchanged table does not increase.
Thus the transfer adds at most $m/S$ to the sum of the two supremum payoffs.
Hence \eqref{eq:general:quantized} gives
\[
    U_{\widehat v}(\widetilde P_A;b) + U_{\widehat v}(\widetilde P_B;1/b)
    \le 1 + 3z/32 + 2/K + m/S.
\]
We also undo the second value rounding: replacing $\widehat v$ by $v^{\rm red}$ changes each term by at most $e/256$ (\cref{sec:general:reduction}).
Using $2/K = e/64$, $m/S \le e/128$ and $3z/32 < e/256$, we obtain
\begin{equation}
    U_{v^{\rm red}}(\widetilde P_A;b) + U_{v^{\rm red}}(\widetilde P_B;1/b) - 1 < e/16.
    \label{eq:general:reduced-error}
\end{equation}
These are the distributions used in the reconstruction below.
We keep their bids normalized to one until the final sampling step, when $B$'s bids are multiplied by $b$.

\subsection{Returning allocations on the original battlefields}
\label{sec:general:lift}

It remains to convert allocations on the reduced battlefields into allocations on the original battlefields.
We work with a distribution $P$ on reduced allocations that spend at most one, such as $\widetilde P_A$ or $\widetilde P_B$, and multiply the resulting allocations by the player's original budget at the end.
The reconstruction uses the values $v'$ from the first rounding in \cref{sec:general:reduction}.
Every battlefield kept individually retains its bid.
If no battlefields were replaced, define $\mathcal L(P) = P$; no reconstruction is needed.
Otherwise, the light battlefields, whose total value is $V$, were replaced by $k = 8/e$ representative battlefields of value $V/k$ each.
We must distribute the bids on these representative battlefields among the original light battlefields.
Let $p_j = v'_j / V$ for each light battlefield $j$, so that these proportions sum to one.

Consider one reduced allocation, with bids $x_0, \ldots, x_{k - 1}$ on the representative battlefields.
Light battlefield $j$ has value $V p_j$, which is $k p_j$ times the value $V / k$ of a representative battlefield.
A natural first attempt gives each light battlefield $j$ the bid $k p_j x_a$, choosing the index $a \in \{0, \ldots, k - 1\}$ independently and uniformly for each $j$.
Since $\sum_j p_j = 1$, the expected total bid on the light battlefields is $\sum_j k p_j \cdot \frac1k \sum_a x_a = \sum_a x_a$.
Choosing these indices independently can nevertheless exceed the budget in an individual allocation.
The following construction enforces the budget in every realization, at the cost of a controlled change in these marginal distributions.

We use a single random shift to coordinate the bids on all light battlefields.
View $[0,1)$ as a circle by identifying its endpoints.
For the chosen allocation, sort a separate copy of the representative bids as $x_{(0)} \le \cdots \le x_{(k - 1)}$.
Place these bids, multiplied by $k$, on equal intervals of the circle:
\[
    Z(t) = k x_{(a)} \quad\text{for } a / k \le t < (a + 1) / k.
\]
Starting at zero, partition the circle into consecutive arcs of lengths $p_j$, following the original order of the light battlefields.
Draw $\omega$ uniformly from $[0,1)$, independently of the allocation, and add it to every arc endpoint modulo one.
Let $I_j(\omega)$ be the rotated arc of light battlefield $j$, and bid
\begin{equation}
    X_j = p_j \mathop{\rm ess\,inf}_{t \in I_j(\omega)} Z(t)
    \label{eq:general:circle}
\end{equation}
on light battlefield $j$.
The essential infimum ignores arc endpoints, so it is the smallest value of $Z$ on an interval $[a / k, (a + 1) / k)$ that overlaps the arc in positive length.
The arcs partition the circle, and each arc's length times the minimum of $Z$ on that arc is at most the integral of $Z$ over the arc.
Consequently,
\[
    \sum_{j \text{ light}} X_j
    \le \sum_{j \text{ light}} \int_{I_j(\omega)} Z(t) \, dt
    = \int_0^1 Z(t) \, dt
    = \sum_{a = 0}^{k - 1} x_a.
\]
Thus every reconstructed allocation satisfies the budget.
Write $\mathcal L(P)$ for the distribution obtained by drawing an allocation from $P$, drawing $\omega$, and applying this construction.
The next lemma bounds the increase in the opponent's supremum payoff caused by the reconstruction.

\begin{lemma}
\label{lem:general:lift}
Suppose the reduction replaced the light battlefields by representative battlefields.
For every distribution $P$ on allocations of the reduced game that spend at most one and every opposing budget $c \ge 0$,
\[
    U_{v'}(\mathcal L(P); c)
    \le U_{v^{\rm red}}(P; c) + \frac{\sum_{j \text{ light}} (v'_j)^2}{V} + \frac Vk
    \le U_{v^{\rm red}}(P; c) + \frac{9 e}{40}.
\]
The representative bids may be dependent and need not have identical marginal distributions.
\end{lemma}

The first error term comes from taking an arc minimum instead of choosing one representative bid uniformly.
Fix a light battlefield $j$ and a bid amount $t \ge 0$.
Since $Z$ is nondecreasing on $[0,1)$, the points where $p_j Z$ is at most $t$ form an interval starting at zero.
A uniformly chosen point lies in this interval with the same probability that a uniformly chosen bid $k p_j x_a$ is at most $t$.
The reconstructed bid $X_j$ is at most $t$ exactly when its arc overlaps this interval in positive length.
An arc of length $p_j$ can overlap the interval even when its starting point is outside it, but this adds at most $p_j$ to the probability.
The same argument applies with strict inequalities.
Thus taking the arc minimum changes each marginal CDF by at most $p_j$, and weighting these errors by $v'_j$ gives the first term.

The second error term comes from comparing an opposing deviation on the original battlefields with deviations on the reduced battlefields.
Fix an opposing allocation $y$ with total bid at most $c$.
Comparing $y_j$ with $k p_j x_a$ is equivalent to comparing $y_j/p_j$ with $k x_a$, since dividing both bids by $p_j$ preserves wins and ties.
Define a random bid $Y$ by choosing light battlefield $j$ with probability $p_j$ and returning $y_j/p_j$.
The payoff of $y$ against the bids $kp_jx_a$ on the light battlefields is $V$ times the expected share of $Y$ against $kx_a$, since $v'_j=Vp_j$.
Here $a$ is uniform and independent of $Y$.
The mean of $Y$ is the total bid $\sum_{j\text{ light}} y_j$ on the light battlefields.

The proof in \cref{app:lem:general:lift:proof} constructs a uniform distribution on a list of $k$ bid amounts that approximates $Y$, without increasing the mean and with CDF error at most $1/k$.
Divide each list entry by $k$ and assign one to each representative battlefield.
Every such assignment respects the opponent's budget, keeping its bids on individually retained battlefields unchanged.
Choosing uniformly among the $k$ cyclic permutations makes each representative battlefield receive each list entry with probability $1/k$, even when some entries are equal.
The average payoff of these reduced deviations is within $V/k$ of the payoff of $y$ against the independently chosen bids $k p_j x_a$ considered above.
These deviations depend only on $y$, so the comparison remains valid after averaging over allocations drawn from $P$.
Their average expected payoff against $P$ is at most $U_{v^{\rm red}}(P;c)$.
Finally, the first error is at most $\max_{j\text{ light}}v'_j \le e/10$, and $V/k \le e/8$, giving the bound $9e/40$.

\paragraph{Finitely many shifts.}
To sample using finitely many random bits, we replace the continuous shift by a uniform choice from a list of $S_{\rm out} = \operatorname{pow2ceil}(64/e)$ rational shifts.
We choose these shifts away from positions where an arc endpoint coincides with a boundary $a/k$ of one of the intervals defining $Z$.
Recall that the first value rounding writes $v'_j = w'_j/Q'$ with positive integer weights $w'_j$.
Hence $p_j = w'_j / W_G$, where $W_G = \sum_{j \text{ light}} w'_j = O(n / e)$.
With $D_{\rm sh} = k W_G$, every shift at which an arc endpoint meets an endpoint of an interval $[a / k, (a + 1) / k)$ is a multiple of $1 / D_{\rm sh}$.
Thus divide the possible shifts into cells $[r / D_{\rm sh}, (r + 1) / D_{\rm sh})$, for $r = 0,\ldots,D_{\rm sh}-1$.
Within the interior of each cell, every arc overlaps the same intervals defining $Z$, so every reconstructed bid is constant.
For $0 \le \ell < S_{\rm out}$, let $\omega_\ell$ be the midpoint of the cell containing the evenly spaced point $(\ell + 1/2)/S_{\rm out}$.
We sample the index $\ell$ uniformly even when different indices give the same shift.

\begin{lemma}
\label{lem:general:finite-shifts}
Fix a reduced allocation and a light battlefield $j$.
Replacing the uniform shift $\omega$ by $\omega_\ell$ for a uniformly random $\ell \in \{0, \ldots, S_{\rm out} - 1\}$ changes each of the probabilities $\Prob(X_j < t)$ and $\Prob(X_j \le t)$ by at most $1 / S_{\rm out}$, for every bid amount $t \ge 0$.
\end{lemma}

For a fixed reduced allocation, light battlefield $j$, and bid amount $t$, the shifts giving $X_j < t$ form an interval on the circle, possibly empty or the whole circle.
Under a continuous uniform shift, this event has probability equal to the interval's length.
The fraction of equally spaced points in a half-open interval differs from its length by at most $1/S_{\rm out}$.
The proof handles the interval endpoints and shows that moving the points to cell midpoints preserves this error bound, both for $X_j < t$ and for $X_j \le t$.
\Cref{app:general:sampling} gives the proof and an $O(n + \log(1/e))$-operation procedure for evaluating all bids of a sampled allocation.

To sample a player's strategy, choose a row $i \in [T]$ of its label table uniformly and decode its bids using the final anchors.
Independently choose $\ell$ uniformly from $\{0, \ldots, S_{\rm out} - 1\}$, and reconstruct the light bids using this allocation and the shift $\omega_\ell$.
The bids on individually retained battlefields come from the same selected row.
Finally, multiply all bids by the player's original budget: one for $A$ and $b$ for $B$.
Since $T$ and $S_{\rm out}$ are powers of two, this uses exactly $\log_2(T S_{\rm out})$ random bits.
If no battlefields were replaced, set $S_{\rm out} = 1$ and omit the shift.

\begin{proof}[Proof of \cref{thm:general}]
\Cref{app:general:boundary} treats the cases $n = 1$ and $b = 1$, so assume $n > 1$ and $b < 1$.
The algorithm reduces the game (\cref{sec:general:reduction}), selects $\rho$ (\cref{lem:general:endpoint}), and computes $P_A$ and $P_B$ (\cref{lem:general:source}).
It then replaces the bids by anchors and adjusts at most one anchor table (\cref{sec:general:transfer}).
It describes each player's strategy by its label table, its final anchor table, and the data for the reconstruction with finite shifts.
With the final anchors, each row of the label table specifies a reduced allocation that spends at most one.
The reconstruction does not increase the total bid, and the final multiplication restores the original budget, so every sampled allocation is feasible.
Let $\mu_A$ and $\mu_B$ be the resulting strategies on the original battlefields.
Here $U$ uses the original value vector $v$.
By \eqref{eq:prelim:security-sum}, each regret is at most $U(\mu_B; 1) + U(\mu_A; b) - 1$, and
\[
    U(\mu_B; 1) + U(\mu_A; b) - 1
    \le \underbrace{\frac{e}{16}}_{\text{reduced game}}
    + \underbrace{2 \cdot \frac{9 e}{40}}_{\text{reconstruction}}
    + \underbrace{2 \cdot \frac{e}{64}}_{\text{finite shifts}}
    + \underbrace{2 \cdot \frac{e}{128}}_{\text{original values}}
    = \frac{179 e}{320}
    < \varepsilon.
\]
The first term is \eqref{eq:general:reduced-error}.
The second applies \cref{lem:general:lift} with opposing budget $c=b$ against $\widetilde P_A$ and $c=1/b$ against $\widetilde P_B$, when light battlefields were replaced.
Multiplying $B$'s bids and the bids of its opponent by $b$ converts the latter comparison to the original budgets without changing payoffs.
If no battlefields were replaced, both the reconstruction error and the finite-shift error are zero.
When finite shifts are used, the third term follows from \cref{lem:general:finite-shifts} and \eqref{eq:prelim:half-tie}, since $V \le 1$ and $1 / S_{\rm out} \le e / 64$.
The last term replaces the values $v'$ by the original values $v$ (\cref{sec:general:reduction}).
The players sample independently, and each sample uses one complete row, which preserves the correlations across battlefields used in the proof.
\Cref{app:general:ledger,app:general:description,app:general:sampling} prove the bounds on computation, storage, description length, and sampling.
\end{proof}

\section{Conclusion}
\label{sec:conclusion}

We give an FPTAS for two-player Colonel Blotto with arbitrary budgets and common battlefield values.
It remains open whether an $\varepsilon$-Nash equilibrium can be computed in time polynomial in the input length and $\log(1/\varepsilon)$.
Our finite-bit coupling algorithm achieves this in the General Lotto coupling regime.
Improving the dependence on $1/\varepsilon$ in the FPTAS is another natural question.

For more than two players, it remains open whether an FPTAS exists, even for three players with common battlefield values.
Another open question is whether an FPTAS exists for two players with arbitrary player-specific values, beyond the sufficient General Lotto coupling condition in \cref{thm:prg:specific}.

\section*{Acknowledgements}

This project has received funding from the European Research Council (ERC) under the European Union's Horizon Europe research and innovation programme (grant agreement No.\ 101198689).
The views presented in this paper are solely the authors' and do not necessarily represent the positions of the European Union or the European Research Council Executive Agency.
Neither the EU nor the granting authority accepts responsibility for these views.

\paragraph{AI methodology.}
We used ChatGPT to help work out proof details from our sketches and ideas, optimize the FPTAS in \cref{sec:general} and improve its running-time bounds, and write preliminary drafts of parts of the manuscript.
We also used ChatGPT to generate the accompanying Lean implementation and formal proofs of feasibility and equilibrium guarantees, available at \url{https://github.com/AbheekG/colonel-blotto-lean}.

\printbibliography

\clearpage
\appendix
\section{Omitted proofs}
\label{app:omitted-proofs}

\subsection{Proof of Lemma~\ref{lem:prelim:cdf}}
\label[appendix]{app:lem:prelim:cdf:proof}
\begin{proof}
Suppose $d(\mu, \nu) \le \delta$.
On each battlefield, the marginal CDFs differ by at most $\delta$; taking limits from below gives the same bound for their left limits.
By \eqref{eq:prelim:half-tie}, changing the opponent's marginal CDF by at most $\delta$ changes the expected share at every fixed bid by at most $\delta$.
Summing with weights $v_i$, which sum to one, and averaging over one's own strategy proves the first assertion.

For a fixed opposing bid $t$, the identity $h(s, t) = 1 - h(t, s)$ gives the same bound when the distribution of one's own bid $s$ changes.
Summing over battlefields and averaging over the opponent's strategy therefore bounds the effect of changing one's own strategy by its marginal CDF distance as well.
Changing the two strategies in turn now gives
\[
    |u_j(\mu'_A, \mu'_B) - u_j(\mu_A, \mu_B)| \le \delta_A + \delta_B,
    \qquad j \in \{A, B\}.
\]
The first assertion also bounds the change in $A$'s supremum deviation payoff by $\delta_B$, and the change in $B$'s by $\delta_A$.
Combining these bounds with \eqref{eq:prelim:regret} proves the claim.
\end{proof}

\subsection{Proof of Theorem~\ref{thm:prg:specific}: recognition and exact equilibrium}
\label[appendix]{app:prg:specific}

\begin{proof}
We recall the cubic equations used in the parameter enumeration of \cite[Proposition~1]{perchet2024algorithmic} and explain how to test the endpoint inequality in \eqref{eq:prg:specific-promise}.
Sort the ratios $v_i/w_i$, grouping equal ratios.
The distinct ratios partition $(0,\infty)$ into open intervals, including those below the smallest ratio and above the largest.
On each interval, the sets $H = \{i:v_i/w_i > \gamma\}$ and $S = [n]\setminus H$ are fixed.
Define
\[
    A_1 = \sum_{i\in H}w_i,\quad
    A_2 = \sum_{i\in S}v_i^2/w_i,\quad
    B_1 = \sum_{i\in H}w_i^2/v_i,\quad
    B_2 = \sum_{i\in S}v_i.
\]
Substituting the marginal parameters from \eqref{eq:prg:specific-marginals} into the two mean equations gives
\[
    2\lambda=\gamma A_1+A_2/\gamma,
    \qquad 2b\lambda=\gamma^2 B_1+B_2.
\]
Eliminating $\lambda$ yields
\begin{equation}
    B_1\gamma^3-b A_1\gamma^2+B_2\gamma-b A_2 = 0,
    \qquad \lambda = \frac{\gamma^2 B_1+B_2}{2b}.
    \label{eq:prg:cubic}
\end{equation}
The polynomial is nonzero: if $H$ is nonempty, then $B_1>0$; otherwise $B_2=\sum_i v_i=1$.
Check its positive roots inside the interval and test each ratio breakpoint in the mean equations as well.
At a breakpoint, the formulas for the summand moving between $H$ and $S$ agree.
Each retained $\gamma$ determines a positive $\lambda$ through \eqref{eq:prg:cubic}.
Thus this procedure finds every positive pair satisfying the mean equations.

As we scan the sorted ratios, each battlefield moves from $H$ to $S$ once, so the four sums require constant work per moved battlefield.
For each candidate pair, the endpoint condition $\max_i\ell_i\le b$ becomes
\[
    \max\!\left\{\max_{i\in S}v_i,\ \gamma\max_{i\in H}w_i\right\}
    \le b\lambda,
\]
where an empty maximum is zero.
In the sorted order, $S$ is a prefix and $H$ a suffix.
Precomputing prefix maxima of $v_i$ and suffix maxima of $w_i$ therefore makes each candidate test take $O(1)$ operations.
Each interval contributes at most three roots, and there are at most $n$ distinct breakpoints, so there are $O(n)$ candidates in total.
Test candidates until one passes; the first positive root need not satisfy the endpoint condition.
If none passes, no positive pair satisfies \eqref{eq:prg:specific-promise}.

This proves the $O(n\log n)$ recognition bound in the model allowing exact roots of polynomials of degree at most three to be obtained at unit cost.
Construct the marginal parameters only for a candidate that passes the test, then store them in $O(n)$ additional work.
All subsequent coupling formulas use arithmetic and comparisons only, so they also apply to these real parameters.

Apply the coupling construction separately to the two players using the accepted parameters.
The mean equations and $\ell_i\le b\le1$ verify \eqref{eq:prg:coupling-condition}, so the resulting strategies $\mu_A$ and $\mu_B$ have the prescribed marginals and spend their budgets exactly.
Since payoffs depend only on the battlefield marginals, these couplings form a Nash equilibrium by \cite[Theorem~1]{perchet2024algorithmic}.
The linear sampling bound follows from \cref{sec:prg:sampler}.
\end{proof}

\section{Finite-bit implementation and complexity}
\label[appendix]{app:computation:bits}

The finite-bit model accounts for the binary representation of both the input and the output.
Let $L$ be the total binary encoding length of the input, including the approximation parameter $\varepsilon$.
To give sharper bounds, write $L=L_0+E$, where $L_0$ is the encoding length of the game input and $E$ is that of $\varepsilon$.
Bit bounds include reading both inputs, arithmetic on their binary representations, writing the output, and generating random bits.
Each finite-bit construction specifies its rational output and bounds any resulting loss in the equilibrium guarantee.

In the bit analyses, $\Rat(W)$ is a nondecreasing superadditive upper bound on the cost of each exact rational arithmetic operation, comparison, integer floor, or greatest-common-divisor computation on operands of $O(W)$ bits.
For the bounds in the main text, we use arithmetic algorithms with $\Rat(W) = \softO(W)$.
The bounds below also allow the conservative choice $\Rat(W) = \softO(W^2)$ for schoolbook arithmetic.

\subsection{Finite-bit implementation of the common-value sampler}
\label[appendix]{app:prg:bits}

To prove \cref{cor:prg:finite}, we first express the coupling for $w\ge(1-p)\ell$ in \cref{lem:prg:four} as a finite mixture of uniform distributions on line segments.
We then approximate the continuous uniforms, bound the bits used for branch choices, and account for operand lengths.
Write $k=\lceil\log_2(1/\varepsilon)\rceil$.
Reading the rational approximation parameter and determining $k$ take $O(E)$ bit operations: compare its numerator and denominator after a binary shift determined by their bit lengths.

\paragraph{Combining the segment couplings.}
Consider the case $w\ge(1-p)\ell$ of \cref{lem:prg:four}.
The first coupling has coordinates $(X_1,X_2,Z_1)$ and the second has coordinates $(Z_2,X_3,X_4)$, with $Z_1=d-Z_2$.
Both auxiliary variables $Z_1,Z_2$ are uniform on $[0,d]$.
Recall that segment $j$ of the first coupling has weight $q_j$ and $Z_1$ endpoint values $r_j,s_j$.
Given $Z_1=z$ away from all endpoint values, the conditional sampler chooses segment $j$ with probability $dq_j/|r_j-s_j|$ when $z$ lies between $r_j,s_j$, and zero otherwise.
Once that segment is chosen, $X_1,X_2$ are affine functions of $z$.
We now combine this rule with the second coupling so that the midpoint estimate below applies to all four coordinates.

Fix a segment of the second coupling, with weight $\beta$, and let $t\in[0,1]$ be its interpolation parameter.
Sampling uniformly on this segment means choosing $t$ uniformly on $[0,1]$.
Along the segment, $Z_2,X_3,X_4$ are affine functions of $t$.
The $Z_2$ coordinate is nonconstant on every segment of positive weight, since otherwise its uniform marginal would have an atom.
As $t$ varies, set $z=d-Z_2$ and subdivide $[0,1]$ wherever $z$ crosses a value $r_j$ or $s_j$.
On the interior of each resulting subinterval, the same segments of the first coupling contain a point with $Z_1=z$, and their conditional probabilities are constant.

Consider one such subinterval $[t_0,t_1]$ and an eligible segment $j$ of the first coupling.
The resulting four coordinates are affine functions of $t$: $X_3,X_4$ come from the fixed second segment, and $X_1,X_2$ are affine functions of $z=d-Z_2$ on segment $j$.
They therefore trace a line segment as $t$ runs from $t_0$ to $t_1$.
The probability of choosing the second segment and landing in this subinterval is $\beta(t_1-t_0)$.
Multiplying by the conditional probability of choosing segment $j$ gives the combined segment weight
\[
    \beta(t_1-t_0)\frac{d q_j}{|r_j-s_j|},
\]
and its endpoints are obtained by evaluating the four output coordinates at $t_0$ and $t_1$.
Summing these weights over the eligible first segments gives $\beta(t_1-t_0)$; summing over all subintervals and second segments then gives one.
The corresponding points in both original segments remain feasible at $t_0$ and $t_1$, so the combined endpoints are feasible as well.
These finitely many subdivision endpoints have probability zero and do not affect the marginal distributions.

We sample the resulting mixture by choosing a segment with its weight and sampling a point uniformly on it.
Each coupling has at most seven segments, so there are at most fourteen endpoint values $r_j,s_j$.
Each segment of the second coupling is therefore divided into at most fifteen intervals, giving at most $7\cdot15\cdot7=735$ combined segments.
Their weights and endpoints use a constant number of rational operations on the parameters.

\paragraph{Approximating the continuous uniforms.}
An $N$-point midpoint uniform chooses each value $(j+1/2)/N$, $0\le j<N$, with probability $1/N$.
For $N=2^K$, this needs exactly $K$ random bits.
Replacing a scalar uniform on a line segment by this midpoint uniform changes the CDF of each nonconstant affine coordinate by at most $1/(2N)$.
Constant coordinates and the sum remain exact.
The estimate survives finite mixtures, proportional splitting, and reflection; for reflection, apply the same CDF estimate to left limits.
It therefore covers the three-variable segment tables and the four-variable segment mixture just constructed.

The only additional estimate concerns the first case of \cref{lem:prg:nested}, whose segment probabilities depend on the sampled location $S\in[0,t]$.
Recall that the target third marginal is $q\Unif[0,t]+(1-q)\Unif[0,c]$ and that $h=q(c-t)>0$.
This case uses one uniform draw to generate $S$ and a second to sample a point on the selected segment.
The randomness used to choose the segment is handled separately below.
Conditional on $S=s$, the third coordinate has distribution $(1-p_s)\delta_s+p_s\Unif[s,c]$, where $p_s=(c-h-2s)/(c-s)$.
For a fixed threshold $z<c$, its conditional CDF $K_s(z)$ is zero for $s>z$, and for $s\le z$ it is
\[
    K_s(z)=(1-p_s)+p_s\frac{z-s}{c-s}
    =1-\frac{(c-h-2s)(c-z)}{(c-s)^2}.
\]
As a function of $s$ on $s\le z$, its derivative is $2(h+s)(c-z)/(c-s)^3\ge0$; after $s$ passes $z$, it is zero.
Thus it has at most one increasing part and one downward jump, giving total variation at most two because its values lie in $[0,1]$.
Composition with the increasing inverse CDF for $S$ preserves this bound.
For a function of total variation $V$ on $[0,1]$, the midpoint rule has error at most $V/(2N)$: decompose it into the difference of two monotone functions and use midpoint discrepancy $1/(2N)$ for each.
Thus replacing the uniform used to generate $S$ contributes CDF error at most $1/N$.
The first two conditional marginals are unchanged for every $S$, so they have no error from this replacement.
Replacing the uniform draw along the selected segment adds at most $1/(2N)$ in each coordinate.

Every remaining case is a mixture, translation, reflection, or proportional split of these constructions.
Consequently, replacing the at most two scalar uniforms in the coupling formulas by $K$-bit midpoint uniforms, while retaining the exact branch probabilities, gives marginal CDF error at most
\begin{equation}
    \nu=\frac{3}{2\cdot2^K}.
    \label{eq:prg:midpoint-error}
\end{equation}
This bound is uniform over all parameters and all outcomes of the probability rounding; it contains no reciprocal length or slack.
All sampled allocations still have the required exact sum.

\paragraph{A bounded number of bits for branch choices.}
For rational success probability $p$, simulate a Bernoulli choice by generating independent random bits and comparing them with the binary expansion of $p$ until the first difference.
Return one if the random bit at that position is smaller, and zero if it is larger.
For dyadic $p$, if the sampled prefix matches its terminating binary expansion, return the Bernoulli outcome zero immediately: every extension of that prefix is at least $p$.
Probabilities zero and one require no bits.
Each generated bit matches the corresponding digit with probability $1/2$, unless the procedure has already stopped.
Conditionally on all earlier random choices, the bit count $Z$ is therefore dominated by a geometric random variable of parameter $1/2$ on $\{1,2,\ldots\}$, so
\[
    \E[(4/3)^Z\mid\text{past}]\le2.
\]
Finite segment mixtures may be sampled by successive conditional Bernoulli choices, discarding zero weights.
The reduction has at most $n-1$ outer choices, followed by a constant number of choices within the three- or four-variable coupling; the formulas above permit the conservative bound $C_*=n+1000$ on their total.
Pad the sequence to $C_*$ choices using choices that need no random bits.
Iterating the conditional moment bound and applying Markov's inequality yields
\[
    \Prob\!\left[\text{total branch bits}>R\right]
    \le 2^{C_*}(3/4)^R.
\]
Impose the shared cap $R=3(C_*+K)$.
If another branch bit is needed after $R$ bits have been used, return $(B,0,\ldots,0)$ for the relevant budget $B$.
Since $2(3/4)^3<1$ and $(3/4)^3<1/2$, the probability of this fallback is at most $2^{-K}$.
Runs with and without the cap agree unless fallback occurs, so the cap changes the output distribution in total variation by at most $2^{-K}$, and hence changes every marginal CDF by at most that amount.

Choose $K=\lceil\log_2(32/\varepsilon)\rceil$.
The resulting marginal distance from each ideal strategy is at most
\[
    \delta\le\nu+2^{-K}=\frac{5}{2\cdot2^K}.
\]
By \cref{lem:prelim:cdf}, each opposing deviation bound increases by at most $\delta$.
The ideal bounds from \cref{sec:prg:security} sum to one, so \eqref{eq:prelim:security-sum} gives regret at most $2\delta<\varepsilon$ for each player.
All paths, including fallback, spend the budget exactly.
Each sampler uses at most $R+2K=O(n+k)$ bits.

\paragraph{Bit complexity and output.}
The initial lengths $\ell_i$ and values $r_i=p_i\ell_i$ can all be written with a common denominator of $O(L_0)$ bits.
For example, the product of the denominators of the input values and budget ratio is such a denominator, since $\ell_i=2v_i$ and $p_i$ is either $1$ or $b$.
A value $r_i$ strictly between zero and $\ell_i$ created by rounding is a sum of initial $r_j$ minus lengths $\ell_j$ on a set of coordinates; the sets merged at each step are disjoint.
Thus rounding the mixture probabilities and grouping keep $O(L_0)$-bit numerators and denominators.
The conditional probabilities in the rounding rule are ratios of differences between these values, so they also have $O(L_0)$-bit numerators and denominators.
The three- or four-variable coupling then uses a constant number of arithmetic operations and at most two $O(k)$-bit midpoints.
Its output coordinates and their proportional splits therefore have $W=O(L_0+k)$ bits.
The deterministic strategy description stores the input parameters and the precisions, together with the fixed sampling procedure; it need not enumerate the finite support.

Computing the strategy description takes $O(n\Rat(L_0)+E)$ bit operations.
A complete bounded draw takes
\[
    O\bigl(n\Rat(W)+(n+k)W\bigr)
\]
bit operations and writes $O(nW)$ bits of rational coordinates.
The second term accounts for comparing random bits with the binary expansions of the rational branch probabilities; each digit uses $O(W)$ bit work.
Using $\Rat(W)=\softO(W)$ and $L=L_0+E$ gives the bounds in the main text.
This proves all assertions of \cref{cor:prg:finite}.

\subsection{Finite-bit implementation for player-specific values}
\label[appendix]{app:prg:specific-bits}

We complete the finite-bit part of \cref{thm:prg:specific} by implementing the parameter search exactly and, when a qualifying pair exists, constructing the samplers.
Unlike the common-value case, the parameters can be irrational even for rational input.
We account for the bit cost of representing these parameters and performing exact arithmetic and comparisons.

\paragraph{Exact algebraic parameters.}
After clearing denominators, each polynomial in \eqref{eq:prg:cubic} has degree at most three and integer coefficients of $O(L_0)$ bits.
Remove repeated factors by taking the squarefree part, then use a Sturm sequence to isolate its real roots in disjoint rational intervals.
For polynomials of this fixed degree, bounds on root magnitudes and separation give polynomial precision and polynomial bit cost in $L_0$.
Retain roots in the relevant ratio interval, and test the rational breakpoints separately as in \cref{app:prg:specific}.

To obtain an irreducible polynomial for each candidate, first detect rational roots.
A rational root of an integer polynomial with leading coefficient $a_d$ has denominator at most $|a_d|$.
Refining an isolating interval to width less than $1/(4|a_d|^2)$ permits rational reconstruction by continued fractions, followed by exact substitution to test the candidate.
Every reducible polynomial of degree two or three has a rational root, so removing the resulting linear factors leaves an irreducible factor for each remaining candidate.

Represent each candidate $\gamma$ by its irreducible polynomial and a rational isolating interval.
All lengths and mixture probabilities computed from it belong to the number field $\mathbb Q(\gamma)$, whose degree $d$ is at most three.
We represent an element of this field by its rational coefficients in the basis $1,\gamma,\ldots,\gamma^{d-1}$.
Addition, multiplication, and reduction modulo the defining polynomial use rational arithmetic; division reduces to a linear system of size at most three.
An element is zero exactly when all its reduced coefficients are zero.

For sign comparisons, suppose that these coefficients and the defining polynomial have binary lengths at most $W$.
The norm of a nonzero field element is a nonzero rational number with $O(W)$-bit numerator and denominator, so its absolute value is at least $2^{-O(W)}$.
The other conjugates have absolute values at most $2^{O(W)}$; since the norm is their product with the chosen real value, that value also has magnitude at least $2^{-O(W)}$.
Refining the isolating interval and evaluating with $O(W)$ bits of precision therefore determines its sign.
The candidate parameters have polynomial coefficient lengths in $L_0$, so these operations test $\max_i\ell_i\le b$ in polynomial time, including equality.
If every candidate fails, reject; otherwise retain one passing pair for the sampling construction below.

\paragraph{Bounded randomness and rational output.}
The positive lengths satisfy $\min_i\ell_i\ge2^{-\operatorname{poly}(L_0)}$ by the preceding bound on nonzero field elements.
Starting from one and halving until the value is at most every $\ell_i$ therefore gives $q_0=2^{-s}\le\min_i\ell_i$ with $s=\operatorname{poly}(L_0)$, using polynomially many exact comparisons.
As in \cref{app:prg:bits}, there are at most $C_*=n+1000$ Bernoulli choices and at most two scalar uniform draws per allocation.
We use $K$ bits for each scalar uniform, $H$ bits for each Bernoulli choice, and $P$ bits after the binary point for the final output coordinates, where
\[
    K=\left\lceil\log_2\frac{48}{\varepsilon}\right\rceil,
    \qquad H=\left\lceil\log_2\frac{32C_*}{\varepsilon}\right\rceil,
    \qquad P=s+\left\lceil\log_2\frac{32}{\varepsilon}\right\rceil.
\]
Use $K$-bit midpoint uniforms for the scalar uniforms in the coupling formulas.
For a branch probability $r\in\mathbb Q(\gamma)$, compare an independent $H$-bit midpoint uniform with $r$ by an exact algebraic sign test.
Return one if the midpoint is less than $r$, and zero otherwise; the success probability differs from $r$ by at most $2^{-H}$.
Compare this sampler with the one using the same $K$-bit midpoint draws but exact branch probabilities.
Until their first differing branch choice, the two runs have the same conditional parameters, so each choice has discrepancy probability at most $2^{-H}$ under a coupling.
Summing over the at most $C_*$ choices bounds the change in the output distribution in total variation by $C_*2^{-H}$.
Each resulting algebraic allocation remains nonnegative and has the required exact sum, since every selected branch is feasible.

Finally replace each coordinate $x\ge0$ by $2^{-P}\lfloor2^P x\rfloor$ using exact algebraic comparisons.
All normalized coordinates lie in $[0,1]$, so binary search takes $O(P)$ comparisons per coordinate.
This rounding decreases coordinates by less than $2^{-P}$ and keeps the total allocation at most the player's budget on every path.
It also leaves every zero coordinate equal to zero.

Before output rounding, \eqref{eq:prg:midpoint-error} and the branch error bound give marginal CDF error at most $3/(2\cdot2^K)+C_*2^{-H}$.
Fix a battlefield $i$, and let $X$ and $X_{\rm round}$ be its coordinate before and after output rounding.
For $t\ge0$, they satisfy
\[
    \Prob[X\le t]\le\Prob[X_{\rm round}\le t]
       \le\Prob[X\le t+2^{-P}].
\]
If $F$ is the ideal marginal CDF, this sandwich shows that rounding increases the CDF error by at most $F(t+2^{-P})-F(t)$.
The difference is the ideal probability of $(t,t+2^{-P}]$, which excludes the atom at zero even when $t=0$.
On $(0,\infty)$ the ideal marginal has density at most $1/\ell_i\le1/q_0$, so this additional error is at most $2^{-P}/q_0$.
For $t<0$, both the ideal and rounded CDFs are zero.
Hence the total marginal error is at most
\[
    \delta\le\frac{3}{2\cdot2^K}+C_*2^{-H}+\frac{2^{-P}}{q_0}
    \le\frac{3\varepsilon}{32}.
\]
The proof of \cref{lem:prelim:cdf} only uses the normalization of the values of the player whose payoff is being bounded.
It therefore applies separately to the value vectors $v$ and $w$, giving regret at most $3\delta<\varepsilon$ for each player.

\paragraph{Bit complexity and output.}
It remains to bound the coefficient lengths reached during sampling, so that the algebraic operations above have polynomial cost.
The coefficients of all initial lengths $\ell_i$ and values $r_i=p_i\ell_i$ can be written with a common denominator of polynomial bit length in the basis of $\mathbb Q(\gamma)$.
Rounding the mixture probabilities and grouping use sums and differences of these quantities on disjoint sets of coordinates, so their coefficient lengths remain polynomial.
Branch probabilities are obtained by division in $\mathbb Q(\gamma)$.
The three- or four-variable coupling uses a constant number of arithmetic operations on these parameters and the midpoint inputs, followed by proportional splitting into the original coordinates.
All resulting coefficients therefore have lengths polynomial in $n,L_0,\log(1/\varepsilon)$.
Root isolation, arithmetic and sign tests, and the $O(nP)$ output comparisons consequently have polynomial cost in $n,L,\log(1/\varepsilon)$, including reading and processing the approximation parameter.
The description consists of the rational input, selected root representation, precisions, and fixed sampler code.
Each allocation uses at most $C_*H+2K$ independent random bits and has polynomial output length.
This completes the finite-bit guarantee.

\section{Minimax, parameters, and candidate bids}
\label[appendix]{app:general:catalogue}

\subsection{Proof of Lemma~\ref{lem:general:minimax}}
\label[appendix]{app:general:minimax}

\begin{proof}
Recall that $m \ge 2$ and $0 < \rho < 1$, and write $v$ for the battlefield values.
Adding unused budget to the bids of a strategy cannot increase the opponent's payoff, and adding unused budget to the bids of a deviation cannot decrease the deviating player's payoff.
Hence $C_1$ and $C_\rho$ do not change if both players are restricted to allocations that spend their full budgets, and we assume this restriction throughout the proof.
Since $u_A + u_B = 1$, the definitions give
\[
    1 - C_1 = \sup_P \inf_Q u_A(P, Q)
    \qquad\text{and}\qquad
    C_\rho = \inf_Q \sup_P u_A(P, Q),
\]
where $P$ ranges over distributions on $A$'s allocations and $Q$ over distributions on $B$'s allocations.
The inequality $1 - C_1 \le C_\rho$ holds in every game, since $\inf_{Q'} u_A(P, Q') \le u_A(P, Q) \le \sup_{P'} u_A(P', Q)$ for all $P$ and $Q$.
We prove the reverse inequality in two steps.

Call a distribution of $A$ \emph{regular} if its marginals have no atoms.
We first show that, against every fixed distribution $Q$ of $B$, the supremum of $u_A(P, Q)$ over regular $P$ equals the supremum over all $P$.
Since $u_A(P, Q)$ is an average of the payoffs of $A$'s allocations, it suffices to approximate the payoff $u_A(x, Q)$ of each allocation $x$ of $A$ by regular distributions.
Suppose first that $x$ has at least two positive coordinates.
For small $\delta > 0$, give each zero coordinate an independent uniform amount in $[\delta^2, 2 \delta^2]$ and subtract the total of these amounts equally from the positive coordinates.
Independently, draw independent uniform variables $V_j$ on $[-1, 1]$ for the positive coordinates, let $\overline V$ be their mean, and add $\delta (V_j - \overline V)$ to each positive coordinate $j$.
For sufficiently small $\delta$, this gives a regular distribution that spends the full budget.
Fix an allocation $y$ of $B$ and a battlefield $j$, and let $\delta \to 0$.
If $x_j > 0$ and $y_j = x_j$, $A$'s new bid differs from $x_j$ by $\delta (V_j - \overline V)$ plus a term of order $\delta^2$; since $V_j - \overline V$ has a symmetric distribution without atoms, $A$ wins with probability tending to $1/2$.
If $x_j = y_j = 0$, $A$ now wins battlefield $j$, and if $x_j \ne y_j$, the comparison is eventually unchanged.
Hence the payoff against $y$ converges to at least $u_A(x, y)$, and bounded convergence gives the same conclusion against $Q$.
If $x$ has only one positive coordinate, it equals $1$ and beats every bid of $B$, which is at most $\rho < 1$.
Give the other coordinates independent uniform amounts in $[\delta^2, 2 \delta^2]$ and subtract their total from the positive coordinate, keeping it above $\rho$; the resulting distribution is regular because $m \ge 2$, and the same argument applies.
This is the only use of $\rho < 1$.

Second, let $\mathcal P_{\rm reg}$ be the nonempty convex set of regular distributions of $A$.
Let $\mathcal Q$ be the set of distributions on $B$'s allocations; with the weak topology, it is compact and convex, since these allocations form a compact simplex.
For a regular $P$, the function $Q \mapsto u_A(P, Q)$ is continuous because $u_A(P, y) = \sum_j v_j (1 - F_{P, j}(y_j))$ is continuous in $y$ when the marginal CDFs $F_{P, j}$ of $P$ are continuous.
The payoff is affine in each player's distribution.
The Kneser--Fan minimax theorem \cite[Theorem~4.2$'$]{sion1958general} applies to two convex sets when the minimizing set is compact and the payoff is affine in each variable and lower semicontinuous in the minimizing variable; it requires no continuity or compactness in the maximizing variable.
Applying it to $\mathcal P_{\rm reg}$ and $\mathcal Q$ and using the first step gives
\[
    C_\rho
    = \inf_{Q \in \mathcal Q} \sup_{P \in \mathcal P_{\rm reg}} u_A(P, Q)
    = \sup_{P \in \mathcal P_{\rm reg}} \inf_{Q \in \mathcal Q} u_A(P, Q)
    \le 1 - C_1.
\]
Together with $1 - C_1 \le C_\rho$, established above, this proves $C_1 + C_\rho = 1$.
\end{proof}

\paragraph{Relation to the criterion of Barelli, Govindan, and Wilson.}
For the game restricted to allocations that spend the full budgets, the perturbation also verifies the one-sided \emph{payoff approachability} condition in \cite[Definition~2.8 and Remark~2.10(3)]{barelli2014majority}.
For every allocation $x$ of $A$ and distribution $Q$ of $B$ in this restricted game, it suffices to find allocations $x_\delta \to x$ that have zero probability of a tie against $Q$ and satisfy $\liminf_{\delta \to 0} u_A(x_\delta, Q) \ge u_A(x, Q)$.
Let $P_\delta$ be the perturbed distribution constructed in the proof.
Its allocations approach $x$ as $\delta \to 0$, and its atomless marginals avoid all atoms of $Q$'s marginals with probability one, since each marginal has at most countably many atoms.
We can therefore choose an allocation $x_\delta$ with zero probability of a tie against $Q$ and with $u_A(x_\delta, Q) \ge u_A(P_\delta, Q) - \delta$.
The payoff bound in the proof gives the required limiting inequality.
The standing regularity assumption of \cite[Assumption~2.1]{barelli2014majority} also holds: for either player, replace an allocation $X$ by $(1 - \eta) X + \eta Z$, where $Z$ is an independent uniform allocation on that player's full-budget simplex and $\eta \in (0, 1)$ tends to zero.
The resulting distributions converge weakly to the distribution of $X$ and have atomless marginals, so they assign zero probability to ties against every fixed opposing allocation.
Thus \cite[Remark~2.10(3)]{barelli2014majority} provides another route to the minimax equality.

\paragraph{Why the cited existence theorems do not apply directly.}
The general existence claims in \cite[Section~2.1]{perchet2024algorithmic} and \cite[Section~2]{boix2021multiplayer} refer to \cite{reny1999existence} and \cite{dasgupta1986existence}, respectively.
The following example shows that their hypotheses can fail here.
Take two battlefields of value $1/2$ each, budgets $1$ and $1/3$, and allocations
\[
    x^* = (1/3, 2/3), \qquad y^* = (1/3, 0), \qquad
    x_t = (1/3 + t, 2/3 - t), \quad 0 < t < 1/3.
\]
At $(x^*, y^*)$, the players' payoffs are $(3/4, 1/4)$, whereas at $(x_t, y^*)$ they are $(1, 0)$.
Thus $(x^*, y^*)$ is not an equilibrium, and as $t \to 0$ the profiles $(x_t, y^*)$ converge to it with payoff vector $(1, 0)$.
Moreover, both bids in $x_t$ exceed $B$'s entire budget, so every feasible strategy of $B$ earns zero against $x_t$.

The mixed-strategy existence theorem in \cite[Corollary~5.2]{reny1999existence} requires \emph{better-reply security} of the mixed extension: whenever strategy profiles converge to a nonequilibrium profile and their payoff vectors converge, some player must have a strategy guaranteeing strictly more than its limiting payoff against every opponent strategy in a neighborhood of the limit.
Here neighborhoods of mixed strategies use the weak topology; pure allocations are identified with the distributions concentrated on them.
In the example, $A$ cannot earn more than its limiting payoff of $1$.
Nor can $B$ guarantee a positive payoff near $x^*$, because every neighborhood contains some $x_t$, against which all its strategies earn zero.
Hence better-reply security fails even when deviations may be mixed.

The example also violates \emph{payoff security} of the mixed extension: at $(x^*, y^*)$, $B$ cannot guarantee a payoff arbitrarily close to its current payoff $1/4$ against all nearby strategies of $A$.
The hypotheses of \cite{dasgupta1986existence} imply this property, as shown in \cite[Section~6.2]{reny1999existence}, so they also fail in this example.
This example does have a pure equilibrium: $A$ bids $(1/2, 1/2)$ and $B$ uses any feasible allocation.
It therefore rules out the direct applications of these existence theorems, not equilibrium existence itself.

\subsection{Parameters and positive value rounding}
\label[appendix]{app:general:parameters}

\Cref{sec:general:reduction} defines both roundings and proves their error bounds.
The first, with scale $256n/e$, has $\ell_1$ error at most $e/128$ and common denominator $Q' = O(n/e)$; its integer weights $w'_j$ are kept for the reconstruction in \cref{sec:general:lift}.
The second, with scale $512m/e$, has $\ell_1$ error at most $e/256$ and common denominator $Q_v < 512m/e + m$.

Here and below, $\operatorname{pow2ceil}(a)$ is the least power of two that is at least $a$; it is computed by repeated doubling.
The parameters for budget transfer and for the candidate bids depend only on $m$ and $e$.
The first group controls budget transfer: $K$ is the number of labels per battlefield, $S$ the score scale, $d$ the number of anchors of nonzero labels, and $H$ the bound on the coefficients of constraint slacks in \cref{app:general:roots}.
The denominator $Q_g$ of the grid used to select $\rho$ gives the bound $\overline q$ on the denominator of $\rho$.
The second group controls the candidate bids: $D$ bounds the denominators of the rational anchors in the proof of \cref{lem:general:candidates}, $C$ is the denominator of the evenly spaced grid contained in $\mathcal S$, and $\Delta_{\rm phys}$ is the increment $\Delta$ that realizes open choices.
The full code set $I$ has $U_{\rm code} = 2^\Lambda$ elements.
Their values are
\begin{align*}
    &z = e/2048, \qquad K = 128/e = 1/(16z), \qquad
      S = \operatorname{pow2ceil}(128m/e), \qquad d = m(K - 1), \\
    &H = (m + 3)^{d + 2}, \qquad Q_g = \operatorname{pow2ceil}(4H^2),
      \qquad \overline q = 2Q_g \le 16H^2, \\
    &J_D = m(64/z - 1) + 1, \qquad D = \overline q(m + 2)^{J_D}, \\
    &C = \operatorname{pow2ceil}(4D^2), \qquad
      \Delta_{\rm phys} = \frac1{4m\overline q C^{m + 2}}, \qquad
      U_{\rm code} = 8C = 2^\Lambda.
\end{align*}
These definitions give $mz < 1$, $\Lambda \ge m$, and $\Lambda = \softO(m/z)$.
The remaining parameters control the iterative computation of \cref{app:general:source} and are fixed from the reduced input before any candidate allocations are selected.
They specify the number of updates, the size and accuracy of CDF summaries, and the precision used to evaluate the updates:
\[
\begin{array}{c|l|l}
\text{parameter}&\text{value}&\text{role}\\\hline
\eta = \alpha&z/2^{14}&\text{update rate; optimization accuracy}\\
\nu&\alpha/128&\text{CDF-summary accuracy}\\
\theta&\alpha/16&\text{loss-rounding mesh}\\
k_{\rm ep}&1/z&\text{iterations per epoch}\\
M_{\rm ep}&\operatorname{pow2ceil}(2^{24}(\Lambda + 2)/z)&\text{number of epochs}\\
T&M_{\rm ep}k_{\rm ep}&\text{total iterations}\\
s&\operatorname{pow2ceil}(4/\nu)&\text{summary size}\\
\xi&\eta\theta/2&\text{decrement of the update base}\\
a_{\rm geom}&1 - \xi&\text{base of the multiplicative updates}\\
G&8T/\theta&\text{maximum accumulated exponent}\\
\mathcal E&16(G + 1)(\Lambda + 2)&\text{factor bounding accumulated rounding error}\\
Q_{\rm prof}&2TQ_v/\theta&\text{common profit denominator}
\end{array}
\]
Choose the mantissa precision $p_{\rm mant}$ by increasing it until $64\mathcal E\, 2^{1 - p_{\rm mant}} \le \nu$.
In particular, $T = \Theta(\Lambda/z^2)$ is a power of two, and $K$ divides $T$ because $k_{\rm ep} = 16K$.

\subsection{Proof of Lemma~\ref{lem:general:candidates}}
\label[appendix]{app:general:comparator}

\begin{proof}
Fix a rational $\rho = p/q$ with $0 < p < q \le \overline q$, where $p$ and $q$ need not be relatively prime, the player's budget $c \in \{1, \rho\}$, and the opponent's budget $c'$.
Let $\mathcal F_D$ be the set of rationals in $[0, 1]$ with reduced denominator at most $D$, where $D$ is specified in \cref{app:general:parameters}.
We follow the two steps described after \cref{lem:general:candidates} and supply the details omitted there.

The first step needs one further fact: each rank $\tau_j$ is uniform on $[0, 1]$.
Recall that, given the allocation $x$, the rank $\tau_j$ is uniform on $[F_j(x_j-), F_j(x_j)]$.
For $u \in (0, 1)$, let $t = Q_j(u)$.
Bids $x_j < t$ satisfy $F_j(x_j) < u$, so they contribute probability $F_j(t-)$ to $\Prob(\tau_j \le u)$.
If $F_j$ has an atom at $t$, the bid $x_j = t$ contributes $u - F_j(t-)$, since $u$ lies in $[F_j(t-), F_j(t)]$; otherwise $F_j(t-) = F_j(t) = u$, and this bid contributes zero.
Bids $x_j > t$ have $\tau_j \ge F_j(t) \ge u$, with equality only on an event of probability zero.
Hence $\Prob(\tau_j \le u) = u$.
The cap in the definition of the label $l_j$ matters only on the event $\tau_j = 1$, which has probability zero.

For the second step, write $a_{jl} = Q_j(l/K')$ for the anchors.
Treat the $m(K' - 1)$ anchors of nonzero labels as variables, keeping $a_{j0} = 0$; their values may still be zero.
Require every equality and strict inequality that the original anchors satisfy between two anchors on the same battlefield, and between an anchor and $0$ or $1$.
Also require, for every tuple $(l_1, \ldots, l_m)$ and each budget in $\{1, \rho\}$, the same relation ($<$, $=$ or $>$) between $\sum_j a_{j l_j}$ and that budget as for the original anchors.
Replace each strict inequality by the corresponding weak inequality with a common margin $t$, add the constraint $0 \le t \le 1$, and maximize $t$.
The original anchors satisfy these constraints with some positive margin, since there are finitely many constraints.
At an optimal vertex, the anchors and $t$ solve a nonsingular subsystem of tight constraints of order at most $J_D = m(K' - 1) + 1$; this holds even if the feasible set is lower-dimensional, since a vertex has a full-rank tight subsystem in the ambient variables.
Each row of this subsystem has coefficients in $\{0, 1, -1\}$, at most $m + 1$ of them nonzero, and a right-hand side whose denominator divides $q$.
Cramer's rule and Hadamard's inequality bound the denominator of each new anchor by $q(m + 2)^{J_D} \le D$, so the new anchors lie in $\mathcal F_D$.

Keep the joint probabilities of the label tuples.
Their feasibility for budget $c$ is unchanged because every anchor sum retains its comparison with $c$.
By \cref{lem:general:responses}, applied with the anchors as thresholds, the opponent's supremum payoff is determined by the joint distribution of the labels, the order of the anchors on each battlefield, including repeated anchors, and the comparison of each anchor sum with $c'$.
All of these are preserved, so replacing the anchors by these rational values preserves the opponent's supremum payoff.
The total increase from $P$ is at most $1/K' = z/64$.
The set of candidate bids is $\mathcal S = \mathcal F_D \cup \{t/C : t = 0, \ldots, C\}$, with the power of two $C$ from \cref{app:general:parameters}.
It contains $\mathcal F_D$, zero and one, and every element has denominator at most $C$, hence encoding length $O(\log C)$.
Since $D = \overline q(m + 2)^{m(K' - 1) + 1}$ and $4D^2 \le C < 8D^2$, the code length $\Lambda = \log_2(8C)$ satisfies
\[
    \Lambda = O\bigl(\log\overline q + (m/z)\log(m + 2)\bigr).
\]
The set $\mathcal S$ depends on the common bound $\overline q$, not on the particular ratio $p/q$.
The distributions $P$ and $P^*$ appear only in the analysis; the algorithm does not construct the linear program above or compute the probabilities of $P^*$.
\end{proof}

\subsection{Ordered codes and exact resource tags}
\label[appendix]{app:general:codes}

The candidate bids form the set $\mathcal S = \mathcal F_D \cup \{t/C : t = 0, 1, \ldots, C\}$ of \cref{app:general:comparator}.
We encode the candidate bids, and the bids $s + \Delta_{\rm phys}$ for $s \in \mathcal S$ that realize open choices, by integers, so that comparisons of codes reproduce comparisons of bids.
For $t = 0, \ldots, C$, the bucket $[t/C, (t + 1)/C)$ receives the codes $4t$, $4t + 1$, $4t + 2$ and $4t + 3$, which represent the grid point $t/C$, the bid $t/C + \Delta_{\rm phys}$, the point of $\mathcal F_D$ strictly inside the bucket if there is one, and that point plus $\Delta_{\rm phys}$.
We show below that a bucket contains at most one point of $\mathcal F_D$ in its interior.
Codes of bids in $\mathcal S$ are called closed, and codes of bids $s + \Delta_{\rm phys}$ are called open.
A code is valid if it represents a bid; candidate allocations use only closed codes, and response allocations use closed and open codes.
The open code of the bid one is invalid because neither budget exceeds one, the codes $4t + 2$ and $4t + 3$ are invalid when bucket $t$ has no interior point of $\mathcal F_D$, and every code above $4C$ is invalid; these last codes only make the size $8C$ of $I = \{0, \ldots, 8C - 1\}$ a power of two.
Invalid codes may carry mass in the auxiliary marginals but never occur in a selected allocation.
Distinct candidate bids differ by at least $1/C^2 > \Delta_{\rm phys}$, so the codes are ordered as the bids they represent.

The denominator of every candidate bid is at most $C$.
For a sum $a$ of at most $m$ candidate bids and a budget $c' \in \{1, p/q\}$, a nonzero difference satisfies
\[
    |a - c'| \ge \frac1{qC^m} > m\Delta_{\rm phys}.
\]
To test budgets without adding the small increments, represent a partial tuple of choices by a resource tag $(a, o)$, where $a$ is the sum of its thresholds and $o$ records whether any choice is open.
Tags add by $(a, o) + (a', o') = (a + a', o\lor o')$ and are ordered lexicographically, with closed before open.
Addition preserves this order, and a completed tag is feasible precisely when
\begin{equation}
    a < c'\quad\text{or}\quad a = c'\text{ and }o = 0,
    \label{eq:general:tags}
\end{equation}
which is condition \eqref{eq:general:affordable}.
The corresponding allocation adds $\Delta_{\rm phys}$ at each open choice, and the displayed gap shows that it satisfies the budget exactly when \eqref{eq:general:tags} holds.
Together with the gap between candidate bids, this verifies that $\Delta = \Delta_{\rm phys}$ has the properties required in \cref{sec:general:source}.

The algorithm never lists $\mathcal S$; it only needs the least valid code in a given interval of codes.
Distinct rationals with reduced denominator at most $D$ differ by at least $D^{-2}$, which exceeds the bucket width $1/C$, so a bucket contains at most one of them in its interior.
If such a fraction exists, its distance to the midpoint $(2t + 1)/(2C)$ of the bucket is less than $1/(2(q')^2)$, where $q' \le D$ is its reduced denominator.
By Legendre's theorem, it is then a convergent of the continued fraction of the midpoint.
A later convergent with denominator at most $D$ would be closer to the midpoint and hence also inside the bucket, contradicting uniqueness.
Therefore take the last convergent with denominator at most $D$, and retain it only if it lies strictly inside the bucket.
To find the least valid code in an interval of codes, inspect the constant number of codes in its first bucket and, if necessary, the grid point of the next bucket.
Report absence if the result lies outside the interval; this also covers intervals that consist only of invalid codes.
The Euclidean algorithm gives $O(\Lambda)$ real-RAM operations and $\softO(\Lambda^2)$ bit operations per query, as justified in \cref{app:general:arithmetic}.

\section{Computing the strategy distributions}
\label[appendix]{app:general:source}

This appendix implements the four updates of \cref{sec:general:source} and proves the accuracy bounds in \cref{lem:general:updates}.
We describe the computation for a fixed budget $c \in \{1, \rho\}$ of the player whose strategy is computed; applying it for both choices gives the two required distributions.
Let $X$ be the candidate allocations for this budget and $Y$ the response allocations for the opponent's budget $c'$, with the codes and the increment $\Delta = \Delta_{\rm phys}$ of \cref{app:general:codes}.
We use the notation of \cref{sec:general:source} for auxiliary marginals, CDFs on codes, the payoff $f$ and the pairing $\lambda \cdot F_p$, set $F_{p, j}(-1) = 0$, and omit the battlefield index in formulas for a single battlefield.
Define $H_{p, j}(a) = (F_{p, j}(a - 1) + F_{p, j}(a))/2$.
With respect to the inner product $\langle g, p \rangle = \sum_j \widehat v_j \sum_a g_j(a) p_j(a)$, the gradient of $f$ is $1 - H_q$ in $p$ and $H_p$ in $q$.

\subsection{Updates and selected allocations}
\label[appendix]{app:general:updates}

For each battlefield $j$, the algorithm maintains the distributions $p_j$ and $q_j$ on $I$ and $\lambda_j$ and $\mu_j$ on $\{+, -\} \times I$; we call these $4m$ distributions the components.
In a signed component $\lambda_j$ or $\mu_j$, we order the pairs with positive sign first and those with negative sign second, each in increasing order of codes, and normalize over both signs together.
For a signed component, define the signed tail
\[
    G_{\lambda, j}(a) = \sum_{a' \ge a} \bigl(\lambda_j(+, a') - \lambda_j(-, a')\bigr).
\]
Then $\lambda \cdot F_p = \langle G_\lambda, p \rangle$, and in particular $\lambda \cdot F_x = \sum_j \widehat v_j G_{\lambda, j}(x_j)$ for an allocation $x$; this is the sum over battlefields that the dynamic program of \cref{app:general:profit} maximizes.
Every component starts as the uniform distribution.
In iteration $t$, before changing any component, the algorithm selects feasible allocations $x_t \in X$ and $y_t \in Y$ that approximately maximize $\lambda_t \cdot F_x$ and $\mu_t \cdot F_y$, respectively.
The losses of the four updates are the gradients of the objectives in \cref{sec:general:source}, negated for the three maximized objectives:
\begin{align}
    g_p(a)& = 1 - H_q(a) + G_\lambda(a), &
    g_q(a)& = -H_p(a) + G_\mu(a), \notag\\
    g_\lambda(s, a)& = s(F_{x_t}(a) - F_{p}(a)), &
    g_\mu(s, a)& = s(F_{y_t}(a) - F_{q}(a)),
    \label{eq:general:losses}
\end{align}
where $p$, $q$, $\lambda$ and $\mu$ stand for the current components $p_t$, $q_t$, $\lambda_t$ and $\mu_t$.
These formulas apply separately to each battlefield; the weight of a battlefield enters through the inner product and is not multiplied again into its normalized update.

The algorithm approximates the terms in \eqref{eq:general:losses} on the mesh $\theta$.
It rounds strict and weak CDFs and the tail probabilities for each sign separately down to this mesh, and then forms their half-sums or differences.
Write $\widehat g_t$ for the resulting losses, computed with the same selected allocations $x_t$ and $y_t$.
Their magnitudes are bounded by $2$, $2$, $1$ and $1$, respectively, and each lies on the mesh $\theta/2$.
The algorithm stores the integer level increment
\begin{equation}
    d_t(a) = 2(\widehat g_t(a) + 2)/\theta \in \{0, \ldots, 8/\theta\}.
    \label{eq:general:levels}
\end{equation}
We call the distribution defined by all preceding updates the \emph{full distribution}, to distinguish it from the summaries used below.
It assigns code $a$ probability proportional to $a_{\rm geom}^{\sum_{i < t} d_i(a)}$.
Since $a_{\rm geom}^{d_t(a)} = e^{-\kappa(\widehat g_t(a) + 2)}$ with $\kappa = -2\log(a_{\rm geom})/\theta \in [\eta, 2\eta]$, each update is a multiplicative-weights update with rate $\kappa$.
The constant $2$, added to make the increments nonnegative, contributes the same factor to every weight of a component and disappears on normalization.
The rate $\kappa$ is used only in the analysis.
The algorithm records $x_t$ as the $t$-th list entry before applying the updates, and it returns the uniform distribution on the $T$ candidate allocations.

To avoid processing all codes in every iteration, the algorithm groups the iterations into epochs of $k_{\rm ep}$ consecutive iterations.
At the start of an epoch, it selects $s$ codes at approximate quantiles of each current component, keeping repeated selections.
Their uniform distribution serves as a summary of that component; \cref{app:general:tree} bounds its CDF error relative to the current component.
For each signed component, the rounded signed tail computed from its summary changes only at the successors of the codes in the summary.
Partition $I$ at these successors and, using \cref{app:general:codes}, cache in each part the least valid code that a candidate allocation (for $\lambda_j$) or a response allocation (for $\mu_j$) can use; a part without such a code supplies no option.
Replacing a bid by the cached code of its part preserves the rounded objective and can only decrease the resource tag, so the cached codes contain an optimal choice for the rounded objective.
The all-zero allocation guarantees that a feasible choice exists.

Each summary keeps its selected codes throughout the epoch, but their weights are updated by the same factors as the full distribution.
Each selected code starts with weight $1/s$; after level increments totaling $L$ have accumulated at its code during the epoch, its unnormalized weight is $a_{\rm geom}^L/s$.
We call this total $L$ the local level, since it includes only the current epoch's updates.
Normalizing these weights answers the CDF queries.
The rounded losses change only at the codes selected for the summaries of $p$ and $q$ and their successors, at the successors of the codes in the signed summaries, at the cached codes of the epoch, and, in signed components, at the boundary between the two signs.
These codes partition the codes of each component into $O(s + k_{\rm ep})$ intervals, which we call cells.
The algorithm maintains exact local levels for each cell and each selected code.
Each iteration evaluates the geometric powers of the current local levels of the selected codes, sums the numerators of their fixed-precision approximations exactly, forms the rounded CDFs and tails, selects the two feasible allocations, records $x_t$, and increments the levels according to \eqref{eq:general:levels}.
At the end of the epoch, each cell's exact cumulative level is added once to the corresponding interval in a binary tree representing the full distribution, and the epoch arrays are discarded.
\Cref{app:general:tree} describes this tree and its updates.
The next epoch constructs its summaries from the distributions with all these updates applied.

\subsection{A fixed profit grid with exact budget tests}
\label[appendix]{app:general:profit}

The two optimizations over feasible allocations must respect the budget exactly while approximating their objectives $\lambda_t\cdot F_x$ and $\mu_t\cdot F_y$.
We keep exact resource tags $(a, o)$, recording the sum of the chosen thresholds and whether any choice is open, as in \cref{app:general:codes}; only objective values are rounded.
These objectives can have both signs.
Let $\widehat G_j$ be the signed tail computed from the current summary of $\lambda_t$ or $\mu_t$ on battlefield $j$, with the two sign probabilities rounded as in \cref{app:general:updates}.
The rounded objective for selecting an allocation is $\sum_j \widehat v_j \widehat G_j(x_j)$.
Adding the constant one gives nonnegative coordinate profits
\[
    \pi_j(a) = \widehat v_j(1 + \widehat G_j(a)).
\]
The profits sum to at most two over any allocation, and they have integer numerators over the fixed denominator $Q_{\rm prof} = 2TQ_v/\theta$.
Every positive profit is therefore at least $1/Q_{\rm prof}$.
This denominator is fixed for the whole iterative computation.

Construct a balanced binary tree with one leaf per battlefield and pad it to $2^{h_0}$ leaves, where $h_0 = \lceil\log_2m\rceil$, using dummy leaves with zero resource and profit.
Set
\[
    w_{\rm prof} = \operatorname{pow2ceil}(4(h_0 + 1)/\alpha), \qquad
    B_{\rm prof} = \min\{h \in \mathbb Z_{\ge 0}:2^h \ge 4Q_{\rm prof}\}, \qquad
    F_{\rm prof} = 2^{B_{\rm prof}}w_{\rm prof}.
\]
The profit grid has numerators zero, all $a 2^h$ for integers $a, h$ with $w_{\rm prof} \le a < 2w_{\rm prof}$ and $0 \le h \le B_{\rm prof}$, and the endpoint $2F_{\rm prof}$, over denominator $F_{\rm prof}$.
It has $J = \softO(1/z)$ entries.
Rounding $x$ down to this grid loses at most the fraction $1/w_{\rm prof}$ whenever $2^{-B_{\rm prof}} \le x \le 2$.

At each leaf and each profit-grid value, retain the least exact resource tag among options rounded to that value, together with a reference to the selected option.
Unreachable entries are explicitly absent.
At an internal node enumerate pairs of reachable states, one from each child, add their represented profits and exact resource tags, round the profit sum down, and retain the least resource tag at each resulting profit-grid value.
Store references to the two child states that attain this tag.
At the root choose the largest profit-grid value whose resource tag satisfies the budget, and follow the stored references to recover its complete allocation.
The resource tags are never rounded.

For any actual tuple of choices on a subtree of height $r$, induction gives a retained state with no larger tag and represented profit at least $(1 - 1/w_{\rm prof})^{r + 1}$ times its true profit.
At a merge, the two children have the same height, tag addition preserves order, and one further rounding supplies the next factor.
Conversely every state has an actual predecessor tuple whose profit is at least its represented value.
No positive sum is rounded to zero: before applying the relative bound at any positive internal sum, the preceding inductive bound is at least
\[
    \frac{(1 - 1/w_{\rm prof})^{h_0 + 1}}{Q_{\rm prof}} \ge \frac{3}{4Q_{\rm prof}} > 2^{-B_{\rm prof}}.
\]
The selected allocation therefore loses at most $2(h_0 + 1)/w_{\rm prof} \le \alpha/2$ in the rounded objective compared with the best feasible allocation.
For a selected response allocation, the algorithm adds $\Delta_{\rm phys}$ at every open choice and checks the sum of the resulting bids against the budget.

The balanced tree also controls the number of bit operations.
At a node whose subtree contains $j$ battlefields, an exact resource tag may require $O(j\Lambda)$ bits; it is not uniformly an $O(\Lambda)$-bit quantity.
There are $O(m/j)$ such nodes at that level, each examining $O(J^2)$ pairs of child states.
Summing over levels gives $\softO(m\Lambda J^2)$ bit operations per optimization, with $\softO(m\Lambda J)$ bits for the states and predecessors.
Pair sums are generated as needed, not stored in a quadratic table.

\subsection{Rounded positive arithmetic and interval trees}
\label[appendix]{app:general:tree}

The CDF summaries must approximate the distributions after all preceding updates, even though we store their weights approximately.
We prove an error bound that depends on the tree depth and accumulated update levels, without an additional factor for the number of times a stored mass was recomputed.
Write $u_{\rm fp} = 2^{1 - p_{\rm mant}}$.
A positive stored number has a $p_{\rm mant}$-bit normalized mantissa, an encoded signed exponent, and a separate marker for zero.
If $x$ is the exact result of a positive addition or multiplication, its stored approximation satisfies
\[
    (1 - u_{\rm fp})x \le \operatorname{fl}(x) \le x.
\]
Positive values are never rounded to zero: the exponent is stored as an integer with the bit-length bound proved below, rather than restricted to a fixed hardware range.
The base $a_{\rm geom} = 1 - \xi$ is exactly representable because the precision choice implies $p_{\rm mant} \ge \log_2(1/\xi) + 20$.
Repeated squaring gives a shared table of the powers $a_{\rm geom}^{2^i}$, rounded down at each multiplication.
To evaluate $a_{\rm geom}^a$ for an integer exponent $0 \le a \le G$, multiply the table entries selected by the binary expansion of $a$, starting at exact one.
Induction over squarings and these multiplications gives
\begin{equation}
    (1 - u_{\rm fp})^a a_{\rm geom}^a \le \operatorname{Pow}(a) \le a_{\rm geom}^a,
    \qquad \operatorname{Pow}(0) = 1.
    \label{eq:general:power}
\end{equation}

Represent each component by a binary tree of intervals of its ordered codes: the root covers all its codes, and each split bisects an interval.
A node $v$ represents an interval $I_v$ and stores an exact nonnegative offset $a_v$, an approximate mass $M_v$, and either two child pointers or a marker declaring its interval uniform.
A uniform leaf means that all its codes have the same accumulated level; its subintervals are created only when an update distinguishes them.
The cached value $M_v$ approximates $B_v$, defined without the node's own offset:
\[
    B_v = |I_v|\quad\text{at a uniform leaf}, \qquad
    B_v = a_{\rm geom}^{a_L}B_L + a_{\rm geom}^{a_R}B_R
       \quad\text{otherwise}.
\]
If $A_v$ is the sum of offsets from the root through $v$, the node's true global mass is $a_{\rm geom}^{A_v}B_v$.
An update covering the node's entire interval adds its integer level only to $a_v$.
A partial update splits a uniform leaf into children with offset zero, retaining the old offset at the parent, then recurses.
To recompute the parent's mass, round each product $\operatorname{Pow}(a_L)M_L$ and $\operatorname{Pow}(a_R)M_R$, then round their sum.
The previously stored parent mass is not used in this recomputation.
No offset is pushed down or redistributed.

Each code's path sum equals its accumulated update level and is at most $G$.
For a node of remaining height $h_v$, structural induction proves
\begin{equation}
    (1 - u_{\rm fp})^{(G + 2)h_v}B_v \le M_v \le B_v.
    \label{eq:general:cache}
\end{equation}
A leaf is exact.
When recomputing an internal node, the children's error bound contributes at most $(G + 2)(h_v - 1)$ to the exponent of $1 - u_{\rm fp}$, the power evaluation contributes at most $G$, and the multiplication and addition contribute two more.
An update covering the node's entire interval leaves $B_v$ and its stored approximation unchanged.
Thus earlier recomputations of stored masses require no additional error allowance.
The approximate total mass used for normalization is $\operatorname{fl}(\operatorname{Pow}(a_{\rm root})M_{\rm root})$, including the root's own offset.

For each quantile target $i/s$, $i = 1, \ldots, s$, descend from the root by repeatedly choosing one of the two code intervals.
Accumulate an approximate prefix mass, compare the prefix through the left child with $i/s$ times the freshly computed total mass, and select the corresponding half.
If the descent reaches a uniform leaf before a single code, continue bisecting its interval and halve its mass without creating stored child nodes.

To evaluate a node's global mass during a query, sum its path offsets exactly and compute $\operatorname{fl}(\operatorname{Pow}(A_v)M_v)$.
Each tree has depth at most $\Lambda + 1$, so \eqref{eq:general:power} and \eqref{eq:general:cache} bound this approximation below by the true mass times $(1 - u_{\rm fp})^{(G + 2)(\Lambda + 1) + G + 1}$.
A prefix sums at most $\Lambda + 2$ such masses, so rounding the additions contributes at most $\Lambda + 1$ further factors of $1 - u_{\rm fp}$.
Allowing one more factor for multiplying the normalizer by $i/s$ gives a total exponent at most $(G + 3)(\Lambda + 2) \le \mathcal E$.
Halving the mass of a uniform leaf is exact in the stored representation.
By $1 - (1 - u_{\rm fp})^r \le r u_{\rm fp}$, all queried global masses, prefixes, normalizers, and scaled normalizers therefore have relative downward error at most $e_0 = \mathcal E u_{\rm fp} \le \nu/64$.
Consequently a normalized comparison has error at most $\delta_q = 2e_0/(1 - e_0) < \nu/16$.
Maintain a bracket $[L, R)$ satisfying
\[
    F_{p}(L - 1) \le i/s + \delta_q, \qquad F_{p}(R - 1) \ge i/s - \delta_q.
\]
Each comparison preserves the appropriate endpoint inequality, even if the rounded comparisons are not monotone.
At the final code $a_i$ these inequalities give
\[
    F_{p}(a_i - 1) \le i/s + \delta_q, \qquad F_{p}(a_i) \ge i/s - \delta_q.
\]
To see why these inequalities control the summary's CDF, fix a code threshold $a$.
A target $i/s < F_p(a)-\delta_q$ must select a code $a_i \le a$, since $a_i > a$ would violate the first inequality.
Conversely, a target $i/s > F_p(a)+\delta_q$ must select $a_i > a$ by the second inequality.
Counting the $s$ equally spaced targets, with repeated selected codes counted separately, bounds the summary's CDF error by $1/s+\delta_q < \nu$.
The codes of a signed component are traversed as one ordered interval, so its two signs share a normalizer.
Zero prefix masses are represented exactly, and $\operatorname{Pow}(0)$ is returned as exact one.

A contiguous range update visits and creates $O(\Lambda)$ nodes.
There are $O(s + k_{\rm ep})$ interval updates per epoch and component, so
\[
    B_{\rm tree} = 1 + 64(M_{\rm ep} + 1)(s + k_{\rm ep} + 4)(\Lambda + 2)
\]
suffices as a deterministic per-tree node bound.
Nodes store offsets, mantissas, exponents, and pointers; interval endpoints are retained only on the current traversal stack instead of being stored as $\Lambda$-bit keys in every node.
Local levels, exponents and addresses use $O(\log(\Lambda + 2) + \log(1/z))$ bits.
The precision is $p_{\rm mant} = O(\log(\Lambda + 2) + \log(1/z))$.
All these bounds hold also during the disjoint interval updates within an epoch.

\subsection{Proof of the accuracy bounds in Lemma~\ref{lem:general:updates}}
\label[appendix]{app:general:regret}

\begin{proof}
At the start of an epoch, the uniformly weighted summary approximates the full distribution to CDF error less than $\nu$.
During the epoch, both distributions are multiplied by the same code-dependent weights and renormalized.
We show that the summary remains accurate after these updates.
The ordered variation $\operatorname{TV}(g)$ is the sum of absolute differences between consecutive values of a function $g$ on the ordered codes of a component.
Each executed loss $\widehat g_t$ has ordered variation at most four.
For the losses of $p$ and $q$, this follows from the monotonicity of CDFs and the tail probabilities for each sign.
For a signed loss, write $g(a) = \ind\{x \le a\} - c(a)$, where $c$ is a rounded CDF with final value one.
Then $g(U_{\rm code} - 1) = 0$ and $\operatorname{TV}(g) \le 2-|g(0)|$ (if $x = 0$, the sharper bound $\operatorname{TV}(g) = g(0)$ holds).
The concatenation $(g, -g)$ has variation $2\operatorname{TV}(g) + |g(0)| \le 4$.
Thus after $l \le k_{\rm ep}$ iterations the local level $L_l$ satisfies
\[
    0 \le L_l \le 8l/\theta, \qquad
    \operatorname{TV}(L_l) \le 8l/\theta.
\]
For the multiplicative reweighting function $\varphi_l = a_{\rm geom}^{L_l}$, Bernoulli's inequality and telescoping adjacent integer powers give
\[
    1 - 4\eta l \le \varphi_l \le 1, \qquad
    \operatorname{TV}(\varphi_l) \le 4\eta l \le 1/4096.
\]
In particular $\varphi_l \ge 31/32$.

Let $\Psi$ be the difference between the summary distribution and the full distribution of a component at the beginning of the epoch; its total mass is zero, and the absolute mass of every prefix is at most $\nu$.
Writing $\langle \varphi, \Psi \rangle = \sum_a \varphi(a)\Psi(a)$ for this single component, summation by parts gives
\[
    |\langle \varphi_l, \Psi\rangle| \le \nu\operatorname{TV}(\varphi_l), \qquad
    |\langle \varphi_l\ind_{(-\infty, a]}, \Psi\rangle| \le \nu(\operatorname{TV}(\varphi_l) + \|\varphi_l\|_\infty).
\]
Both distributions have total reweighted mass at least $31/32$, so normalization bounds the CDF distance between them by $34\nu/31$.
Fresh power evaluations in \eqref{eq:general:power} add less than $\nu/128$ after normalization.
The local powers lie in $[1/2, 1]$, so their numerators can all be summed exactly over denominator $2^{p_{\rm mant}}$; exact one uses numerator $2^{p_{\rm mant}}$.
The reweighted summary therefore has CDF error less than $5\nu$ relative to the full distribution at that iteration.
The bounds hold for the sequence of losses generated by the algorithm, even though each loss depends on earlier updates.

Each signed tail is the difference between the probabilities of two intervals in the ordered codes, one for each sign.
The preceding error, followed by rounding each term separately, implies
\[
    \|g_t - \widehat g_t\|_\infty \le \epsilon_g = 25\nu + 3\theta = 49\alpha/128.
\]
Comparing the objectives computed from rounded and exact tail probabilities at both the optimal and returned allocations, and applying \cref{app:general:profit}, gives
\[
    \max_{x \in X}\lambda_t \cdot F_{x} - \lambda_t \cdot F_{x_t} \le \beta_{\rm opt},
    \qquad \beta_{\rm opt} = \alpha + 40\nu + 4\theta = 25\alpha/16,
\]
and the same bound for $Y, \mu_t, y_t$.
Battlefield values sum to one, so neither error gains a factor $m$.

With the rate $\kappa \in [\eta, 2\eta]$ from \cref{app:general:updates}, the normalized update of a component $p$ with loss $\widehat g$ is $p^+(a) = p(a)e^{-\kappa\widehat g(a)}/Z$, where $Z$ normalizes.
Write $\operatorname{KL}(p\|q) = \sum_x p(x)\log(p(x)/q(x))$ for relative entropy.
For any test distribution $p^\star$ and $|\widehat g| \le M_g$, the inequalities $e^x \le 1 + x + x^2$ for $|x| \le 1/2$ and $\log x \le x - 1$ give
\[
        \operatorname{KL}(p^\star\|p) - \operatorname{KL}(p^\star\|p^+) \ge \kappa\langle\widehat g, p -
        p^\star\rangle - \kappa^2 M_g^2.
\]
The four squared magnitude bounds sum to $4 + 4 + 1 + 1 = 10$.
For a uniform initial distribution $p_1$ on at most $2^{\Lambda + 1}$ choices, $\operatorname{KL}(p^\star\|p_1) \le \Lambda + 1$ for every test distribution $p^\star$.
The four families, weighted by battlefield values summing to one, therefore contribute at most $4(\Lambda + 1)$ to the initial relative-entropy bound.
Summing the four inequalities over $T$ iterations and replacing executed losses by their true versions bounds their total average regret by
\[
    \frac{4(\Lambda + 1)}{\eta T} + 20\eta + 8\epsilon_g.
\]
The logarithm and exponential here are analysis only, not algorithmic primitives.

Set
\[
    R=\frac{4(\Lambda+1)}{\eta T}+20\eta+8\epsilon_g.
\]
The preceding argument bounds the sum of the four average regrets by $R$ uniformly over their fixed comparison distributions.
The exact-objective optimization bound above supplies $\beta_{\rm opt}$ for both feasible allocations.
Substituting the parameters of \cref{app:general:parameters} gives
\[
    R+2\beta_{\rm opt}<\frac{1443}{16}\alpha<z/64.
\]
This proves the accuracy assertions of \cref{lem:general:updates}.
\Cref{app:general:ledger} proves its computation time and space assertions.
The proof of \cref{lem:general:source} in the main text derives the guarantee for the retained complete allocations from these bounds.
\end{proof}

\section{Exact transfer between budget ratios}
\label[appendix]{app:general:transfer}

This appendix gives the budget-transfer construction used in \cref{sec:general:transfer}.
We first specify which combinations of labels the new anchors must make unaffordable and prove the resulting payoff bound.
We then choose the auxiliary ratio $\rho$ and show how to compute suitable anchors.

Fix the player whose anchor table must change, with label table $(l_{ij})$, normalized anchor table $u$, and opposing budgets $\beta_0 < \beta$ at ratios $\rho$ and $b$.
The bids selected by each row of the label table spend at most one under $u$, and each label has probability $1/K$ on every battlefield.
The new anchors will preserve the order of the labels, but need not preserve equalities between their old anchors.
The parameters $K$, $S$, $d$, $H$, $Q_g$ and $\overline q$ are those in \cref{app:general:parameters}; in particular, $d = m(K-1)$ and $H = (m+3)^{d+2}$.

\subsection{Payoff scores and the constraints on anchors}
\label[appendix]{app:general:scores}
We measure the opponent's payoff from beating combinations of labels by integer scores.
Let $S = \operatorname{pow2ceil}(128 m / e)$ be the least power of two that is at least $128 m / e$, and for label $l$ on battlefield $j$ let
\[
    s_{jl} = \left\lfloor \frac{S \widehat v_j (l + 1)}{K} \right\rfloor.
\]
A bid above the anchors of labels $0, \ldots, l$ on battlefield $j$ wins the value $\widehat v_j$ with probability at least $(l + 1) / K$, and $s_{jl} / S$ rounds this payoff down.
A tuple $(l_1, \ldots, l_m)$ with one label per battlefield has score $\sum_j s_{j l_j}$ and anchor sum $\sum_j u_{j l_j}$.
Let $h_*$ be the largest score of a tuple whose anchor sum is strictly less than $\beta_0$.
For such a tuple, the opponent can bid slightly more than $u_{j l_j}$ on every battlefield $j$ within the budget $\beta_0$.
This deviation beats every bid with label $0, \ldots, l_j$ on battlefield $j$, even if some of their anchors coincide.
Its payoff is therefore at least $\sum_j \widehat v_j (l_j + 1) / K \ge \sum_j s_{j l_j} / S$.
Hence
\begin{equation}
    h_* / S \le U_{\widehat v}(u; \beta_0).
    \label{eq:general:lower-score}
\end{equation}
A dynamic program over the battlefields computes, for each attainable score, the least anchor sum of a tuple with that score, and hence $h_*$ (\cref{app:general:initializer}).

The new anchor table $u'$ should keep every allocation specified by the label table within the budget and make every tuple with score greater than $h_*$ unaffordable for the opponent's budget $\beta$.
For $\gamma \in \R$, let $\mathcal P(\gamma)$ be the set of anchor tables $u'$ satisfying the following constraints:
\begin{enumerate}
    \item on each battlefield $j$, $u'_{j0} = 0$ and $u'_{jl} \le u'_{j, l + 1} \le 1$ for $0 \le l < K - 1$;
    \item every row $i$ of the label table satisfies $\sum_j u'_{j, l_{ij}} \le 1$;
    \item every tuple with score greater than $h_*$ has anchor sum at least $\gamma$.
\end{enumerate}
Each $\mathcal P(\gamma)$ is a polytope in the $d = m (K - 1)$ anchors of nonzero labels.
The current table $u$ lies in $\mathcal P(\beta_0)$, since by the definition of $h_*$, every tuple with score greater than $h_*$ has anchor sum at least $\beta_0$.
We want a table in $\mathcal P(\gamma)$ for some $\gamma > \beta$.

\begin{lemma}
\label{lem:general:transfer}
Let $u'$ be an anchor table in $\mathcal P(\gamma)$ for some $\gamma > \beta$, or the all-zero table if no tuple has score greater than $h_*$.
Then every allocation specified by the label table spends at most one under $u'$, and
\[
    U_{\widehat v}(u'; \beta) \le U_{\widehat v}(u; \beta_0) + m / S.
\]
\end{lemma}

\begin{proof}
The second constraint keeps every allocation within the budget, and the all-zero table does so trivially.
Suppose first that $u' \in \mathcal P(\gamma)$ with $\gamma > \beta$, and fix a deviation $y$ with $\sum_j y_j \le \beta$.
On battlefield $j$, let $l_j$ be the largest label whose new anchor is at most $y_j$.
Such a label exists because label zero has anchor zero.
Since the new anchors are nondecreasing in the label, the player's bid on battlefield $j$ is at most $y_j$ exactly when its label is at most $l_j$, which has probability $(l_j + 1) / K$.
The opponent's share of battlefield $j$ is at most this probability.
The tuple $(l_1, \ldots, l_m)$ has anchor sum at most $\sum_j y_j \le \beta < \gamma$, so by the third constraint its score is at most $h_*$.
Since $\widehat v_j (l_j + 1) / K < (s_{j l_j} + 1) / S$ for every $j$, the payoff of $y$ is at most
\[
    \sum_j \widehat v_j \frac{l_j + 1}{K} < \frac{h_* + m}{S} \le U_{\widehat v}(u; \beta_0) + \frac mS,
\]
where the last inequality is \eqref{eq:general:lower-score}.
This holds for every feasible deviation, including deviations that tie with some anchors, and hence for the supremum.
If no tuple has score greater than $h_*$, then $h_* = \sum_j s_{j, K - 1} > S - m$, since $s_{j, K - 1} > S \widehat v_j - 1$.
Thus $(h_* + m) / S > 1$, and the bound holds because every payoff is at most one.
\end{proof}

We call the third family of inequalities defining $\mathcal P(\gamma)$ the \emph{score constraints}.
The order and row-budget inequalities are the \emph{static constraints}, since they do not depend on $\gamma$.
There may be exponentially many score constraints.
A dynamic program tests all of them by finding, for each attainable score, the least anchor sum of a tuple with that score.

The following lemma supplies both the choice of $\rho$ and the required anchor table.
Its ratio is chosen before the lists of allocations are computed; the subsequent anchor adjustment uses those lists.

\begin{lemma}
\label{lem:general:endpoint}
Using only $b$, $m$ and $e$, the algorithm selects a rational $\rho \in (0, 1)$ with denominator at most $\overline q = 2 \operatorname{pow2ceil}(4 H^2)$, and determines which anchor table, if any, must change.
In particular, $\log \overline q = O(d \log(m + 3)) = O((m / e) \log(m + 2))$.
If a table must change, then for the rows and scores later obtained at this $\rho$, the algorithm either finds that no tuple has score greater than $h_*$ or computes an anchor table in $\mathcal P(\gamma)$ for some $\gamma > \beta$, whose entries have $O(\log H)$ bits.
\end{lemma}

\subsection{A uniform bound on changes of feasibility}
\label[appendix]{app:general:roots}

Write every constraint as $a_i u' \le b_i^0 + \gamma b_i^1$, where $u'$ is the vector of the $d$ variable anchors.
Its coefficient row $a_i$ is integral with $\|a_i\|_1 \le \max\{m, 2\}$, and $b_i^0, b_i^1 \in \{-1, 0, 1\}$.
For any $d$ constraints whose coefficient rows form a nonsingular matrix $M$, solving them with equality gives the table
\[
    u_M(\gamma) = M^{-1}(b_M^0 + \gamma b_M^1).
\]
The slack of constraint $i$ at this table is $b_i^0 + \gamma b_i^1 - a_i u_M(\gamma)$.
Multiplying it by $\det M$ gives an affine function $c_i^0 + c_i^1 \gamma$, whose coefficients are, up to sign, integer determinants of order $d + 1$ formed from $M$, the row $a_i$, and the right-hand sides.
The rows of these determinants have norms at most $m + 2$, so Hadamard's inequality gives
\[
    |c_i^0|, |c_i^1| < (m + 2)^{d + 1} < H.
\]
The same bound covers the minors and cofactors of these constraint matrices.
A slack that is not identically zero can vanish only at a rational $\gamma$ whose numerator and denominator have absolute value less than $H$.
We call such roots \emph{critical values} of the opposing budget.
For player $A$, where $\gamma = t$ is the budget ratio, a slack that is not constant in $t$ vanishes only at $t = -c_i^0/c_i^1$.
For player $B$, substituting $\gamma = 1/t$ and multiplying by $t > 0$ gives the root $t = -c_i^1/c_i^0$.
Both roots have reduced denominator at most $H$; we call them the \emph{critical ratios}.
If the relevant coefficient is zero, the slack has no positive root unless it is identically zero, and no division is needed.
These bounds hold for every label table and score threshold, so they are available before either distribution is computed.

The root bound lets us continue a feasible vertex as the opposing budget changes.
Suppose $u_M(\gamma_0)$ is feasible and $\gamma_0$ is not critical.
Every constraint slack is affine in $\gamma$, so it is either identically zero or strictly positive at $\gamma_0$.
If no critical value lies strictly between $\gamma_0$ and $\gamma_1$, no positive slack can become negative before $\gamma_1$.
By continuity, every slack is nonnegative at $\gamma_1$, even if $\gamma_1$ itself is critical.
Thus $u_M(\gamma_1)$ is feasible.
The remaining construction selects such an interval and finds a defining matrix $M$ without enumerating the score constraints.

\subsection{Choosing the auxiliary budget ratio}
\label[appendix]{app:general:classifier}

Construct the integer $j = \lfloor Q_g b\rfloor$ and the rationals
\[
    b_- = j/Q_g, \qquad b_+ = (j + 1)/Q_g, \qquad b_\circ = (2j + 1)/(2Q_g).
\]
The closed interval $[b_-, b_+]$ contains $b$, has width $1/Q_g \le 1/(4H^2)$, and therefore contains at most one rational with denominator at most $H$.
Call it $\zeta$ if it exists.
The midpoint $b_\circ$ has reduced denominator $2Q_g > H$.
If $\zeta$ exists, its distance from $b_\circ$ is less than $1/(2\operatorname{denom}(\zeta)^2)$, so by Legendre's theorem it is a convergent of the continued fraction of $b_\circ$.
A later convergent with denominator at most $H$ would be closer to $b_\circ$ and hence also in $[b_-, b_+]$, contradicting uniqueness.
Thus compute the last convergent of $b_\circ$ with denominator at most $H$ and test its membership in the \emph{closed} interval.

The rule below selects $\rho$ and, when an anchor table must change, a target ratio $t$ on the other side of $b$.
We will continue the anchors from $\rho$ to $t$, so that the resulting opposing budget exceeds the one required at $b$.
After normalizing the player's budget to one, the opposing budget at $t$ is $t$ for player $A$ and $1/t$ for player $B$ when $t > 0$.
Apply the following cases in their displayed order.
\[
\begin{array}{c|c|c|c}
\text{condition}&\rho&\text{table changed}&t\\\hline
b = b_-&b_-&\text{none}&--\\
\zeta\text{ exists}, \ b = \zeta&\zeta&\text{none}&--\\
\zeta\text{ exists}, \ b < \zeta&b_-&A&\zeta\\
\zeta\text{ exists}, \ b > \zeta&b_+&B&\zeta\\
\zeta\text{ absent}, \ b = b_\circ&b_\circ&\text{none}&--\\
\zeta\text{ absent}, \ b > b_\circ&b_\circ&A&b_+\\
\zeta\text{ absent}, \ b < b_\circ&b_\circ&B&b_-
\end{array}
\]
Every auxiliary budget ratio selected by this rule lies strictly between zero and one.
For example, $b < \zeta$ cannot have $b_- = 0$, since zero and $\zeta$ would be two rationals with denominator at most $H$ in $[b_-, b_+]$; the same argument excludes $b_+ = 1$ when $b > \zeta$.
If $\zeta$ is absent, neither endpoint of the interval is zero or one.
The selected ratio is represented as $\rho = p/q$ with known integers $0 < p < q \le \overline q$, which need not be relatively prime.
In a case that changes a table, the reduced denominator of $\rho$ is greater than $H$, so the opponent's budget $\beta_0$, which is $\rho$ for player $A$ and $1/\rho$ for player $B$, is not critical.
No critical ratio lies strictly between this $\rho$ and $t$.
The ratio $t$ gives $\gamma = t > b$ when $A$'s anchor table changes and $\gamma = 1/t > 1/b$ when $B$'s anchor table changes, provided $t > 0$ in the latter case.
The case $t = 0$, which can occur only for player $B$, is handled in \cref{app:general:initializer} before a reciprocal is formed.

The input $b$ is used only in the initial quotient and in a constant number of comparisons with constructed rationals of $O(\log H)$ bits.
The continued-fraction computation uses $O(\log H)$ real-RAM operations and $\softO(\log^2 H)$ bit operations on integers of that length.

\subsection{Computing the score threshold and a feasible starting point}
\label[appendix]{app:general:initializer}

The score dynamic program processes the battlefields in order and stores, for every attainable partial score, the least partial anchor sum under the normalized anchor table $u$, together with the label that attains it.
After the last battlefield, $h_*$ is the largest score whose least anchor sum is strictly less than $\beta_0$; it exists because the tuple of zero labels has anchor sum $0 < \beta_0$.
This takes $O(mKS)$ arithmetic operations on anchor sums of up to $O(m\Lambda)$ bits.
By the definition of $h_*$, the table $u$ lies in $\mathcal P(\beta_0)$.

Let $s_{\max} = \sum_j s_{j, K - 1}$.
If $h_* = s_{\max}$, no tuple has score greater than $h_*$, and the algorithm returns the all-zero anchor table, as allowed by \cref{lem:general:transfer}.
In particular, this case occurs whenever the table of player $B$ changes with $t = 0$.
Indeed, the selection rule then chooses $\rho = 1/Q_g$, every bid of player $B$ before normalization is at most $\rho$, and any choice of one unnormalized anchor per battlefield sums to at most $m\rho < 1$, below player $A$'s budget.
All tuples are therefore strictly affordable, and $h_* = s_{\max}$.
Thus, if a tuple has score greater than $h_*$ when $B$'s table changes, then $t > 0$.

In the remaining case, set
\begin{equation}
    N = 4mHpq, \qquad \tau = m/N, \qquad
    \gamma_0 = \beta_0 - \tau, \qquad
    u^0_{jl} = \lfloor Nu_{jl}\rfloor/N.
    \label{eq:general:initializer}
\end{equation}
Both $N\rho = 4mHp^2$ and $N/\rho = 4mHq^2$ are integers, even when $p$ and $q$ are not relatively prime.
Rounding down preserves the zero anchors, the nondecreasing order of the anchors, the bounds $0 \le u^0_{jl} \le 1$, and the budget of every row.
The anchor sum of a tuple with score greater than $h_*$ decreases by less than $m/N = \tau$, so it stays above $\gamma_0$.
Hence $u^0 \in \mathcal P(\gamma_0)$.

By \cref{app:general:roots}, every critical value of $\gamma$ is a rational whose numerator and denominator have absolute value less than $H$.
Write $\beta_0 = p_\beta/q_\beta$, using $(p, q)$ or $(q, p)$ as appropriate.
For every critical value $p'/q'$ different from $\beta_0$,
\[
    |\beta_0 - p'/q'| \ge 1/(q_\beta H) \ge 4\tau.
\]
Since $\beta_0$ is not critical, no value in $[\gamma_0, \beta_0]$ is critical; moreover, $\gamma_0 > 0$.
This small decrease of the threshold lets the computation start from anchors with the common denominator $N$ while keeping all static and score constraints.

\subsection{Finding a defining constraint basis}
\label[appendix]{app:general:basis}

We now find a vertex of $\mathcal P(\gamma_0)$, starting from $u^0$.
Number the $d$ variable anchors as coordinates $1, \ldots, d$.
At stage $j$, the algorithm retains $j$ linearly independent constraints that are tight at the current point and whose coefficient columns $1, \ldots, j$ form an invertible matrix $M_j$.
The current point keeps the coordinates after $j$ at their values in $u^0$ and solves the retained equations for the first $j$ coordinates; at stage zero it is $u^0$.
The algorithm moves from this point in the direction in which coordinate $j + 1$ has velocity one, the later coordinates have velocity zero, and the first $j$ velocities keep the retained constraints tight.
Every slack is affine along this ray, and the feasible part of the ray is a closed interval of length at most one, because coordinate $j + 1$ is at most one.
At its right endpoint, some constraint whose slack decreases along the ray is tight; otherwise a further step would be feasible.
The derivative of this slack is, up to sign, the Schur complement of the new constraint row in column $j + 1$, so it is nonzero, and appending the row keeps the retained rows independent, even if the step length is zero.
After $d$ stages, $d$ independent constraints are tight at a feasible point, which is therefore a vertex.
No full-dimensionality assumption is used.

To find the endpoint of a ray, first scan the static constraints once.
This scan costs $C_s = O(d + mT)$ and gives the largest step $\lambda_s$ that these constraints allow, together with a static constraint that is tight there.
At $\lambda_s$, test the score constraints with the score dynamic program over all scores $0, \ldots, S$; they hold exactly when the least anchor sum at each attainable score greater than $h_*$ is at least $\gamma_0$.
This test at a fixed rational point costs $C_r = \softO(mKS)$ arithmetic operations.
If it passes, the static constraint found by the scan is the required tight constraint.
Otherwise, keep the bracket $[0, \lambda_s]$: its left endpoint is feasible, while its right endpoint violates a score constraint.
Every point in this bracket satisfies the static constraints, so later tests need only check the score constraints.

Within this bracket, each anchor is affine in the step length.
Process the score dynamic program one battlefield at a time.
Once the minimizing choices for the preceding battlefields are fixed on the current open bracket, each attainable target score has at most $K$ candidate predecessors, each giving an affine function of the step length.
Enumerate the pairwise crossings of these functions in the bracket, and sort and deduplicate them.
A binary search through the sorted crossings and the bracket endpoints, using the feasibility test at each tested point, finds adjacent points of which the left is feasible and the right is not; the bracket shrinks to the open interval between them.
On this interval no ordering of candidates changes, so evaluating at an interior point identifies each minimizing affine function and its predecessor.
Proceed to the next battlefield.
After all battlefields, apply the same procedure to the crossings of the least anchor sum at each attainable score greater than $h_*$ with $\gamma_0$.

On the final open interval, all predecessor choices and all comparisons with $\gamma_0$ are fixed.
Its right endpoint is infeasible and its left endpoint is feasible, so by continuity the interval is infeasible and the endpoint of the ray is its left endpoint.
The fixed predecessors give a tuple with score greater than $h_*$ whose score constraint is violated on the interval.
This constraint is tight at the left endpoint, and its slack decreases along the ray, so it is the required tight constraint.
Ties between identical affine functions are broken by a fixed order; equality in a score constraint is feasible.
This procedure also covers repeated crossings and a step length of zero.
Each battlefield uses $O(SK^2)$ crossings and $O(\log(mSK + 2))$ feasibility tests.
Thus finding one ray endpoint costs
\begin{equation}
    \softO(C_s + mC_r + mSK^2).
    \label{eq:general:ray-cost}
\end{equation}
Only one batch of crossings and one dynamic-programming table need to be stored at a time.

To avoid recomputing an inverse at each stage, the algorithm maintains $\det M_j$ and the adjugate $\operatorname{adj} M_j$.
When $M_{j + 1}$ is obtained from $M_j$ by appending a column $\mathbf g$, a row $\mathbf h^\top$ and a corner entry $\kappa$,
\[
    \det M_{j + 1} = \kappa\det M_j - \mathbf h^\top(\operatorname{adj} M_j)\mathbf g,
\]
\[
    \operatorname{adj} M_{j + 1} = \begin{pmatrix}
      \bigl(\det M_{j + 1}\operatorname{adj} M_j + (\operatorname{adj} M_j\,\mathbf g)(\mathbf h^\top\operatorname{adj} M_j)\bigr)/\det M_j&-\operatorname{adj} M_j\,\mathbf g\\
      -\mathbf h^\top\operatorname{adj} M_j&\det M_j
    \end{pmatrix}.
\]
These divisions are exact integer divisions, and $\det M_0 = 1$ for the empty matrix.
The matrix-vector products, these updates, and the right-hand sides of the retained constraints cost $O(j^2 + jd)$ operations at stage $j$, $O(d^3)$ in total, with $O(d^2)$ storage for the matrices.
The current point at stage $j$ has denominator dividing $N\det M_j$, and the velocities have denominator dividing $\det M_j$.
After clearing common denominators, crossing points are ratios of integer coefficients of the affine differences being compared.
The bounds on minors and on the number of terms in each affine sum control the bit lengths of these coefficients, and the tested midpoints combine two such crossing points.
Thus all operands have bit length
\[
    \Omega = O(\log N + d\log(m + 2) + \log(d + m + K + S + T + 2)) = O(\Lambda).
\]
Temporary points are discarded after testing, and the current point of each new stage is solved afresh from the original constraint rows and $u^0$, so denominators do not accumulate across stages.

\subsection{Evaluating the basis at the endpoint}
\label[appendix]{app:general:continuation}

The final basis gives the table $u_M(\gamma) = M^{-1}(b_M^0 + \gamma b_M^1)$, which is feasible at $\gamma = \gamma_0$.
No value in $[\gamma_0, \beta_0]$ is critical (\cref{app:general:initializer}), and no value strictly between $\beta_0$ and the value of $\gamma$ corresponding to $t$ is critical (\cref{app:general:classifier}).
By the continuation argument in \cref{app:general:roots}, every slack is nonnegative at the target threshold.
The table evaluated there therefore satisfies every static and score constraint.
The algorithm also verifies these constraints at the vertex and at the endpoint, with one static scan and one evaluation of the score dynamic program each.

For $A$'s anchor table the resulting threshold is $\gamma = t > b$, and for $B$'s anchor table it is $\gamma = 1/t > 1/b$.
This strict inequality is the hypothesis of \cref{lem:general:transfer}, even if some anchors coincide at the endpoint.
Each entry's denominator divides $\operatorname{denom}(t)\det M$ for $A$ and $\operatorname{numer}(t)\det M$ for $B$.
The scale $N$ does not appear in this expression, because the right-hand sides of the constraints contain only constants and the threshold.
Each entry of the new anchor table therefore has bit length $O(\log H + \log\overline q + \log(d + m + K + S)) = O(\log H)$.
This proves \cref{lem:general:endpoint}; every row still uses the same label on each battlefield.

\section{Output, sampling, and complete complexity}
\label[appendix]{app:general:cost}

\subsection{Proof of Lemma~\ref{lem:general:lift}}
\label[appendix]{app:lem:general:lift:proof}

\begin{proof}
Fix an allocation $x$ on the reduced battlefields, and let $W$ be uniform on $[0, 1)$, so that $Z(W)$ equals $k x_a$ for a uniformly random index $a$.
We first compare $X_j / p_j$ with $Z(W)$.
For a threshold $\theta$, the set $\{t : Z(t) < \theta\}$ is an interval $[0, F)$, since $Z$ is nondecreasing, and $F = \Prob(Z(W) < \theta)$.
The same holds for the set $\{t : Z(t) \le \theta\}$, with $\Prob(Z(W) \le \theta)$ in place of $F$.
The essential infimum of $Z$ over an arc is less than $\theta$ exactly when the arc overlaps $[0, F)$ in positive length.
As $\omega$ is uniform, the starting point of the arc is uniform on the circle, so this event has probability $\min\{1, F + p_j\}$ if $F > 0$ and probability zero otherwise.
Thus, for both strict and weak thresholds, the CDF of $X_j / p_j$ exceeds that of $Z(W)$ by at most $p_j$.
By \eqref{eq:prelim:half-tie}, replacing each bid $X_j$ by $p_j Z(W) = k p_j x_a$, with an independent uniform index for each light battlefield, changes the opponent's expected payoff by at most $\sum_{j \text{ light}} v'_j p_j = \sum_{j \text{ light}} (v'_j)^2 / V$.

Now fix a deviation $y$ on the original battlefields with $\sum_j y_j \le c$.
Against the bids $k p_j x_a$, the opponent's expected payoff on the light battlefields is
\[
    \sum_{j \text{ light}} v'_j \cdot \frac1k \sum_{a = 0}^{k - 1} h(y_j, k p_j x_a)
    = V \sum_{j \text{ light}} p_j \cdot \frac1k \sum_{a = 0}^{k - 1} h(y_j / p_j, k x_a),
\]
since $v'_j = V p_j$ and dividing both bids by $p_j$ preserves their comparison.
Define a random bid $Y$ by choosing light battlefield $j$ with probability $p_j$ and returning $y_j/p_j$.
The right-hand side is $V$ times the expected share of $Y$ against $k x_a$ for an independent uniformly random index $a$.
The mean of $Y$ is $\sum_{j \text{ light}} y_j$.
Let $q_0 = 0$ and, for $1 \le l < k$, let $q_l = \inf\{t : \Prob(Y \le t) \ge l / k\}$ be the quantile of $Y$ at level $l / k$.
Their average is a lower Riemann sum for the mean of $Y$, so $\frac1k \sum_l q_l \le \sum_{j \text{ light}} y_j$.
The CDF of the uniform distribution on $q_0, \ldots, q_{k - 1}$ differs from that of $Y$ by at most $1 / k$, for both strict and weak thresholds.
Replacing $Y$ by a uniformly random $q_l$ therefore changes the payoff above by at most $V / k$.

It remains to realize the resulting payoff by deviations in the reduced game.
For each $s \in \{0, \ldots, k - 1\}$, consider the reduced deviation that bids $q_{(a + s) \bmod k} / k$ on representative battlefield $a$ and $y_j$ on each battlefield kept individually.
Its total bid is at most $\sum_j y_j \le c$, so it is feasible, and it does not depend on the player's allocation.
Each representative battlefield has value $V / k$, so the average payoff of these $k$ deviations on the representative battlefields is
\[
    \frac1k \sum_{s = 0}^{k - 1} \frac Vk \sum_{a = 0}^{k - 1} h\bigl(q_{(a + s) \bmod k} / k, x_a\bigr)
    = \frac{V}{k^2} \sum_{a = 0}^{k - 1} \sum_{l = 0}^{k - 1} h(q_l, k x_a),
\]
which is exactly the payoff above after replacing $Y$ by a uniformly random $q_l$.
On the battlefields kept individually, $y$ and these deviations have the same payoff.
Averaging over $x \sim P$, the expected payoff of $y$ against $\mathcal L(P)$ is therefore at most the average payoff of the $k$ reduced deviations against $P$, plus $\sum_{j \text{ light}} (v'_j)^2 / V + V / k$.
Since each of the $k$ reduced deviations is feasible, their average payoff against $P$ is at most $U_{v^{\rm red}}(P; c)$.
Taking the supremum over $y$ proves the first inequality.
The second follows from $\sum_{j \text{ light}} (v'_j)^2 / V \le \max_{j \text{ light}} v'_j \le e / 10$ and $V / k \le 1 / k = e / 8$.
\end{proof}

\subsection{Finite shifts and evaluation of allocations}
\label[appendix]{app:general:sampling}

The shifts defined in \cref{sec:general:lift} are computed exactly by
\begin{equation}
    t_\ell = \left\lfloor \frac{(2 \ell + 1) D_{\rm sh}}{2 S_{\rm out}} \right\rfloor,
    \qquad
    \omega_\ell = \frac{t_\ell + 1/2}{D_{\rm sh}},
    \qquad 0 \le \ell < S_{\rm out}.
    \label{eq:general:finite-shifts}
\end{equation}
Here $t_\ell$ is the index of the cell containing $(\ell + 1/2)/S_{\rm out}$, and $\omega_\ell$ is its midpoint.

\begin{proof}[Proof of \cref{lem:general:finite-shifts}]
Fix a reduced allocation, a light battlefield and a threshold.
The shifts for which the reconstructed bid is below (or at most) the threshold form an empty, full, or cyclic interval whose endpoints are multiples of $1 / D_{\rm sh}$.
Replace this event by the union of the half-open intervals $[a / D_{\rm sh}, (a + 1) / D_{\rm sh})$ whose interiors lie in it.
The union is again an empty, full, or half-open cyclic interval, and it differs from the event only at interval boundaries.
Under a continuous uniform shift, the event therefore has probability equal to the length of the union.
The selected shifts $\omega_\ell$ are midpoints of the half-open intervals, so they lie in the event exactly when they lie in the union.
Moreover, $\omega_\ell$ lies in the union exactly when the evenly spaced point $(\ell + 1/2) / S_{\rm out}$ does, since both lie in the same half-open interval, even when the point is an interval boundary.
Under the finite shifts, the event therefore has probability equal to the fraction of evenly spaced points in the union.
We finish by showing that this fraction differs from the length of the union by at most $1/S_{\rm out}$.

Recall that the light battlefields have proportions $p_j = w'_j/W_G$ and that $D_{\rm sh} = kW_G$.
For an integer $a \in [0, D_{\rm sh}]$,
\[
    t_\ell < a\quad\Longleftrightarrow\quad
       \ell + 1/2 < S_{\rm out}a/D_{\rm sh}.
\]
For any $r \in [0, S_{\rm out}]$, the number of indices $\ell$ with $\ell + 1/2 < r$ differs from $r$ by at most one half.
Subtracting two such counts shows that, for integers $0 \le a \le a' \le D_{\rm sh}$, the fraction of indices $\ell$ with $a \le t_\ell < a'$ differs from $(a' - a)/D_{\rm sh}$ by at most $1/S_{\rm out}$.
A cyclic interval that wraps around zero is the complement of an ordinary one and satisfies the same bound.
Since $\omega_\ell \in [a/D_{\rm sh}, a'/D_{\rm sh})$ exactly when $a \le t_\ell < a'$, this gives the required bound for the half-open union.
It holds regardless of repeated shifts, zero bids, or equal representative bids, and averaging over rows preserves it.

\end{proof}

After the transfer, the algorithm sorts the bids on the representative battlefields separately in every row, keeping references to their battlefields and breaking ties by battlefield index.
Sorting the labels alone would not suffice, since the same label may denote different anchors on different battlefields.
For each row, the stored order of the representative bids and the bids on the battlefields kept individually come from the same allocation.

The arc minima can be evaluated in one ordered sweep.
Multiply positions on the circle by $2D_{\rm sh}$, and store the cumulative arc boundaries
\[
    \sigma_0 = 0, \qquad \sigma_j = 2k\sum_{h \le j}w'_h, \qquad \sigma_g = 2D_{\rm sh},
\]
where the light battlefields are numbered $1, \ldots, g$ in their original order.
For a selected shift, compute $t_\ell$ once and set $o = 2t_\ell + 1$, which is the position of $\omega_\ell$ in these units.
Before reduction modulo $2D_{\rm sh}$, the arc of light battlefield $j$ has endpoints $L_j = o + \sigma_{j - 1}$ and $R_j = o + \sigma_j$.
These endpoints are odd integers, whereas the boundaries of the intervals on which $Z$ is constant are multiples of $2W_G$ and hence even.
These $k$ intervals become bins of width $2W_G$ in the new units.
Maintain a bin counter $a$ and the next bin boundary, initially $a = 0$ and $2W_G$, and advance them while $L_j$ exceeds the next boundary.
Across the whole draw there are fewer than $2k$ advances.
If $L_j < 2D_{\rm sh} < R_j$, the arc wraps around zero and contains the smallest representative bid $x_{(0)}$, so use it; otherwise use the bid $x_{(a)}$ of bin $a$ if $a < k$ and $x_{(a - k)}$ otherwise.
Multiply this bid by $kw'_j/W_G = kp_j$.
Since $Z$ is nondecreasing, a nonwrapping arc attains its minimum in the leftmost bin it touches.
This evaluates \eqref{eq:general:circle} exactly for each selected row $i$ and shift $\omega_\ell$, using only the stored order and one scan through the bins.
Finally, multiply all coordinates by the original budget of this player.

There are $T$ rows and $S_{\rm out}$ shifts, both powers of two, so exactly $\log_2(TS_{\rm out})$ independent random bits choose a row and a shift independently and uniformly.
In the real-RAM model, each random bit is obtained by comparing a fresh uniform sample with $1/2$.
The two players use independent random bits.
Since $k < n$ whenever representative battlefields are used and $D_{\rm sh} = O(n/e^2)$, computing the shift, choosing the row, sweeping through the bins, and evaluating the coordinates cost
\[
    O(n + k + \log(D_{\rm sh} + 1) + \log T + \log S_{\rm out}) = O(n + \log(1/e)).
\]
Without representative battlefields, $S_{\rm out} = 1$, and the shift computation and the sweep are omitted.

\subsection{One battlefield and equal budgets}
\label[appendix]{app:general:boundary}

For one battlefield, both players spend their full budgets there, giving an exact equilibrium.
For equal budgets and $n > 1$, choose $e = 2^{-a}$ with $\varepsilon/2 < e \le \varepsilon$ as in \cref{sec:general:reduction}, and normalize the common budget to one.
Round the values with the first rounding of \cref{sec:general:reduction}, obtaining positive values $v'_j$ with common denominator $O(n/e)$ and $\Delta_{\rm eq} = \sum_j |v'_j - v_j| \le e/128$.
If the largest rounded value is at least $1/2$, both players bid one on that battlefield and zero elsewhere, and each receives payoff $1/2$.
A deviation that bids less than one on that battlefield loses it and wins at most the remaining value, which is at most $1/2$; a deviation that bids one there has no budget left and coincides with the equilibrium allocation.
Thus this is an exact equilibrium of the game with values $v'$.

Otherwise, every rounded value is less than $1/2$, so condition~\eqref{eq:prg:promise} holds with $b = 1$, and the General Lotto marginals of both players are $\Unif[0, 2v'_j]$.
Apply the grouping construction of \cref{sec:prg:reduction} to the lengths $2v'_j < 1$, whose sum is two.
It gives two or three groups, each of length at most one.
For two groups, both lengths are one, and we use $(U, 1 - U)$ with $U \sim \Unif[0, 1]$; for three groups, we use \cref{lem:prg:three} with $p = 1$.
The result is a mixture of at most seven line segments with rational mixture probabilities, and every allocation on these segments spends exactly one.
Splitting each group proportionally gives the uniform marginals on $[0, 2v'_j]$, so both players using this distribution form an exact equilibrium by \cref{sec:prg:security}.
Now replace the uniform parameter on each segment by a uniform choice among the $N_{\rm mid} = 32/e$ midpoints $(i + 1/2)/N_{\rm mid}$, $0 \le i < N_{\rm mid}$.
For a nonconstant coordinate, the set of parameters at which the coordinate is at most a threshold is an interval, so its CDF changes by at most $1/N_{\rm mid}$; a constant coordinate is unchanged.
After mixing the segments, each marginal CDF has the same error bound.
Both players still use the same distribution, so by symmetry each expected payoff is $1/2$, and by \cref{lem:prelim:cdf} no deviation obtains more than $1/2 + 1/N_{\rm mid}$.
Thus each regret in the game with values $v'$ is at most $1/N_{\rm mid} = e/32$, and restoring the original values gives regret at most $e/32 + 2\Delta_{\rm eq} \le 3e/64 < \varepsilon$.

Store the group labels and proportional coefficients, the at most seven segment endpoints and mixture probabilities, the number of midpoints, and the original common budget.
The rational entries other than the original budget have $O(\log(n/e))$ bits: the rounded values and group lengths have that encoding length, and each segment formula uses only a constant number of arithmetic operations.
No table of $nN_{\rm mid}$ coordinates is expanded.
The mixture probabilities have a common denominator of $O(\log(n/e))$ bits.
In the real-RAM model, a uniform sample and a constant number of comparisons with the cumulative mixture probabilities choose the segment, after which random bits choose the midpoint uniformly.
In the bit model, draw a binary integer with the bit length of the common denominator, and reject it if it is not smaller than the denominator.
Each trial succeeds with probability at least $1/2$, so choosing a segment with exactly its prescribed probability uses expected $O(\log(n/e))$ bits and bit operations.
Every completed draw is feasible, and computing and writing its coordinates costs $\softO(n\, \Rat(L_0 + \log(n/e)))$ bit operations.

\subsection{Implementing arithmetic in the stated models}
\label[appendix]{app:general:arithmetic}

We implement the operations used above from the real-RAM primitives of \cref{sec:prelim:computation}, together with indexed storage and control.
In particular, integer quotients and continued fractions are computed by arithmetic and comparisons; logarithms and exponentials in the regret proof are not evaluated by the algorithm.
To construct $\lfloor x/a\rfloor$ for $x \ge 0, a > 0$, repeatedly double the pair $(a, 1)$ until its first entry exceeds $x$.
Starting with remainder $x$ and quotient zero, traverse the pairs backwards: whenever the first entry is at most the current remainder, subtract it from the remainder and add the second entry to the quotient.
After each pair is processed, the remainder is smaller than its first entry.
This gives the quotient and remainder in $O(1 + \log(\lfloor x/a\rfloor + 1))$ operations without reading the binary encoding of $x$.
Powers of two, exponent indices and loop guards are constructed by doubling or repeated squaring.

For the Euclidean algorithm on constructed integers of $O(\Lambda)$ bits, there are $O(\Lambda)$ quotients because the convergent denominators grow at least as fast as the Fibonacci sequence.
The product of the positive quotients is at most $2^{O(\Lambda)}$, so computing them by the preceding doubling procedure costs $O(\Lambda)$ operations in total.
Convergent updates have the same real-RAM bound.
Their operands have $O(\Lambda)$ bits, yielding $\softO(\Lambda^2)$ bit operations and $\softO(\Lambda)$ temporary bits under the arithmetic convention of \cref{app:computation:bits}.
These integers are constructed by the algorithm; this procedure does not require recovering the binary encoding of the original ratio $b$ in the real-RAM model.

To implement the rounded arithmetic of \cref{app:general:tree}, multiply the $p_{\rm mant}$-bit mantissas and normalize by constructed powers of two.
For addition compare the encoded exponents; if their gap exceeds $p_{\rm mant} + 2$, dropping the smaller term meets the relative rounding bound, and otherwise align only $O(p_{\rm mant})$ positions.
Quotients recover at most $p_{\rm mant}$ significant digits.
Comparisons first use the encoded exponents and then align the mantissas when necessary, without expanding a small weight into a rational with a large denominator.
Each such arithmetic operation costs $\softO(1)$ real-RAM operations at the chosen precision and $\softO(p_{\rm mant})$ bit operations.
Each power evaluation uses $O(\log(G + 1))$ of these operations.
A signed exponent of magnitude at most $\Lambda + G + p_{\rm mant} + 8$ suffices, and its encoding has logarithmic length.

In the bit model all exact integer and rational operations are charged at the actual operand lengths using $\Rat(w) = \softO(w)$.
Input rounding computes each $\lceil Mv_j\rceil$ independently.
Balanced rational summation combines disjoint input subsets at each level; the sum of operand lengths at a level is $O(L_0 + n\log(n/e))$.
Thus original value processing costs $\softO(L_0 + E + n)$ bit operations, rather than repeatedly charging the longest input rational in every battlefield.
Selecting the auxiliary budget ratio uses the original ratio $b$ in one quotient and a constant number of comparisons with constructed rationals of $O(\log H)$ bits.
Computing the dyadic $e$ uses its binary input in the bit model, or repeated comparisons in the real-RAM model.
The original budgets are retained as shared exact scales until output.

\subsection{Computation and simultaneous storage}
\label[appendix]{app:general:ledger}

We first account for the stages using the bid encoding length $\Lambda$, then substitute $\Lambda = \softO(m/z)$ to obtain the theorem's bounds.
All bounds include the sequential computations for both budgets $c \in \{1, \rho\}$.
We write $R$, $B$, and $W$ for real-RAM operations, bit operations, and simultaneously stored bits, with subscripts indicating the stage.
The number of epochs is $\Theta(\Lambda/z)$, the epoch length and summary size are $\Theta(1/z)$, and $T = \Theta(\Lambda/z^2)$.
Fresh local powers, exact summary sums and rounded losses cost $\softO(m\Lambda/z^3)$ real-RAM operations and $\softO(m\Lambda^2/z^3)$ bit operations.
Over all $T$ iterations, the two optimizations performed in each iteration cost $\softO(m\Lambda/z^4)$ real-RAM operations and $\softO(m\Lambda^2/z^4)$ bit operations, using the balanced tree and the resource bit lengths at each level from \cref{app:general:profit}.

Quantile queries, valid-code queries, and the interval updates at epoch boundaries cost $\softO(m\Lambda^2/z^2)$ real-RAM operations and $\softO(m\Lambda^3/z^2)$ bit operations.
The additional factor in the number of bit operations accounts for arithmetic on $\Lambda$-bit codes and for continued-fraction queries using the Euclidean algorithm.
The sparse trees use $\softO(m\Lambda^2/z^2)$ bits; each traversal's endpoint stack uses $O(\Lambda^2)$ temporary bits.
The $T$ closed candidate rows use $O(mT\Lambda)$ bits.
Response coordinates after adding the open-bid increment can have $O(m\Lambda)$ bits; checking the sum of the bids against the budget contributes $\softO(m^2T\Lambda)$ bit operations and $O(m^2\Lambda)$ temporary bits.
Decoding candidate coordinates contributes $\softO(mT\Lambda^2)$ bit operations.
Using $\Lambda \ge m$ absorbs both terms into
\begin{equation}
\begin{aligned}
    R_{\rm src}& = \softO(m\Lambda^2/z^2 + m\Lambda/z^4), \\
    B_{\rm src}& = \softO(m\Lambda^3/z^2 + m\Lambda^2/z^4), \\
    W_{\rm src}& = \softO(m\Lambda^2/z^2).
\end{aligned}
\label{eq:general:source-cost}
\end{equation}

Next decode and sort every bid on each battlefield to form the $K$ quantile blocks, retain the complete rows of labels, and check every stored row's budget.
The trees used during the iterative computation can be discarded before this step; the bounds below include decoding the stored bids again.
Normalize the $mK$ anchors and compute the score threshold $h_*$ using the strict budget test, before rounding anchors down to multiples of $1/N$.
With $\lambda_N = O(\log N) = O(\Lambda)$, the decoding, quantization, and initialization costs satisfy
\begin{align*}
    R_{\rm rec}& = \softO(mT\Lambda + mKS + mT + d\lambda_N + 1), \\
        B_{\rm rec}& = \softO(mT\Lambda^2 + m^2KS\Lambda + m^2T\Lambda + d(\Lambda + \lambda_N)\lambda_N +
        \lambda_N), \\
    W_{\rm rec}& = \softO(mT\Lambda + Sm\Lambda + mS\log(K + 1) + mK\Lambda + d\lambda_N + \Lambda).
\end{align*}
The anchor sums in the computation of $h_*$ have $O(m\Lambda)$ bits.
For an anchor $u_{jl} = P/Q$, rounding it to a multiple of $1/N$ computes $\lfloor NP/Q\rfloor$ using operands of bit length $O(\Lambda + \lambda_N)$.
Using $K = O(1/z)$, $S = O(m/z)$, $d = O(m/z)$, $T = \Theta(\Lambda/z^2)$ and $\Lambda \ge m$ gives
\[
    R_{\rm rec} = \softO(m\Lambda^2/z^2), \qquad
    B_{\rm rec} = \softO(m\Lambda^3/z^2), \qquad
    W_{\rm rec} = \softO(m\Lambda^2/z^2).
\]

For the endpoint, use $C_s = O(d + mT)$ and $C_r = \softO(mKS)$ from \cref{app:general:basis}.
The complete arithmetic count, including all static scans and final checks of the defining inequalities, is bounded up to logarithms by
\[
    F_{\rm end} = d^3 + dC_s + dmC_r + dmSK^2 + (C_s + C_r).
\]
These computations take $\softO(F_{\rm end}\Omega)$ bit operations.
The simultaneous storage required is
\[
    \softO((d^2 + C_r + SK^2)\Omega + mT\log(K + 1) + mK\Lambda).
\]
There are $d$ rays, with one full static scan and $O(m)$ crossing batches per ray, and the basis computations use $O(d^3)$ arithmetic operations in total.
With $\Omega = O(\Lambda)$ this becomes
\begin{align*}
    R_{\rm end}& = \softO(m^2\Lambda/z^3 + m^4/z^3 + m^3/z^4), \\
    B_{\rm end}& = \softO((m^2\Lambda/z^3 + m^4/z^3 + m^3/z^4)\Lambda), \\
    W_{\rm end}& = \softO((m^2/z^2 + m/z^3)\Lambda + m\Lambda/z^2).
\end{align*}

Constructing the parameters and selecting $\rho$ cost $\softO(n + \Lambda + m/z)$ real-RAM operations and $\softO(L_0 + E + n + \Lambda^2 + m\Lambda + 1/z)$ bit operations.
Computing the final anchor tables and battlefield maps and sorting the representative bids costs $\softO(n + mT + mK)$ real-RAM operations and
\[
    \softO\bigl(n\log(n/e) + mT\log(mK + 1) + (Tk + mK)\Lambda\bigr)
\]
bit operations, where the $Tk$ term is omitted if no representative battlefields are used.
The sorting compares the final bids and stores battlefield references rather than copies of their rational coordinates.

The following table substitutes $\Lambda = \softO(m/z)$ into the same construction's component bounds.
Input work is listed separately and all entries suppress logarithms.
\begin{center}
\begin{tabular}{lccc}
\toprule
Stage&Real-RAM operations&Bit operations&Working bits\\
\midrule
Distributions at $\rho$&$m^3/z^4 + m^2/z^5$&$m^4/z^5 + m^3/z^6$&$m^3/z^4$\\
Quantization and initialization&$m^3/z^4$&$m^4/z^5$&$m^3/z^4$\\
Basis and endpoint&$m^4/z^3 + m^3/z^4$&$m^5/z^4 + m^4/z^5$&$m^3/z^3 + m^2/z^4$\\
Selecting $\rho$&$m/z$&$m^2/z^2$&$m/z$\\
Final sorting and maps&$n + m^2/z^3$&$n + m^3/z^4$&description size\\
\bottomrule
\end{tabular}
\end{center}
Since $mz < 1$, the total is
\[
    \softO(n + m^2/z^5), \qquad
    \softO(L_0 + E + m^3/z^6), \qquad
    \softO(L_0 + E + m^3/z^4)
\]
for real-RAM operations, bit operations, and working bits, respectively.
Since $\varepsilon/2 < e \le \varepsilon$, we have $z = \Theta(\varepsilon)$ and $m \le \min\{n, 18/e\} \le 36n_*$; substituting gives the computation and space bounds in \cref{thm:general}.

For the storage bound, retain the input and first table of candidate allocations while computing the other player's distribution, releasing the trees and epoch arrays of each computation after use.
During quantization, stored coordinates, sort indices, both label tables and anchor tables coexist; finish the checks on the unrounded coordinates before releasing the stored coordinates.
The endpoint computation retains both label tables and the unchanged anchor tables, storage for one constraint basis, and one score dynamic-programming table and batch of crossings.
Release these workspaces before sorting the final representative bids.
The stored strategy description is included throughout, and no table enumerating all $TS_{\rm out}$ allocations is constructed.
The largest simultaneous storage requirement in this schedule gives the claimed working-space bound.

\subsection{Description length and rational output}
\label[appendix]{app:general:description}

For each player, the output stores a table of $mT$ labels specifying the retained allocation rows and an anchor table containing the $K$ anchors for each of the $m$ battlefields, with the player's budget normalized to one.
Each row specifies its bid on a battlefield by a label in that battlefield's anchor table, so the anchor is stored once even when several rows use it.
When representative battlefields are used, the output also stores $Tk$ references specifying the order of their bids in each row.
For the original battlefields, it records which were kept individually and which were replaced, together with the positive integer weights used in their reconstruction.
Finally, it stores the finite-shift rule and the original budget scales.
Its size is
\begin{align*}
    O\bigl(&L_0 + E + n\log(n/e) + mT\log(K + 1)\\
           & + Tk\log(mK + 1) + mK\Lambda\bigr) = \softO(L_0 + E + n + n_*^2/\varepsilon^3).
\end{align*}
The term involving $k$ is absent if no representative battlefields are used.
The description defines an allocation for each of the $TS_{\rm out}$ pairs consisting of a row index $i$ and a shift index $\ell$; different pairs may give the same allocation.
Enumerating these allocations must evaluate and write $nTS_{\rm out}$ rational coordinates per player, and is a separate operation.

A coordinate computed from one selected row combines an anchor of $O(\Lambda)$ bits, a light battlefield's proportion of $O(\log(n/e))$ bits if needed, and the player's original budget of at most $L_0$ bits.
The bit length of each operand and output coordinate is at most
\[
    W_{\rm emit} = O(L_0 + \Lambda + \log(n/e)).
\]
Thus a complete draw costs
\[
    \softO\bigl(n\, \Rat(W_{\rm emit}) + \log^2(D_{\rm sh} + 1) + \log(1/e)\bigr)
\]
bit operations, including writing every coordinate and generating the prescribed bits.
The shift term is omitted if no representative battlefields are used.
The description stores each original budget once; the sampling bound charges its use in every emitted coordinate.
The cases of one battlefield or equal budgets have the separate output and randomness bounds in \cref{app:general:boundary}.

\end{document}